\documentclass[a4paper,11pt]{article}
\usepackage{jheppub} 
\usepackage{soul}
\usepackage[normalem]{ulem}
\usepackage{accents}
\usepackage{bbold}
\usepackage{physics}
\allowdisplaybreaks[4]
\usepackage[T1]{fontenc}
\usepackage{xcolor}
\usepackage{appendix}

\definecolor{hotpink}{rgb}{1.0, 0.41, 0.71}

\title{\boldmath Charged rotating quantum black holes}

\author[a]{Dyuman Bhattacharya,}
\author[b]{Robie A.~Hennigar,}
\author[a,c]{Robert B.~Mann,}
\author[a,c,d]{Ming Zhang}
 \affiliation[a]{Department of Physics and Astronomy, University of Waterloo,\\
Waterloo, Ontario N2L 3G1, Canada}
\affiliation[b]{Centre for Particle Theory, Department of Mathematical Sciences, Durham University, \\
Durham DH1 3LE, United Kingdom}
\affiliation[c]{Perimeter Institute for Theoretical Physics, \\  Waterloo, Ontario N2L 2Y5, Canada}
\affiliation[d]{Department of Physics, Jiangxi Normal University,\\ Nanchang 330022, China}

\emailAdd{d7bhatta@uwaterloo.ca}
\emailAdd{robie.a.hennigar@durham.ac.uk}
\emailAdd{rbmann@uwaterloo.ca}
\emailAdd{mingzhang@jxnu.edu.cn}
\abstract{We investigate the thermodynamic and holographic properties of charged and rotating quantum black holes in a doubly holographic braneworld setup. These quantum black holes are derived from the anti-de Sitter C-metric and are exact solutions to a semi-classical gravitational theory which incorporates all orders of the backreaction of quantum fields on spacetime. The inclusion of both charge and rotation extends and generalizes previous studies. The thermodynamics and critical behavior of the black holes are examined from the bulk, brane, and boundary perspectives, and we demonstrate that the inclusion of either charge or rotation removes the re-entrant phase transitions seen in the neutral-static case. The critical exponents of the system are calculated using numerical methods and found to differ from the standard mean field theory values for the neutral-static black holes' re-entrant phase transitions, but in agreement with mean-field theory for the phase transitions of the black holes with charge and rotation. Additionally, to test the validity of the semiclassical treatment, we study a mass-gap energy scale $M_{\rm gap}$ to identify regimes where quantum fluctuations of spacetime geometry are expected to become significant and speculate about a connection with weak cosmic censorship gedankenexperiments. We also generalize the quantum Penrose inequality and the quantum reverse isoperimetric inequality to include charge and rotation. Finally, we compute a renormalized gyromagnetic ratio and analyze it in the limit of large backreaction.  
}

\begin{document}

\maketitle

\flushbottom
\section{Introduction}

Three-dimensional asymptotically anti-de Sitter (AdS) quantum black holes \cite{Emparan:2002px, Panella:2024sor} in the braneworld \cite{Randall:1999ee} have attracted widespread attention and discussion. These braneworld black holes \cite{Gregory:2004vt,Gregory:2008rf}, including  quantum Bañados-Teitelboim-Zanelli (BTZ) black holes \cite{Emparan:1999wa,Emparan:1999fd} as well as their rotating \cite{Emparan:1999fd,Emparan:2020znc} and charged counterparts \cite{Climent:2024nuj,Feng:2024uia}, are induced from the four-dimensional AdS C-metric \cite{Kinnersley:1970zw,Plebanski:1976gy,Dias:2002mi,Griffiths:2005qp,Anabalon:2018qfv,Gregory:2019dtq} in Einstein gravity and are localized on tensional end-of-the-world (ETW) branes \cite{Randall:1999vf,Karch:2000ct,Karch:2000gx}. In braneworld holography \cite{deHaro:2000wj}, the ETW brane acts as an infrared cutoff in the bulk and incorporates the ultraviolet cutoff of the conformal field theory (CFT) on the boundary, thus realizing holographic renormalization \cite{Kraus:1999di,Emparan:1999pm,deHaro:2000vlm,Skenderis:2002wp,Papadimitriou:2004ap,Lewandowski:2004yr,Mann:1999pc}. In this braneworld scenario, $\mathrm{AdS}_4$ Einstein gravity governs classical dynamics in the bulk and is localized on a semi-classical higher-curvature modified effective theory representing quantum dynamics on the two-brane \cite{Emparan:2002px}. In other words, classical asymptotically AdS black holes that solve the classical field equations induce codimension-one quantum black holes that exactly solve the semi-classical field equations \cite{Gubser:1999vj,Skenderis:1999nb}, with the holographic CFT incorporating the quantum effects of the planar limit of a large $N$ expansion \cite{Aharony:1999ti}.

As exact solutions to the effective theory, quantum BTZ black holes in the braneworld have been shown to exhibit a wide range of remarkable properties. For example, they explain the origin of extended thermodynamics \cite{Kubiznak:2014zwa,Kubiznak:2016qmn}, where the cosmological constant is treated as a variable \cite{Kastor:2009wy,Dolan:2010ha,Chamblin:1999tk,Kubiznak:2012wp} in terms of the brane tension \cite{Frassino:2022zaz}. They also strengthen the weak cosmic censorship conjecture via quantum backreaction \cite{Frassino:2024fin}. Furthermore, a quantum Penrose inequality—valid to all orders of backreaction—was found, demonstrating the robustness of cosmic censorship even in nonperturbative semiclassical gravity \cite{Frassino:2024bjg}. In the same study, a quantum version of the classical reverse isoperimetric inequality \cite{Cvetic:2010jb,Hennigar:2014cfa,Amo:2023bbo} was also proposed. Furthermore, superradiance induced by backreaction from quantum conformal matter was discovered \cite{Cartwright:2025fay}, leading to a renewed interpretation of the Gubser-Mitra conjecture on black hole stability \cite{Gubser:2000mm}. Other investigations have explored various properties of quantum BTZ black holes, including critical phenomena and thermodynamic geometry \cite{HosseiniMansoori:2024bfi}, thermodynamic topology \cite{Wu:2024txe}, the Kramer escape rate \cite{Xu:2024iji}, the analytic structure of correlators in the dual brane field theory \cite{Cartwright:2024iwc}, and the applicability of swampland constraints \cite{Anastasi:2025puv}. Properties of the quantum black holes on the de Sitter brane have also been studied in \cite{Emparan:2022ijy,Panella:2023lsi,Climent:2024wol}.

Thus far, research on the chemistry of quantum black holes has been confined to either (un)charged or (non)rotating cases. These studies have uncovered many fascinating phenomena, such as the quantum Penrose inequality and the quantum reverse isoperimetric inequality \cite{Frassino:2024bjg}, the construction of doubly holographic thermodynamics \cite{Frassino:2022zaz,Feng:2024uia}, and various phase transitions \cite{Kudoh:2004ub,Frassino:2023wpc,Climent:2024wol}. However, these results do not provide the full picture. In this article, we aim to complete the final piece of the puzzle by investigating the chemical properties of a quantum black hole that possesses both charge and rotation given in section \ref{fji39p4ejr}.

 The unique properties of  quantum black holes  stem from the double holography proposal \cite{Almheiri:2019hni,Geng:2020qvw,Almheiri:2020cfm,Geng:2020fxl,Chen:2020jvn,Ling:2020laa,Neuenfeld:2021wbl,Geng:2021mic,Karch:2022rvr,Myers:2024zhb}, which can be used to study the page curve and black hole information paradox \cite{Hawking:1975vcx,Hawking:1976ra,Page:1993wv,Page:2013dx} based on the quantum extremal  island prescription \cite{Faulkner:2013ana,Engelhardt:2014gca,Penington:2019npb,Almheiri:2019psf,Almheiri:2019qdq,Chen:2020uac,Chen:2020hmv,Hernandez:2020nem,Grimaldi:2022suv,Suzuki:2022xwv,Takayanagi:2025ula} beyond the classical extremal surfaces \cite{Ryu:2006bv,Ryu:2006ef,Hubeny:2007xt,Casini:2011kv,Dong:2013qoa}. That is, the   braneworld may be viewed from three perspectives  on equal footing: the bulk, the brane, and the boundary. From the bulk perspective, the theory we deal with is  that of classical gravity in $\text{AdS}_{d+1}$, coupled to  an ETW brane (that is asymptotically $\text{AdS}_{d}$) with tension $\tau$. From the brane perspective (which is intermediate between the bulk and boundary), the theory is  one of dynamical gravity coupled to a $\text{CFT}_{d}$ with a UV cutoff. This cutoff CFT on the brane communicates with a boundary $\text{CFT}_{d}$ ($\text{BCFT}_{d}$) \cite{Takayanagi:2011zk,Fujita:2011fp,Geng:2022dua} on the conformal boundary of the bulk space via  transparent boundary conditions. Finally, from the boundary perspective, the object of interest is the $\text{BCFT}_{d}$.  This perspective is viable when the gravitational theory on the ETW brane is itself dual to a defect $\text{CFT}_{d-1}$ ($\text{DCFT}_{d-1}$) \cite{Yamaguchi:2002pa}, as described previously. Therefore, the boundary perspective is  that of a CFT living on the fixed conformal boundary (as opposed to the dynamical ETW brane), which is coupled to the conformal defect.  In \cite{Feng:2024uia},  doubly holographic black hole thermodynamics was constructed for  charged quantum black holes based on the idea of viewing the brane tension as a thermodynamic variable  \cite{Frassino:2022zaz}.  The first laws and  mass (energy) formulas from  the bulk, brane, and boundary were worked out. In section \ref{je93p8923} of the present paper, we  continue to extend this study  to  the charged and rotating case. Furthermore, we  analyze the phase transitions and criticality of the quantum black holes. Critical exponents  are derived, and the elimination of the re-entrant phase transition for the charged or rotating quantum black holes  is demonstrated.  In particular, we demonstrate that  charged, rotating, and charged rotating quantum black holes do not exhibit re-entrant phase transitions. The critical exponents of  these re-entrant phase transitions are calculated numerically and shown to be not of the standard mean field type, while those associated with non-re-entrant phase transitions are shown to be identical to those of classical charged $\text{AdS}$ black holes investigated in \cite{Kubiznak:2012wp}. Our work serves as  a further extension of the study of holographic black hole chemistry \cite{Karch:2015rpa,Visser:2021eqk,Cong:2021fnf,Cong:2021jgb,Ahmed:2023snm,Ahmed:2023dnh,Gong:2023ywu,Mann:2024sru}.

This semi-classical gravitational theory on the brane is equivalent to classical $\text{AdS}_{d+1}$ gravity coupled to a brane that obeys the Israel junction conditions \cite{Israel:1966rt}. Solving the classical bulk field equations with the appropriate brane boundary condition corresponds to finding exact solutions of the semi-classical gravitational theory on the brane. Braneworld holography is therefore an effective way  to study the problem of backreaction in semi-classical gravity outside of perturbation theory. It is essential to note that classical $\text{AdS}_{d+1}$ black hole spacetimes induce a corresponding black hole on the ETW brane, which includes all orders of quantum backreaction \cite{Emparan:2002px}. These are appropriately referred to as quantum black holes. Examination of braneworld quantum black holes reveals that backreaction can create black holes where there were none previously. That is to say, quantum corrections due to backreaction induce a black hole horizon \cite{Panella:2024sor}. 
 However, it is necessary to consider the validity of the holographic braneworld quantum black hole  within the semi-classical gravitational theory. We will elaborate on this point in section \ref{jmvciope}. We will consider the entropy of   quantum black holes in the near-extremal region and study the behavior of an energy scale $M_{\text{gap}}$, which characterizes the validity of the semi-classical approximation. Connections to  weak cosmic censorship will also be clarified.

 There are two well-known inequalities among thermodynamic quantities for  classical black holes: the Penrose inequality \cite{Penrose:1973um}, including its generalized version \cite{Dain:2013qia}, and the reverse isoperimetric inequality \cite{Cvetic:2010jb}. The quantum version  of the former was proposed in \cite{Bousso:2019var, Bousso:2019bkg},  though limited to the perturbative regime. Quantum black holes non-perturbatively incorporate the backreaction of the quantum field to all orders. It is therefore reasonable to ask whether a quantum Penrose inequality and a quantum reverse isoperimetric inequality can be constructed via such braneworld geometry. While such quantum inequalities were first  derived in \cite{Frassino:2024bjg} for the uncharged and non-rotating quantum BTZ black hole, we  will extend these results to the charged and rotating case in section \ref{jfoip3928}.

In section \ref{ejkrio34}, we will investigate  the gyromagnetic ratio $g$ for charged, rotating quantum black holes. For normal matter, the gyromagnetic ratio is a fundamental quantity characterizing its interaction with an external magnetic field and relates the magnetic moment of a particle to its angular momentum. For a classical macroscopic object with charge and rotation, we have $g = 1$; for a quantum elementary particle such as the electron, we have $g = 2$ at tree level, as required by Quantum Electrodynamics, the Standard Model, and the perturbative sector of open string theory \cite{Argyres:1989cu}, with some radiative corrections via loop diagrams in quantum field theory that  shift $g$ to a value slightly greater than 2. We can also define the gyromagnetic ratio for a charged   rotating black hole. A well-known intriguing observation is that for the Kerr-Newman black hole, as a solution of classical Einstein gravity, this ratio is exactly 2—the same as that of the electron predicted from the Dirac equation \cite{Carter:1968rr,Reina:1975rt}. Using the Killing isometries of the Kerr black hole, Wald constructed a Maxwell field in which a stationary and axisymmetric black hole is immersed, and then proved that the gyromagnetic ratio of a slightly charged stationary and axisymmetric black hole must be 2 \cite{Wald:1974np}. The Kerr-Newman AdS black hole  also has a gyromagnetic ratio of 2 \cite{Aliev:2007qi}. However not all charged and rotating black holes are endowed with the same gyromagnetic ratio. Using the same method of a test Maxwell field,   the five-dimensional weakly charged Myers-Perry black hole \cite{Myers:1986un} was shown to have a gyromagnetic ratio of 3 \cite{Aliev:2004ec}, and   the gyromagnetic ratio of higher-dimensional Kerr-AdS black holes was found to depend on the spacetime dimension and specific parameters.  To date, no study has addressed the gyromagnetic ratio of three-dimensional black holes
—except for the quantum one we explore in this paper. We will study the gyromagnetic ratio of this quantum black hole in section \ref{ejkrio34}, and we will show how the backreaction of quantum matter affects the ratio.
Section \ref{jfi389} is devoted to 
a discussion of our results and our conclusions.

Our conventions for the symbols and variables throughout the paper are aligned with those used in \cite{Emparan:2020znc,Panella:2024sor}.

\section{Quantum black holes with charge and rotation}\label{fji39p4ejr}
\subsection{Solution and braneworld embedding}
We begin with the four-dimensional Einstein-Maxwell action with the following conventions, 
\begin{equation}\label{meip93ojfmip32}
    I_{\text{bulk}}=\frac{1}{16\pi G_{4}}\int_{\mathcal{M}}\mathrm{d}^{4}x\sqrt{-g}\left(\mathcal{R}+\frac{6}{\ell_{4}^2}\right)-\frac{1}{4g^{2}_{4}}\int_{\mathcal{M}}\mathrm{d}^{4}x\sqrt{-g}F_{ab}F^{ab}\,,
\end{equation}
where $\mathcal{R}$ is the Ricci scalar of the bulk spacetime with metric $g_{ab}$, the cosmological constant $\Lambda_4$ is related with the AdS length $\ell_3$ via $3/\ell^{2}_{4}$, and the curvature of the electromagnetic field $\bf{F}$ is coupled with the Einstein term via a dimensionless coupling constant $g_4$, which is related with a length scale $\ell_\star$ via $\ell^{2}_{\star}\equiv 16\pi G_{4}/g^{2}_{4}$. Here we are interested in the charged and rotating C-metric~\cite{debever1971type,Plebanski:1976gy,griffiths2009exact}, which is a solution of the above Einstein-Maxwell action with metric and gauge field
\begin{equation}\label{eq:metric2}
\begin{aligned}
{\rm d} s^2=
\frac{1}{\Omega^{2}}&\left[  -\frac{H(r)}{\Sigma(x, r)}\left(\mathrm{d}t+a x^2 \mathrm{d} \phi\right)^2+\frac{\Sigma(x, r)}{H(r)} \mathrm{d} r^2\right. \\
& \left.\,\,+r^2\left(\frac{\Sigma(x, r)}{G(x)} \mathrm{d} x^2+\frac{G(x)}{\Sigma(x, r)}\left(\mathrm{d} \phi-\frac{a}{r^2} \mathrm{d} t\right)^2\right)\right]\,,
\end{aligned}
\end{equation}
where
\begin{equation}
H(r)=\frac{a^2}{r^2}+\kappa +\frac{\ell^2 q^2}{r^2}+\frac{r^2}{\ell_3^2}-\frac{\ell \mu }{r}\,,
\end{equation}
\begin{equation}
\label{Geqn}
G(x)=\frac{a^2 x^4}{\ell_3^2}-q^2 x^4-\mu  x^3-\kappa  x^2+1\,,
\end{equation}
\begin{equation}
\Sigma(x, r)=1+\frac{a^2 x^2}{r^2}\,,
\end{equation}
\begin{equation}
    \Omega=1+\frac{xr}{\ell}\,.
\end{equation}
The gauge field takes the form
\begin{equation}\label{eq:potential x}
    A_\mu dx^\mu=\frac{2}{\ell_\star}\left(-\frac{\ell q r}{a^2 x^2+r^2}\,,\,0\,,\,0\,,\,-\frac{a \ell q r x^2}{a^2 x^2+r^2}\right)\,.
\end{equation}
The metric \eqref{eq:metric2} and gauge field \eqref{eq:potential x} satisfy the Einstein equation 
\begin{equation}\label{fj930p4jpfi394}
R_{ab}+\frac{\ell_\star^2}{2}{F}_a{}^c {F}_{cb}+\frac{\ell_\star^2}{8}{F}^{cd} {F}_{cd}g_{ab}=-3 \left(\frac{1}{\ell^2}+\frac{1}{\ell_3^2}\right)g_{ab}=-\frac{3}{\ell_{4}^{2}}g_{ab}
\end{equation}
and the Maxwell equation 
\begin{equation}\label{jfp9348}
    \nabla\cdot \bf{{F}}=0\,.
\end{equation}
In the above solution, $\mu$ encodes the mass of the black hole, $\ell$ is the inverse of the acceleration for the C-metric, $\ell_{3}$  is related with (but not equal to) the cosmological constant, $\kappa$ parametrizes the topology of the black hole horizon, $q$ is the charge parameter, $a$ is the angular momentum parameter.

The parameter ranges are determined by ensuring that the metric signature remains invariant in the domain of communication.
The inner and outer horizons $r_{\mp}$ are determined by $H(r_{\mp})=0$, and the conformal boundary $r_c$ is determined by $\Omega=0$. The condition we have stated corresponds to $H(r)$ not changing signs between $r_{+}<r<r_{c}$. On the other hand, to ensure proper metric signature we must have $G(x)\geq0$. While the metric is fairly complicated, this constraint on $G(x)$ leads to nothing conceptually new when compared with the neutral-rotating or static charged metrics considered in previous works. That is, we take the range of $x$ to be
\begin{equation}
    0\leq x\leq x_{1}\,,
\end{equation}
where $x_{1}$ is the smallest positive root of $G(x)$. We shall also consider the mass parameter $\mu$ to be positive and take it to be a `derived' parameter. That is, solving for $\mu$ by setting $G(x_1) = 0$ we obtain
\begin{equation}
    \mu=\frac{-\left(q^{2}-a^{2}/\ell_{3}^{2}\right)x_{1}^{4}-\kappa x_{1}^{2}+1}{x_{1}^{3}}\geq0\,.
\end{equation}
 
On the other hand, since $G(x_1) = 0$, the space pinches off at this point as an axis. Hence, to ensure the absence of conical singularities, we must periodically identify the azimuthal coordinate $\phi \in [0, 2 \pi \Delta]$ where
\begin{equation}
    \Delta=\frac{2}{|G'(x_1)|}=\frac{2}{\left|-2\kappa x_{1}-3\mu x_{1}^{2}-4\left(q^{2}-a^{2}/\ell_{3}^{2}\right)x_{1}^{3}\right|}\,.
\end{equation}

Having made these global considerations, it is possible to apply the braneworld `algorithm' to construct the corresponding three-dimensional quantum black hole. To do this, we first must couple the Einstein-Maxwell action to a brane, which we take to be purely tensional with tension $\tau$. The total action therefore have three pieces, 
% We can now use the braneworld formulation and construct the rotating charged quantum black hole by placing an $\text{AdS}_{3}$ brane with tension $\tau$ in the bulk spacetime with metric (\ref{eq:metric2}). The total action has three components, 
\begin{equation}
    I=I_{\text{bulk}}[\mathcal{M}]+I_{\text{GHY}}[\mathcal{\partial\mathcal{M}}]+I_{\text{brane}}[\mathcal{B}]\,,
\end{equation}
which, respectively, are  the Einstein-Maxwell bulk term  given in \eqref{meip93ojfmip32}, the Gibbons-Hawking-York (GHY) boundary term (which is necessary for the variational problem to be well defined), 
\begin{equation}
    I_{\text{GHY}}=\frac{1}{8\pi G_{4}}\int_{\partial\mathcal{M}}\mathrm{d}^{3}x\sqrt{-h}K\,,
\end{equation}
where $K$ is the extrinsic curvature scalar and $h$ is the determinant of the induced metric on the brane $\mathcal{B}$, 
and the brane term,
\begin{equation}
    I_{\text{brane}}=-\tau\int_{\mathcal{B}}\mathrm{d}^{3}x\sqrt{-h}\,,
\end{equation}
where $\tau$ is the tension of the brane.

The variation of the bulk and boundary parts of this action yield the Einstein-Maxwell equations plus the junction conditions at the brane interface. In the case of the C-metric written in the chart~\eqref{eq:metric2}, the surface $x = 0$ is always totally umbilic, i.e. satisfying $K_{ab} \propto h_{ab}$. Hence, such a surface automatically solves the junction conditions arising from a purely tensional brane. It is then a simple matter to relate the brane tension to the parameters of the solution. 
We have \cite{Israel:1966rt, Kudoh:2004ub}
\begin{equation}
    [K_{ab}-Kh_{ab}]=8\pi G_{4}T_{ab}\,,
\end{equation}
where 
\begin{equation}
    T_{ab}=-\frac{2}{\sqrt{-h}}\frac{\delta I_{\text{brane}}}{\delta h^{ab}}=-\tau h_{ab}
\end{equation}
 is the brane stress tensor  and $[F(z)]$ measures the discontinuity  
 $$
 [F]=\lim_{\epsilon\rightarrow 0} 
 \left(F|_{z+\epsilon}-F|_{z-\epsilon}
 \right)
 $$ 
 of $F$ at $z$ \cite{Kudoh:2004ub}. Then, using the fact that 
% In C-metrics, the surface formed by $x=0$ is umbilic \cite{Emparan:2020znc}. This means that the extrinsic curvature $K_{ab}$ and the metric induced on the brane $h_{ab}$ are directly proportional to one another
\begin{equation}
    K_{ab}=-\frac{1}{\ell}h_{ab}
\end{equation}
we find the tension of the brane is given by
 \begin{equation}\label{eq:tension}
     \tau=\frac{1}{2\pi G_{4}\ell}\,.
 \end{equation}

On the brane the induced metric is 
\begin{equation}\label{eq:brane metric}
    \mathrm{d}s^{2}=-H(r)\mathrm{d}t^{2}+\frac{1}{H(r)}\mathrm{d}r^{2}+r^{2}\left(\mathrm{d}\phi-\frac{a}{r}\mathrm{d}t\right)^{2}\,,
\end{equation}
  where $H(r)$ is given by
\begin{equation}
    H(r)=\frac{r^{2}}{\ell_{3}^{2}}+\kappa-\frac{\mu\ell}{r}+\frac{q^{2}\ell^{2}}{r^2}+\frac{a^{2}}{r^{2}}\, .
\end{equation}
In order for the azimuthal coordinate to have the range $[0, 2\pi]$, and take on the canonical asymptotic form of a rotating AdS black hole, we 
define~\cite{Panella:2024sor}
\begin{equation}
    a=\frac{\ell_{3}\bar{a}}{x_{1}^{2}}\,, \quad r_{s}=\ell_{3}\frac{\bar{a}\Delta}{x_{1}}\sqrt{2-\kappa x_{1}^{2}}\,.
\end{equation}
and employ the coordinate transformations  
\begin{equation}\label{eq:coordinate transformation}
    t=\Delta (\bar{t}-\bar{a}\ell_{3}\bar{\phi})\,,\quad r=\sqrt{\frac{\bar{r}^{2}-r_{s}^{2}}{(1-\bar{a}^{2})\Delta^{2}}}\,,\quad \phi=\Delta(\bar{\phi}-\bar{a}\bar{t}/\ell_{3})\,,
\end{equation}
yielding 
\begin{equation}
    \begin{aligned}
        \mathrm{d}s^{2}=&-\left(\frac{\bar{r}^{2}}{\ell_{3}^{2}}+\frac{q^{2}\ell^{2}\Delta^{2}}{r^{2}}-\Delta^{2}\left(\frac{4\bar{a}^{2}}{x_{1}^{2}}-(1+\bar{a}^{2})\kappa\right)-\frac{\ell\Delta^{2}\mu}{r}\right)\mathrm{d}\bar{t}^{2}\\&+\frac{\bar{r}^{2}}{(\bar{a}^{2}-1)^{2}\Delta^{4}r^{2}}\left(\kappa+\frac{r^{2}}{\ell_{3}^{2}}+\frac{q^2\ell^2x_{1}^2+\bar{a}^{2}\ell_{3}^{2}}{x_{1}^{4}r^{2}}-\frac{\ell\mu}{r}\right)^{-1}\mathrm{d}\bar{r}^2\\&+\left(\bar{r}^{2}-\bar{a}^{2}\Delta^{2}\ell_{3}^{2}\left(\frac{q^2\ell^2}{r^2}-\frac{\ell\mu}{r}\right)\right)\mathrm{d}\bar{\phi}^{2}\\&-2\bar{a}x_{1}\ell_{3}\Delta^{2}(q^2x_1+\mu)\left(1+\frac{\ell}{r x_{1}}-\frac{q^2\ell(r x_{1}+\ell)}{r^{2}x_{1}(q^{2}x_{1}+\mu)}\right)\mathrm{d}\bar{t}\mathrm{d}\bar{\phi}
    \end{aligned}
\end{equation}
for the metric on the brane,
where $r=r(\bar{r})$
from \eqref{eq:coordinate transformation}.

Examining the $\bar{t}\bar{t}$-component of the metric (specifically, taking note of the term that is independent of radius $\bar{r}$), allows us to make the following identification for the black hole mass,
\begin{equation}\label{eq:mass}
    8\mathcal{G}_{3}M=-\kappa\Delta^{2}\left(1+\bar{a}^{2}-\frac{4\bar{a}^{2}}{\kappa x_{1}^{2}}\right)\,,
\end{equation}
where $\mathcal{G}_3\equiv G_{4}/2\ell$ is the renormalized Newton's constant \cite{Feng:2024uia}, which can be related to the effective three-dimensional Newton's constant $G_3$ on the brane as
\begin{equation}\label{fj398p40w9}
    G_{3}=\frac{1}{2\ell_{4}}G_{4}=\frac{\ell}{\ell_{4}}\mathcal{G}_{3}\,.
\end{equation}
Note that the effect of electric charge on the mass is hidden in $\Delta$ as there is functionally no difference between the mass (\ref{eq:mass}) of our charged rotating black hole, and the mass of the neutral-rotating black hole. The mass $M$ of the black hole here is related to the conical deficit angle \cite{Emparan:1999wa,Emparan:1999fd}, as confirmed in \cite{Kudoh:2004ub} via the Euclidean formulation. The result should be the same if one uses the Ashtekar-Das conformal mass \cite{Ashtekar:1999jx,Das:2000cu} or the holographic stress tensor \cite{Rangamani:2009xk}, as studied in \cite{Appels:2016uha,Anabalon:2018qfv} for the four-dimensional accelerating black holes when suitably adapted to account for the presence of the brane, though it would be interesting to confirm this.

 The angular momentum is then found from the $\bar{t}\bar{\phi}$ element in the limit of $\bar{r}\rightarrow\infty$,
\begin{equation}\label{eq:angular momentum}
    4\mathcal{G}_{3}J=\bar{a}x_{1}\ell_{3}\Delta^{2}(q^2x_1+\mu)\,.
\end{equation}

\noindent Unlike the expression for the mass, the angular momentum explicitly depends on the charge parameter $q$. We can then rewrite the metric in terms of the mass and angular momentum as
\begin{equation}\label{condensed metric}
    \begin{aligned}
       ds^{2}=-&\left(\frac{\bar{r}}{\ell_{3}^{2}}-8\mathcal{G}_{3} M-\ell\Delta^{2}\left(\frac{\mu}{r}-\frac{q^{2}\ell}{r^{2}}\right)\right)\mathrm{d}\bar{t}^{2}+\left(\bar{r}^{2}+\bar{a}^{2}\ell_{3}^{2}\ell\Delta^{2}\left(\frac{\mu}{r}-\frac{q^2\ell}{r^2}\right)\right)\mathrm{d}\bar{\phi}^{2}\\
       -&8\mathcal{G}_{3}J\left(1+\frac{\ell}{x_{1}r}-\frac{q^{2}\ell(x_{1}r+\ell)}{r^{2}x_{1}(q^{2}x_{1}+\mu)}\right)\mathrm{d}\bar{t}\mathrm{d}\bar{\phi}\\
       +&\left(\frac{\bar{r}^{2}}{\ell_{3}^{2}}-8\mathcal{G}_{3}M+\frac{(4\mathcal{G}_{3}J)^{2}}{\bar{r^{2}}}+(\bar{a}^{2}-1)^{2}\ell\Delta^{4}\frac{(q^{2}\ell-r\mu)}{\bar{r}^{2}}\right)^{-1}\mathrm{d}\bar{r}^{2}\,.
    \end{aligned}
\end{equation}

The above three-dimensional metric is a solution of the semiclassical Einstein equations on the brane. More specifically, it solves the following effective action:
\begin{equation}
I=I_{\mathrm{CFT}}[\mathcal{B}]+I_{\text {Bgrav }}[\mathcal{B}]\,,
\end{equation} 
where $I_{\mathrm{CFT}}[\mathcal{B}]$ stands for the action of the holographic $\mathrm{CFT}_3$ on the brane and $I_{\text {Bgrav }}[\mathcal{B}]$ is the gravitational action
\begin{equation}\label{foeki9jr3p48}
\begin{aligned}
I_{\text {Bgrav }}= & \frac{\ell_4}{8 \pi \mathrm{G}_4} \int \mathrm{~d}^3 x \sqrt{-h}\left[\frac{4}{\ell_4^2}\left(1-\frac{\ell_4}{\ell}\right)+R+\ell_4^2\left(\frac{3}{8} R^2-R_{a b} R^{a b}\right)+\cdots\right] + I_{\rm EM}\,,
\end{aligned}
\end{equation}
where $I_{\rm EM}$ represents the induced electromagnetic action on the brane, which was first given in \cite{Taylor:2000xw} and reads
\begin{align} \label{maxwell_counterterms}
    I_{\text{EM}} =  \dfrac{\ell_4 \ell_\star^2}{8\pi G_4}  \int & {\rm d}^3x \sqrt{-h} \bigg[ - \dfrac{5}{16}\tilde{F}^2 + \ell_4^2 \bigg(\dfrac{1}{288}R\tilde{F}^2 - \dfrac{5}{8} R^a_b \tilde{F}_{ac}\tilde{F}^{bc}  \nonumber \\ 
     & + \dfrac{3}{98}\tilde{F}^{ab}\left( \nabla_b\nabla^c \tilde{F}_{ca} - \nabla_a\nabla^c \tilde{F}_{cb} \right) + \dfrac{5}{24}  \nabla_a\tilde{F}^{ab}\nabla_c{\tilde{F}^c}_b \bigg) + \mathcal{O}(\ell_4^3 )\bigg] \, ,
\end{align}
where $\widetilde{\mathbf{F}}=\mathrm{d} \bar{\mathcal{A}}$ with $\bar{\mathcal{A}}$ the induced gauge potential on the brane

From the brane effective action we can directly read off the three-dimensional coupling constants in terms of their four-dimensional counterparts. For the three-dimensional Newton constant we already did this earlier in Eq.~\eqref{fj398p40w9}. The brane theory also has a three-dimensional cosmological constant $\Lambda_3$ that can be defined by the $\text{AdS}_{3}$ length scale $\ell_3$ via 
\begin{equation}
    \Lambda_{3}=-\frac{1}{\ell_{3}^{2}}\,.
\end{equation}
Likewise, the cosmological constant $\Lambda_4$ in the four-dimensional bulk theory is defined by the $\text{AdS}_{4}$ length scale $\ell_4$ as
\begin{equation}
    \Lambda_{4}=-\frac{1}{\ell_{4}^{2}}\,.
\end{equation}
We may define an effective cosmological length scale $L_3$  from the cosmological constant term in the low-energy effective action of the higher-curvature gravitational theory on the brane as
\begin{equation}\label{jf398p4imof}
    \frac{1}{L_{3}^{2}}=\frac{2}{\ell_{4}^{2}}\left(1-\frac{\ell_{4}}{\ell}\right)=\frac{1}{\ell_3^2}\left[1+\frac{\ell^2}{4 \ell_3^2}+\mathcal{O}\left(\frac{\ell^4}{\ell_3^4}\right)\right]\,,
\end{equation}
where in the last step we used  (\ref{fj930p4jpfi394}) and expanded to quadratic order in $\ell$, neglecting the other terms. The difference between $L_{3}$ and $\ell_{3}$ is that $\ell_{3}$ receives contributions from higher curvature terms \cite{Emparan:2020znc}. 
It is also useful to define a three-dimensional gauge coupling constant $g_3$ to describe the non-linear electromagnetic field on the brane  through
\begin{equation}\label{jfeopi3948jur}
    g_{3}^{2}=\frac{2g_{4}^{2}}{5\ell_{4}}\,.
\end{equation}
Writing the brane action in terms of these intrinsically three-dimensional quantities we have 
\begin{equation}
\begin{aligned}
I_{\text {Bgrav }}= & \frac{1}{16 \pi \mathrm{G}_3} \int \mathrm{~d}^3 x \sqrt{-h}\left[\frac{2}{L_3^2}+R+\ell^2\left(\frac{3}{8} R^2-R_{a b} R^{a b}\right)\right] \\
& +\int \mathrm{d}^3 x \sqrt{-h}\bigg[-\frac{1}{4 g_3^2} \widetilde{\mathbf{F}}^2+\frac{4 \ell^2}{5 g_3^2} \bigg( \frac{1}{288} R \widetilde{\mathbf{F}}^2-\frac{5}{8} R_b^a \widetilde{F}_{a c} \widetilde{F}^{b c}+\frac{5}{24}(\nabla \cdot \widetilde{\mathbf{F}})^2 
\nonumber
\\
&+\frac{3}{98} \widetilde{F}^{a b}\left(\nabla_b \nabla^c \widetilde{F}_{c a}-\nabla_a \nabla^c \widetilde{F}_{c b}\right) \bigg) + \cdots\bigg]\,.
\end{aligned}
\end{equation}

\subsection{Thermodynamic quantities}

 Note that the presence or absence of electric charge does not change the functional form of (\ref{eq:brane metric}), only the function $H(r)$. This is in contrast to the black hole's spin, since the rotation parameter $a$ explicitly enters the metric, outside of $H(r)$. Because of this, many of the parameters derived from the metric, including the angular velocity and temperature of the black hole's outer horizon, have the same functional form as in the neutral-rotating quantum BTZ black hole, and their calculation can be carried over from the analysis performed in \cite{Emparan:2020znc} with no change.

The original form for the induced metric on the brane (\ref{eq:brane metric}) manifestly had the Killing vectors $\partial_{t}$ and $\partial_{\phi}$. Under the coordinate transformation (\ref{eq:coordinate transformation}), these Killing vectors are written as
\begin{equation}\label{eq:Killing transformation}
    \begin{aligned}
        \frac{\partial}{\partial t}&=\frac{1}{\Delta(1-\bar{a}^{2})}\left(\frac{\partial}{\partial\bar{t}}+\frac{\bar{a}}{\ell_{3}}\frac{\partial}{\partial\bar{\phi}}\right)\,,\\
        \frac{\partial}{\partial\phi}&=\frac{1}{\Delta(1-\bar{a}^{2})}\left(\frac{\partial}{\partial\bar{\phi}}+\bar{a}\ell_{3}\frac{\partial}{\partial\bar{t}}\right)\,,
    \end{aligned}
\end{equation}
as shown in \cite{Emparan:2020znc,Panella:2024sor}. The outer horizon of the black hole corresponds to those points with the coordinate $r=r_{+}$, where $r_{+}$ is the largest positive root of $H(r)$. It is the Killing horizon of the Killing vector
\begin{equation}
    k=\frac{\partial}{\partial t}+\frac{a}{r_{+}^{2}}\frac{\partial}{\partial\phi}\,.
\end{equation}
Given (\ref{eq:Killing transformation}), we can see that the canonically normalized Killing vector $\bar{k}$, which generates the outer horizon, takes the form
\begin{equation}
    \bar{k}=\frac{\Delta(1-\bar{a}^{2})}{1+\frac{a^{2}x_{1}^{2}}{r_{+}^{2}}}k=\frac{\partial}{\partial\bar{t}}+\Omega_H\frac{\partial}{\partial\bar{\phi}}\,,
\end{equation}
where $\Omega_H$, the angular velocity of the outer horizon, is
\begin{equation}
    \Omega_H\equiv\frac{a}{r_{+}^{2}+a^{2}x_{1}^{2}}\left(1+\frac{r_{+}^{2}x_{1}^{2}}{\ell_{3}^{2}}\right)\,.
\end{equation}
The temperature of the outer horizon, relative to $\bar{k}$ is
\begin{equation}
    \begin{aligned}
    T=\frac{\Delta(1-\bar{a}^{2})}{1+\frac{a^{2}x_{1}^{2}}{r_{+}^{2}}}\frac{H'(r_{+})}{4\pi}=\frac{\Delta  \left(2r_{+}^{4}-2 \ell_{3}^{2} \left(a^2+q^2 \ell ^2\right)+\mu  r_{+} \ell  \ell_{3}^{2}\right)}{4 \pi  r_{+} \ell_{3}^{2} \left(a^2 x_{1}^{2}+r_{+}^{2}\right)}\,.
    \end{aligned}
\end{equation}
To calculate the entropy of the quantum charged rotating black hole, we compute the entropy of the bulk horizon using the bulk metric (\ref{eq:metric2}). The Bekenstein-Hawking entropy is given by 
\begin{equation}
    \begin{aligned}\label{eq:entropy}
        S_{\text{gen}}=&\frac{\text{Area}(r_{+})}{4G_{4}}\\
        =&\frac{1}{2G_{4}}\int_{0}^{2\pi\Delta}\mathrm{d}\phi\int_{0}^{x_{1}}\mathrm{d}x\frac{r_{+}^{2}\ell^{2}}{(\ell+r_{+}x)^{2}}\left(1+\frac{a^{2}x_{1}^{2}}{r_{+}^{2}}\right)\\
        =&\frac{\pi}{G_{4}}\Delta\frac{\ell x_{1}(r_{+}^{2}+a^{2}x_{1}^{2})}{\ell+r_{+}x_{1}}\,.
    \end{aligned}
\end{equation}
This  can be interpreted as a generalized entropy that incorporates the backreaction of quantum fields on the black hole spacetime, to all orders of perturbation theory. 

The electric charge of the black hole on the brane is identified with the charge of the bulk, which is given by
\begin{equation}\label{eq:charge}
        Q=\frac{1}{g_{4}^{2}}\int\ast{\bf{F}}=\frac{4 \pi  \Delta  q x_{1} \ell }{g_{4}^{2}\ell_{\star}}\,.
\end{equation}
To calculate the conjugate electric potential, we first see that, on the brane (located at $x=0$), the gauge field (\ref{eq:potential x}) becomes
\begin{equation}\label{eq:gauge field on brane}
    \bar{\mathcal{A}}_{\mu}\mathrm{d}\bar{x}^{\mu}=-\frac{2\ell q\Delta}{\ell_{\star}r(\bar{r})}\mathrm{d}\bar{t}+\frac{2\bar{a}\ell_{3}\ell q\Delta}{\ell_{\star}r(\bar{r})}\mathrm{d}\bar{\phi}\,,
\end{equation}
where $r(\bar{r})$, is given by the coordinate transformation (\ref{eq:coordinate transformation}). The electric potential conjugate to the electric charge on the brane is given by the difference in the inner product between this one-form and the Killing vector
\begin{equation}
    \zeta^{\mu}=\frac{\partial}{\partial\bar{t}}+\frac{\bar{a}}{\ell_{3}}\frac{\partial}{\partial\bar{\phi}}\,,
\end{equation}
evaluated at $r\rightarrow\infty$ and $r=r_{+}$. This gives
\begin{equation}\label{eq:potential}
\Phi=\bar{\mathcal{A}}_{\mu}\zeta^{\mu}\big|_{r=r_{+}}^{r\rightarrow\infty}=(1-\bar{a}^{2})\frac{2\ell q\Delta}{\ell_{\ast}r_{+}}\,.
\end{equation}
We now possess several of the physical parameters of the charged rotating quantum black hole.
\subsubsection{Brane expressions}
We can write the expressions found in the preceding subsection as functions of dimensionless variables as
\begin{equation}\label{eq:dimensionless variables}
    z=\frac{\ell_{3}}{r_{+}x_{1}}\,,\quad \nu=\frac{\ell}{\ell_{3}}\,,\quad \chi=qx_{1}^{2}\,, \quad \alpha=\frac{\bar{a}}{\sqrt{-\kappa}x_{1}}\,,
\end{equation}
where $\chi$ and $\alpha$ are the new parameters for electric charge and spin respectively. By combining the above definitions with $H(r_{+})=0$ and $G(x_{1})=0$, we obtain the expressions for $x_{1}^{2}$, $r_{+}^{2}$ and $\mu x_{1}$ as
\begin{equation}\label{eq:dimensionless expressions}
    \begin{aligned}
        x_{1}^{2}=&\frac{\nu  z^3 \left(\chi ^2 (\nu  z+1)-1\right)+1}{\kappa  z^2 \left(\alpha ^2 z^2-\left(\alpha ^2+1\right) \nu  z-1\right)}\,,\\
        r_{+}^{2}=&\frac{\kappa {\ell_{3}}^2 \left(\alpha ^2 z^2-\left(\alpha ^2+1\right) \nu  z-1\right)}{\nu  z^3 \left(\chi ^2 (\nu  z+1)-1\right)+1}\,,\\
        \mu x_{1}=&\frac{\kappa  \left(\left(z^2+1\right) \left(\alpha ^2 \left(z^2-1\right)-1\right)-\chi ^2 z^2 \left(z^2 \left(\alpha ^2 \left(\nu ^2+1\right)+\nu ^2\right)-1\right)\right)}{\nu  z^3 \left(\chi ^2 (\nu  z+1)-1\right)+1}\,.
    \end{aligned}
\end{equation}
It is convenient to define
 \cite{Feng:2024uia}  
\begin{equation}
    \kappa=-\text{sgn}(\nu^{2}\chi^{2}z^{4}+\nu\chi^{2}z^{3}-\nu z^{3}+1)\,,
\end{equation}
where $\text{sgn}$ is the sign function. Since  $r_{+}^{2} > 0$,
 the rotation parameter $\alpha$ is bounded from above as
\begin{equation}\label{eq:upper bound}
    \alpha^{2}\leq \frac{1+\nu z}{z(z-\nu)}\,,
\end{equation}
which is the same restriction found for the neutral-rotating black hole in \cite{Emparan:2020znc}. Given the definition in (\ref{eq:dimensionless variables}), we can see that the case of $\kappa=+1$ corresponds to imaginary values of $\alpha$. These states correspond to naked closed timelike curves, and are not black hole solutions \cite{Emparan:2020znc}.
In the braneworld holographic setup we are working in, the $\text{AdS}_{3}$ brane serves as a holographic renormalization surface for the asymptotic $\text{CFT}_{3}$, and cuts off the UV modes of the field. The central charge of the $\text{CFT}_{3}$ is given by
\begin{align}
    c_{3}=\frac{\ell_{4}^{2}}{G_{4}}=\frac{\nu\ell_{3}}{2G_{3}\sqrt{\nu^{2}+1}}\,,
\end{align}
as expressed in terms of the $\text{AdS}_{3}$ radius $\ell_{3}$.  The dependence of the central charge on the backreaction parameter $\ell$ suggests that it may be regarded as a thermodynamic variable of the system.
We employ the extended thermodynamic phase space treatment 
of black hole chemistry \cite{Kubiznak:2016qmn}
in which the pressure $P_3$ of the system on the brane is identified with the cosmological constant $\Lambda_{3}$ as
\begin{equation}
    P_{3}=-\frac{\Lambda_{3}}{8\pi G_{3}}=\frac{\sqrt{\nu^{2}+1}}{4\pi G_{3}\ell_{3}^{2}(\sqrt{\nu^{2}+1}+1)}\,.
\end{equation}
With (\ref{eq:dimensionless expressions}) at hand, we can find the mass (\ref{eq:mass}), entropy (\ref{eq:entropy}), angular momentum (\ref{eq:angular momentum}), and charge (\ref{eq:charge}) of the quantum charged rotating black hole on the brane in terms of our dimensionless variables respectively as
\begin{align}
    M=&\begin{aligned}[t]
        &\frac{1}{2 G_3}( -\alpha^2+z^2\left(\chi^2+z^2\left(\nu^2 \chi^2-\alpha^2\left(\left(\nu^2+1\right) \chi^2+3\right)\right)\right. \\
        & \left.\left.+2 \nu z\left(2 \alpha^2+\chi^2+1\right)+3\right)+1\right)^{-2} \\
        \times & {\left[\sqrt { \nu ^ { 2 } + 1 } ( \nu z ^ { 3 } ( \chi ^ { 2 } ( \nu z + 1 ) - 1 ) + 1 ) \left(\alpha^2+z^2\left(4 \alpha^4 z(\nu-z)\right.\right.\right.} \\
        & \left.\left.\left.+\alpha^2\left(z\left(4 \nu+\nu \chi^2(\nu z+1)-z\right)+4\right)+\nu z+1\right)\right)\right]\,,
    \end{aligned}\\
     S_{\mathrm{gen}}=& \begin{aligned}[t]
        &-\frac{\pi}{G_3}\left[\sqrt{\nu^2+1} z \ell_3\left(\alpha^2\left(z^2\left(\nu \chi^2 z-1\right)+1\right)+1\right)\right] \\
        \times & {\left[\left(\alpha^2+z^2\left(-\chi^2+z\left(-2 \nu\left(2 \alpha^2+\chi^2+1\right)\right.\right.\right.\right.} \\
        & \left.\left.\left.\left.+\chi^2 z\left(\left(\alpha^2-1\right) \nu^2+\alpha^2\right)+3 \alpha^2 z\right)-3\right)-1\right)\right]^{-1}\,,
    \end{aligned}\\
    J=& \begin{aligned}[t]
        & \frac{1}{G_3}\left[\alpha z \ell_3\left(\alpha^2-\alpha^2 z^4+\left(\alpha^2+1\right) \nu \chi^2 z^3(\nu z+1)+z^2+1\right)\right. \\
        & \left.\times \sqrt{\left(\nu^2+1\right)\left(z\left(\nu+\alpha^2(\nu-z)\right)+1\right)\left(\nu z^3\left(\chi^2(\nu z+1)-1\right)+1\right)}\right] \\
        & {\left[-\alpha^2+z^2\left(\chi^2+z^2\left(\nu^2 \chi^2-\alpha^2\left(\left(\nu^2+1\right)     \chi^2+3\right)\right)\right.\right.} \\
        & \left.\left.+2 \nu z\left(2 \alpha^2+\chi^2+1\right)+3\right)+1\right]^{-2}\,,
    \end{aligned}\\
    Q=& \begin{aligned}[t]
        - & {\left[4 \sqrt{\pi} \chi z^2 \sqrt{\nu^2+1}\left(z\left(\nu+\alpha^2(\nu-z)\right)+1\right)\right] } \\
        \times & {\left[\sqrt { 5 G _ { 3 } } g _ { 3 } \left(\alpha^2+z^2\left(-\chi^2+z\left(-2 \nu\left(2 \alpha^2+\chi^2+1\right)\right.\right.\right.\right.} \\
        & \left.\left.\left.\left.+\chi^2 z\left(\left(\alpha^2-1\right) \nu^2+\alpha^2\right)+3 \alpha^2 z\right)-3\right)-1\right)\right]^{-1}\,.
    \end{aligned}
\end{align}
Given these quantities, we can obtain the temperature $T$, angular velocity $\Omega_H$, electric potential $\Phi$, three-dimensional thermodynamic volume $V_{3}$ and three-dimensional chemical potential $\mu_{3}$ as
\begin{align}
    T=&\begin{aligned}[t]
        &\left(\frac{\partial M}{\partial S_{\text{gen}}}\right)_{P_{3},c_{3},Q,J}\\
        =&\bigg[\big(z\big(\nu  \chi ^2 z^2 \left(z^2 \left(\alpha ^2 \left(\nu ^2+1\right)+\nu ^2\right)+2 \nu  z+1\right)\\
        &\quad\quad-\nu  \left(\alpha ^2 \left(z^4+3\right)+z^2+3\right)+4 \alpha ^2 z\big)-2\big)\\
        &\quad\quad\times\left(\alpha ^2+z^2 \left(\alpha ^2 z \left(-2 \nu +\nu  \chi ^2 (\nu  z+1)+z\right)-\nu  z-1\right)\right)\bigg]\\
        &\times\bigg[2 \pi  z \ell_{3} (\nu  z+1) \left(\alpha ^2 \left(z^2 \left(\nu  \chi ^2 z-1\right)+1\right)+1\right)\\
        &\quad\quad\times\big(\alpha ^2+z^2 \big(-\chi ^2+z \big(-2 \nu  \big(2 \alpha ^2+\chi ^2+1\big)\\
        &\quad\quad+\chi ^2 z \left(\left(\alpha ^2-1\right) \nu ^2+\alpha ^2\right)+3 \alpha ^2 z\big)-3\big)-1\big)\bigg]^{-1}\,,
    \end{aligned}\\
    \Phi=& \begin{aligned}[t]
        &\left(\frac{\partial M}{\partial Q}\right)_{S_{\text{gen}},P_{3},c_{3},J}\\
        =&-\bigg[\sqrt{5} g_{3} \nu  \chi  z \left(\alpha ^2 z^2-\left(\alpha ^2+1\right) \nu  z-1\right)\\
        &\quad\quad\times\left(\alpha ^2+z^2 \left(\alpha ^2 z \left(-2 \nu +\nu  \chi ^2 (\nu  z+1)+z\right)-\nu  z-1\right)\right)\bigg]\\
        &\times\bigg[2 \sqrt{G_{3}\pi} (\nu  z+1) \left(\alpha ^2 \left(z^2 \left(\nu  \chi ^2 z-1\right)+1\right)+1\right)\\
        &\quad\quad\times\big(\alpha ^2+z^2 \big(-\chi ^2+z \big(-2 \nu  \left(2 \alpha ^2+\chi ^2+1\right)\\
        &\quad\quad\quad+\chi ^2 z \left(\left(\alpha ^2-1\right) \nu ^2+\alpha ^2\right)+3 \alpha ^2 z\big)-3\big)-1\big)\bigg]^{-1}\,,
    \end{aligned}\\
    \Omega_H=& \begin{aligned}[t]
        &\left(\frac{\partial M}{\partial J}\right)_{S_{\text{gen}},P_{3},c_{3},Q}\\
        =&\frac{\alpha  \left(z^2+1\right) \sqrt{\left(z \left(\nu +\alpha ^2 (\nu -z)\right)+1\right) \left(\nu  z^3 \left(\chi ^2 (\nu  z+1)-1\right)+1\right)}}{z \ell_{3} (\nu  z+1) \left(\alpha ^2 \left(z^2 \left(\nu  \chi ^2 z-1\right)+1\right)+1\right)}\,,
    \end{aligned}\\
    V_{3}=& \begin{aligned}[t]
        &\left(\frac{\partial M}{\partial P_{3}}\right)_{S_{\text{gen}},c_{3},Q,J}\nonumber\\
        =&\frac{4 \pi  z^2 \ell _3^2 \left(\alpha ^2 \left(z^2-1\right)-1\right)^2}{\left(\alpha ^2+\left(\chi ^2+3\right) z^2 \left(\alpha ^2 z^2-1\right)-1\right)^2}-\bigg[4 \nu \left(\alpha ^2+\left(\chi ^2+3\right) z^2 \left(\alpha ^2 z^2-1\right)-1\right)^{-3}\\
        &\quad\quad\times\bigg(\pi  z \ell _3^2 \left(\alpha ^2+\left(\chi ^2-1\right) z^2 \left(\alpha ^2 z^2-1\right)+1\right)\\
        &\quad\quad\times(\alpha ^4-\alpha ^2+2 \alpha ^4 \left(\chi ^2+3\right) z^6-\alpha ^2 z^4 \left(\left(\alpha ^2+2\right) \chi ^2+\alpha ^2+8\right)\\
        &\quad\quad-z^2 \left(2 \alpha ^4+\alpha ^2 \left(\chi ^2+3\right)-2\right))\bigg)\bigg]+\mathcal{O}\left(\nu ^2\right)\,,
    \end{aligned}
\end{align}
\begin{align}
    \mu_{3}=&\begin{aligned}[t]
        &\left(\frac{\partial M}{\partial c_{3}}\right)_{S_{\text{gen}},P_{3},Q,J}\\
        =&-\bigg[\nu  \big(4 z \left(\alpha ^2 \left(z^4+1\right)-z^2\right) \big(\chi ^2 z^2 (\alpha  z-1) (\alpha  z+1) \left(\alpha ^2 \left(z^2+1\right)-1\right)\\
        &\quad\quad-\left(z^2-1\right) \left(\alpha ^2 \left(z^2-1\right)-1\right)^2\big) \big(\left(-\alpha ^2-\left(\left(\chi ^2+3\right) z^2 \left(\alpha ^2 z^2-1\right)\right)+1\right)\\
        &\quad\quad\times\left(\alpha ^2 z \left(\left(\chi ^2-1\right) z^2+1\right)+z\right)+4 z^3 \left(2 \alpha ^2+\chi ^2+1\right) \left(1-\alpha ^2 \left(z^2-1\right)\right)\big)\\
        &\quad\quad+\left(1-\alpha ^2 \left(z^2-1\right)\right) \big(4 \alpha ^4+4 \alpha ^2-4 \left(2 \alpha ^6+\alpha ^4\right) z^{10}\\
        &\quad\quad+\left(34 \alpha ^6+28 \alpha ^4+8 \alpha ^2\right) z^8-2 \left(19 \alpha ^6+35 \alpha ^4+12 \alpha ^2+2\right) z^6\\
        &\quad\quad+\left(22 \alpha ^6+44 \alpha ^4+38 \alpha ^2+4\right) z^4-2 \left(5 \alpha ^6+9 \alpha ^4+5 \alpha ^2+3\right) z^2\\
        &\quad\quad-4 \chi ^2 z^4 \left(\alpha ^2 \left(z^4+1\right)-z^2\right) \left(\alpha ^2 \left(5-4 z^2\right)+\alpha ^4 \left(z^4-9 z^2+2\right)+3\right)\\
        &\quad\quad-4 \alpha ^2 \chi ^4 z^6 (\alpha  z-1) (\alpha  z+1) \left(-z^2+\alpha ^2 \left(z^4+z^2+2\right)-1\right)\big)\\
        &\quad\quad\times\left(-\alpha ^2-\left(\left(\chi ^2+3\right) z^2 \left(\alpha ^2 z^2-1\right)\right)+1\right)\big)-4 z \left(1-\alpha ^2 \left(z^2-1\right)\right)\\
        &\quad\quad\times\left(\alpha ^2 \left(z^4+1\right)-z^2\right) \Big(\chi ^2 z^2 (\alpha  z-1) (\alpha  z+1) \left(\alpha ^2 \left(z^2+1\right)-1\right)\\
        &\quad\quad-\left(z^2-1\right) \left(\alpha ^2 \left(z^2-1\right)-1\right)^2\Big) \left(-\alpha ^2-\left(\left(\chi ^2+3\right) z^2 \left(\alpha ^2 z^2-1\right)\right)+1\right)\bigg]\\
        &\times\bigg[{4 \ell _3 \left(\alpha ^2 \left(z^2-1\right)-1\right)^2 \left(\alpha ^2+\left(\chi ^2+3\right) z^2 \left(\alpha ^2 z^2-1\right)-1\right)^3}\bigg]^{-1}+\mathcal{O}(\nu^2)\,,
    \end{aligned}
\end{align}
 keeping the three-dimensional Newton's constant $G_{3}$ fixed.  We have written $V_{3}$ and $\mu_{3}$ only to the first order in the backreaction parameter $\nu$, owing to their size;  the complete expressions are given in  appendix \ref{section:A}. These thermodynamic expressions (when using the complete forms of $V_{3}$ and $\mu_{3}$) satisfy the first law of black hole thermodynamics 
\begin{equation}\label{eq:first law on the brane}
    \mathrm{d}M=T\mathrm{d}S_{\text{gen}}+\Phi \mathrm{d}Q+\Omega_H \mathrm{d}J+V_{3}\mathrm{d}P_{3}+\mu_{3}\mathrm{d}c_{3}
\end{equation}
on the brane and the corresponding Smarr relation
\begin{equation}
    0=TS_{\text{gen}}-2P_{3}V_{3}+\mu_{3}c_{3}+\Omega_H J\,.
\end{equation}
The mass $M$ and charge $Q$ are absent from the Smarr relation since $G_{3}M$ and $Q$ are dimensionless. Here we have taken the central charge as a variable, as was done in the standard holographic black hole chemistry \cite{Karch:2015rpa}; see \cite{Mancilla:2024spp} for other solutions.

\subsubsection{Bulk expressions}

From the bulk perspective, there is a black hole sliced by an ETW brane with tension (\ref{eq:tension}), which is a thermodynamic variable. The charged rotating quantum black hole on the brane is associated with the classical solution of the bulk gravitational theory coupled to the brane, and contains all orders of backreaction from the $\text{CFT}_{3}$ on the brane. Just as in the brane perspective, we work in the extended thermodynamic phase space, which means that 
\begin{equation}
    P_{4}=-\frac{\Lambda_{4}}{8\pi G_{4}}=\frac{3(\nu^{2}+1)}{8\pi G_{4}\ell^{2}}\,,
\end{equation}
where 
the four-dimensional bulk pressure $P_{4}$ is a variable. Tuning the brane tension $\tau$ is done by altering the position of the brane in the bulk (with the rest following from the Israel junction conditions). The thermodynamic variable conjugate to $\tau$ will be referred to as the thermodynamic area of the brane, and will be denoted as $A_{\tau}$. 

Given our previous expressions (\ref{eq:mass}), (\ref{eq:entropy}), (\ref{eq:charge}), and (\ref{eq:angular momentum}), the charged rotating quantum black hole's mass $M$, entropy $S$, electric charge $Q$, and spin $J$ can be written as
\begin{align}
    M=&\begin{aligned}[t]
        &\frac{\ell}{G_{4}}\bigg[\left(\nu  z^3 \left(\chi ^2 (\nu  z+1)-1\right)+1\right) \big(\alpha ^2+z^2 \big(4 \alpha ^4 z (\nu -z)\\
        &\quad\quad+\alpha ^2 \left(z \left(4 \nu +\nu  \chi ^2 (\nu  z+1)-z\right)+4\right)+\nu  z+1\big)\big)\bigg]\\
        &\times\bigg[\big(-\alpha ^2+z^2 \big(\chi ^2+z^2 \left(\nu ^2 \chi ^2-\alpha ^2 \left(\left(\nu ^2+1\right) \chi ^2+3\right)\right)\\
        &\quad\quad+2 \nu  z \left(2 \alpha ^2+\chi ^2+1\right)+3\big)+1\big)\bigg]^{-2}\,,
    \end{aligned}\\
    S_{\text{gen}}=& \begin{aligned}[t]
        &-\frac{2\pi\ell^{2}}{G_{4}}\bigg[z\left(\alpha ^2 \left(z^2 \left(\nu  \chi ^2 z-1\right)+1\right)+1\right)\bigg]\\
        &\times\bigg[\nu  \big(\alpha ^2+z^2 \big(-\chi ^2+z \big(-2 \nu  \left(2 \alpha ^2+\chi ^2+1\right)\\
        &\quad\quad+\chi ^2 z \left(\left(\alpha ^2-1\right) \nu ^2+\alpha ^2\right)+3 \alpha ^2 z\big)-3\big)-1\big)\bigg]^{-1}\,,
    \end{aligned}\\
    Q=&\begin{aligned}[t]
        &-\frac{4\ell}{g_{4}}\sqrt{\frac{\pi}{G_{4}}}\bigg[\chi  z^2 \left(z \left(\nu +\alpha ^2 (\nu -z)\right)+1\right)\bigg]\\
        &\times\bigg[\alpha ^2+z^2 \big(-\chi ^2+z \big(-2 \nu  \left(2 \alpha ^2+\chi ^2+1\right)\\
        &\quad\quad+\chi ^2 z \left(\left(\alpha ^2-1\right) \nu ^2+\alpha ^2\right)+3 \alpha ^2 z\big)-3\big)-1\bigg]^{-1}\,,
    \end{aligned}\\
    J=&\begin{aligned}[t]
        &\frac{2\ell^{2}}{G_{4}}\bigg[\alpha  z \left(\alpha ^2-\alpha ^2 z^4+\left(\alpha ^2+1\right) \nu  \chi ^2 z^3 (\nu  z+1)+z^2+1\right)\\
        &\quad\quad\times\sqrt{\left(\nu ^2+1\right) \left(z \left(\nu +\alpha ^2 (\nu -z)\right)+1\right) \left(\nu  z^3 \left(\chi ^2 (\nu  z+1)-1\right)+1\right)}\bigg]\\
        &\times\bigg[ \nu  \sqrt{\nu ^2+1} \big(-\alpha ^2+z^2 \big(\chi ^2+z^2 \left(\nu ^2 \chi ^2-\alpha ^2 \left(\left(\nu ^2+1\right) \chi ^2+3\right)\right)\\
        &\quad\quad+2 \nu  z \left(2 \alpha ^2+\chi ^2+1\right)+3\big)+1\big)^2\bigg]^{-1}\,,
    \end{aligned}
\end{align}
where $G_{4}$ is the four-dimensional Newton constant.

The above expressions allow us to write down the bulk  temperature $T$, electric potential $\Phi$, angular velocity $\Omega_H$, thermodynamic volume $V_{4}$, and thermodynamic area $A_{\tau}$ as, respectively
\begin{align}
    T=& \begin{aligned}[t]
        &\left(\frac{\partial M}{\partial S_{\text{gen}}}\right)_{P_{4},\tau,Q,J}\\
        =&-\bigg[\nu  \big(z \big(\nu  \chi ^2 z^2 \left(z^2 \left(\alpha ^2 \left(\nu ^2+1\right)+\nu ^2\right)+2 \nu  z+1\right)\\
        &\quad\quad-\nu  \left(\alpha ^2 \left(z^4+3\right)+z^2+3\right)+4 \alpha ^2 z\big)-2\big)\\
        &\quad\quad\times\left(\alpha ^2+z^2 \left(\alpha ^2 z \left(-2 \nu +\nu  \chi ^2 (\nu  z+1)+z\right)-\nu  z-1\right)\right)\bigg]\\
        &\times\bigg[2 \pi  z \ell  (\nu  z+1) \left(\alpha ^2 \left(z^2 \left(\nu  \chi ^2 z-1\right)+1\right)+1\right)\\
        &\quad\quad\times\big(\alpha ^2+z^2 \big(-\chi ^2+z \big(-2 \nu  \left(2 \alpha ^2+\chi ^2+1\right)\\
        &\quad\quad+\chi ^2 z \left(\left(\alpha ^2-1\right) \nu ^2+\alpha ^2\right)+3 \alpha ^2 z\big)-3\big)-1\big)\bigg]^{-1}\,,
    \end{aligned}\\
    \Phi=&\begin{aligned}[t]
        &\left(\frac{\partial M}{\partial Q}\right)_{S_{\text{gen}},P_{4},\tau,J}\\
        =&\frac{g_{4}}{\sqrt{\pi G_{4}}}\bigg[\nu  \chi  z \left(z \left(\nu +\alpha ^2 (\nu -z)\right)+1\right)\\
        &\quad\quad\quad\quad\times\left(\alpha ^2+z^2 \left(\alpha ^2 z \left(-2 \nu +\nu  \chi ^2 (\nu  z+1)+z\right)-\nu  z-1\right)\right)\bigg]\\
        &\times\bigg[(\nu  z+1) \left(\alpha ^2 \left(z^2 \left(\nu  \chi ^2 z-1\right)+1\right)+1\right)\\
        &\quad\quad\times\big(\alpha ^2+z^2 \big(-\chi ^2+z \big(-2 \nu  \left(2 \alpha ^2+\chi ^2+1\right)\\
        &\quad\quad+\chi ^2 z \left(\left(\alpha ^2-1\right) \nu ^2+\alpha ^2\right)+3 \alpha ^2 z\big)-3\big)-1\big)\bigg]^{-1}\,,
    \end{aligned}\\
    \Omega_H=&\begin{aligned}[t]
        &\left(\frac{\partial M}{\partial J}\right)_{S_{\text{gen}},P_{4},\tau,Q}\\
        =&\frac{\alpha  \nu  \left(z^2+1\right) \sqrt{\left(z \left(\nu +\alpha ^2 (\nu -z)\right)+1\right) \left(\nu  z^3 \left(\chi ^2 (\nu  z+1)-1\right)+1\right)}}{z \ell  (\nu  z+1) \left(\alpha ^2 \left(z^2 \left(\nu  \chi ^2 z-1\right)+1\right)+1\right)}\,,
    \end{aligned}
\end{align}
\newpage
\begin{align}
    V_{4}=&\begin{aligned}[t]
        &\left(\frac{\partial M}{\partial P_{4}}\right)_{S_{\text{gen}},\tau,Q,J}\\
        =&\bigg[8 \pi  z \ell ^3 \Big(\alpha ^2 \nu ^2 \chi ^4 z^5 (\nu  z+1) \left(\nu  z \left(\left(\alpha ^2+1\right) \nu  z+2\right)+1\right)\\
        &\quad+\left(\alpha ^2 \left(z^2-1\right)-1\right) \big(\nu  \left(\alpha ^2+z^2 \left(\alpha ^2 \left(3 z^2-2\right)-3\right)\right)\\
        &\quad+z \left(\alpha ^2 \left(z^2-1\right)-1\right)+\nu ^2 z^3 \left(\alpha ^2 \left(z^2-3\right)-2\right)\big)\\
        &\quad+\alpha ^2 \nu  \chi ^2 z^2 \Big(z \Big(-\left(\left(\alpha ^2+1\right) \nu ^3 z^4\right)+\nu  \left(-\alpha ^2 \left(1-2 z^2\right)^2+4 z^2+1\right)\\
        &\quad+z \left(\alpha ^2 \left(1-2 z^2\right)+2\right)+\nu ^2 z^3 \left(1-\alpha ^2 \left(z^2-4\right)\right)\Big)+1\Big)\Big)\bigg]\\
        &\times\bigg[3 \nu ^2 (\nu  z+1) \big(\alpha ^2+z^2 \big(-\chi ^2+z \big(-2 \nu  \left(2 \alpha ^2+\chi ^2+1\right)\\
        &\quad\quad+\chi ^2 z \left(\left(\alpha ^2-1\right) \nu ^2+\alpha ^2\right)+3 \alpha ^2 z\big)-3\big)-1\big)^2\bigg]^{-1}\,,
    \end{aligned}
    \end{align}
    \begin{align}
    A_{\tau}=& \begin{aligned}[t]
        &\left(\frac{\partial M}{\partial \tau}\right)_{S_{\text{gen}},P_{4},Q,J}\\
        =&\bigg[2 \pi  \ell ^2 \Big(-\alpha ^2 \nu ^2+\alpha ^2 \nu ^2 z^8 \left(\left(\nu ^2+2\right) \chi ^2+1\right) \left(2 \alpha ^2-\nu ^2 \chi ^2\right)\\
        &\quad+\nu  z^7 \big(2 \alpha ^4 \left(-3 \nu ^2-\left(\nu ^4+\nu ^2-2\right) \chi ^2-2\right)\\
        &\quad-\alpha ^2 \nu ^2 \left(2 \left(\nu ^2+2\right) \chi ^4-\left(\nu ^2-1\right) \chi ^2+3\right)+\nu ^4 \chi ^2\big)\\
        &\quad+z^6 \Big(-2 \alpha ^4 \left(\nu ^4 \left(\chi ^2-2\right)+\nu ^2 \left(3 \chi ^2-2\right)+1\right)\\
        &\quad-\alpha ^2 \nu ^2 \left(\nu ^2 \left(\chi ^4+\chi ^2-4\right)+2 \left(\chi ^2+1\right)^2\right)+\nu ^4 \left(2 \chi ^2+1\right)\Big)\\
        &\quad+\nu  z^5 \left(2 \alpha ^4 \left(\left(\nu ^2-1\right) \chi ^2+3\right)+\alpha ^2 \left(6 \nu ^2-4 \chi ^2+8\right)+\nu ^2 \left(\chi ^2+3\right)\right)\\
        &\quad+\alpha ^2 z^4 \left(2 \alpha ^2 \left(\nu ^2 \left(\chi ^2-1\right)+2\right)-2 \nu ^4 \chi ^2-7 \nu ^2+4\right)\\
        &\quad+\nu  z^3 \left(-2 \alpha ^4 \left(\nu ^2+2\right)+\alpha ^2 \left(-2 \left(\nu ^2+1\right) \chi ^2+\nu ^2-6\right)+\nu ^2-4\right)\\
        &\quad+z^2 \left(-2 \alpha ^2 \left(\alpha ^2+\nu ^2+2\right)+\nu ^2-2\right)+2 \alpha ^2 \left(\alpha ^2+1\right) \nu  z\Big)\bigg]\\
        &\times\bigg[\nu ^2 \Big(-\alpha ^2+z^2 \big(\chi ^2+z^2 \left(\nu ^2 \chi ^2-\alpha ^2 \left(\left(\nu ^2+1\right) \chi ^2+3\right)\right)\\
        &\quad\quad+2 \nu  z \left(2 \alpha ^2+\chi ^2+1\right)+3\big)+1\Big)^2\bigg]^{-1}\,,
    \end{aligned}
\end{align}
where just as in the brane perspective, we keep Newton's constant fixed. The above expressions satisfy the first law
\begin{equation}
    \mathrm{d}M=T\mathrm{d}S_{\text{gen}}+\Phi \mathrm{d}Q+\Omega_H \mathrm{d}J+V_{4}\mathrm{d}P_{4}+A_{\tau}\mathrm{d}\tau\,,
\end{equation}
and the Smarr relation
\begin{equation}
    M=2TS-2V_{4}P_{4}+2\Omega_H J+\Phi Q-A_{\tau}\tau\,.
\end{equation}
The latter can be derived from the former by applying Euler's theorem for  homogeneous functions \cite{Kastor:2009wy,Frassino:2022zaz,Hale:2025veb}.

\subsubsection{Boundary expressions}

The double holography prescription states that the $\text{DCFT}_{2}$ (resulting from the intersection of the ETW brane in the bulk with the boundary) coupled to the $\text{BCFT}_{3}$, provides another description of the thermodynamics of the quantum charged rotating black hole on the brane. The degrees of freedom of the $\mathrm{DCFT}_{2}$ are related to the three-dimensional Newtonian constant $G_{3}$ and the three-dimensional cosmological constant $L_{3}$ via the holographic correspondence formula \cite{Brown:1986nw}
\begin{equation}
    c_{2}=\frac{3L_{3}}{2G_{3}}=\frac{3\ell}{2G_{3}\sqrt{2\nu^{2}-2\sqrt{\nu^{2}+1}+2}}\,.
\end{equation}
The holographic dictionary tells us that the energy $E$, entropy $\mathcal{S}$, electric charge $\mathcal{Q}$, and volume $V_{2}$ of the $\text{DCFT}_{2}$ are given by
\begin{align}
    &E=M= \begin{aligned}[t]
        &\frac{1}{2G_{3}}\big(-\alpha ^2+z^2 \big(\chi ^2+z^2 \left(\nu ^2 \chi ^2-\alpha ^2 \left(\left(\nu ^2+1\right) \chi ^2+3\right)\right)\\
        &\quad\quad\quad+2 \nu  z \left(2 \alpha ^2+\chi ^2+1\right)+3\big)+1\big)^{-2}\\
        &\times\bigg[\sqrt{\nu ^2+1} \left(\nu  z^3 \left(\chi ^2 (\nu  z+1)-1\right)+1\right) \big(\alpha ^2+z^2 \big(4 \alpha ^4 z (\nu -z)\\
        &\quad\quad+\alpha ^2 \left(z \left(4 \nu +\nu  \chi ^2 (\nu  z+1)-z\right)+4\right)+\nu  z+1\big)\big)\bigg]\,,
    \end{aligned}\\
    &\mathcal{S}=S_{\text{gen}}=
    \begin{aligned}[t]
        &-\bigg[\pi  \sqrt{\nu ^2+1} z \ell  \left(\alpha ^2 \left(z^2 \left(\nu  \chi ^2 z-1\right)+1\right)+1\right)\bigg]\\
        &\times\bigg[G_{3} \nu  \big(\alpha ^2+z^2 \big(-\chi ^2+z \big(-2 \nu  \left(2 \alpha ^2+\chi ^2+1\right)\\
        &\quad\quad+\chi ^2 z \left(\left(\alpha ^2-1\right) \nu ^2+\alpha ^2\right)+3 \alpha ^2 z\big)-3\big)-1\big)\bigg]^{-1}\,,
    \end{aligned}\\
    &\mathcal{Q}=QL_{3}= \begin{aligned}[t]
        &-\bigg[2 \sqrt{2 \pi } \sqrt{\nu ^2+1} \chi  z^2 \ell  \left(z \left(\nu +\alpha ^2 (\nu -z)\right)+1\right)\bigg]\\
        &\times\bigg[g_{3} \sqrt{5 G_{3}} \sqrt{\nu ^2-\sqrt{\nu ^2+1}+1}\\
        &\quad\quad\times\big(\alpha ^2+z^2 \big(-\chi ^2+z \big(-2 \nu  \left(2 \alpha ^2+\chi ^2+1\right)\\
        &\quad\quad\quad+\chi ^2 z \left(\left(\alpha ^2-1\right) \nu ^2+\alpha ^2\right)+3 \alpha ^2 z\big)-3\big)-1\big)\bigg]^{-1}\,,
    \end{aligned}\\
    &\mathcal{J}= \begin{aligned}[t]
        &\bigg[\alpha  z \ell  \left(\alpha ^2-\alpha ^2 z^4+\left(\alpha ^2+1\right) \nu  \chi ^2 z^3 (\nu  z+1)+z^2+1\right)\\
        &\quad\times\sqrt{\left(\nu ^2+1\right) \left(z \left(\nu +\alpha ^2 (\nu -z)\right)+1\right) \left(\nu  z^3 \left(\chi ^2 (\nu  z+1)-1\right)+1\right)}\bigg]\\
        &\times\bigg[G_{3} \nu  \big(-\alpha ^2+z^2 \big(\chi ^2+z^2 \left(\nu ^2 \chi ^2-\alpha ^2 \left(\left(\nu ^2+1\right) \chi ^2+3\right)\right)\\
        &\quad\quad+2 \nu  z \left(2 \alpha ^2+\chi ^2+1\right)+3\big)+1\big)^2\bigg]^{-1}\,,
    \end{aligned}\\
    &V_{2}=2\pi L_{3}= \begin{aligned}[t]
        &\frac{2\pi\ell}{G_{3}\sqrt{2\nu^{2}-2\sqrt{\nu^{2}+1}+2}}\,.
    \end{aligned}
\end{align}
The form of $E$ for the $\text{DCFT}_{2}$ was chosen to match the brane expression for the mass, and the charge $\mathcal{Q}$ is rescaled by $L_{3}$. The temperature $T$, electric potential $\Phi$, angular velocity $\Omega_H$, two-dimensional thermodynamic pressure $P_{2}$, and two-dimensional chemical potential $\mu_{2}$ are 
\begin{align}
    &T=\begin{aligned}[t]
        &\left(\frac{\partial E}{\partial \mathcal{S}}\right)_{\ell,V_{2},c_{2},\mathcal{Q},\mathcal{J}}\\
        =&-\bigg[\nu  \big(z \big(\nu  \chi ^2 z^2 \left(z^2 \left(\alpha ^2 \left(\nu ^2+1\right)+\nu ^2\right)+2 \nu  z+1\right)\\
        &\quad\quad-\nu  \left(\alpha ^2 \left(z^4+3\right)+z^2+3\right)+4 \alpha ^2 z\big)-2\big)\\
        &\quad\quad\times\left(\alpha ^2+z^2 \left(\alpha ^2 z \left(-2 \nu +\nu  \chi ^2 (\nu  z+1)+z\right)-\nu  z-1\right)\right)\bigg]\\
        &\times\bigg[2 \pi  z \ell  (\nu  z+1) \left(\alpha ^2 \left(z^2 \left(\nu  \chi ^2 z-1\right)+1\right)+1\right)\\
        &\quad\quad\times\big(\alpha ^2+z^2 \big(-\chi ^2+z \big(-2 \nu  \left(2 \alpha ^2+\chi ^2+1\right)\\
        &\quad\quad+\chi ^2 z \left(\left(\alpha ^2-1\right) \nu ^2+\alpha ^2\right)+3 \alpha ^2 z\big)-3\big)-1\big)\bigg]^{-1}\,,
    \end{aligned}\\
    &\Phi=\begin{aligned}[t]
        &\left(\frac{\partial E}{\partial \mathcal{Q}}\right)_{\ell,\mathcal{S},V_{2},c_{2},\mathcal{J}}\\
        =&\bigg[\sqrt{5} g_{3} \nu  \chi  z \sqrt{\nu ^2-\sqrt{\nu ^2+1}+1} \left(z \left(\nu +\alpha ^2 (\nu -z)\right)+1\right)\\
        &\quad\quad\times\left(\alpha ^2+z^2 \left(\alpha ^2 z \left(-2 \nu +\nu  \chi ^2 (\nu  z+1)+z\right)-\nu  z-1\right)\right)\bigg]\\
        &\times\bigg[\sqrt{2\pi G_{3}} \ell  (\nu  z+1) \left(\alpha ^2 \left(z^2 \left(\nu  \chi ^2 z-1\right)+1\right)+1\right)\\
        &\quad\quad\times\big(\alpha ^2+z^2 \big(-\chi ^2+z \big(-2 \nu  \left(2 \alpha ^2+\chi ^2+1\right)\\
        &\quad\quad+\chi ^2 z \left(\left(\alpha ^2-1\right) \nu ^2+\alpha ^2\right)+3 \alpha ^2 z\big)-3\big)-1\big)\bigg]^{-1}\,,
    \end{aligned}\\
    &\Omega_H=\begin{aligned}[t]
        &\left(\frac{\partial E}{\partial \mathcal{J}}\right)_{\ell,\mathcal{S},V_{2},c_{2},\mathcal{Q}}\\
        =&\frac{\alpha  \nu  \left(z^2+1\right) \sqrt{\left(z \left(\nu +\alpha ^2 (\nu -z)\right)+1\right) \left(\nu  z^3 \left(\chi ^2 (\nu  z+1)-1\right)+1\right)}}{z \ell  (\nu  z+1) \left(\alpha ^2 \left(z^2 \left(\nu  \chi ^2 z-1\right)+1\right)+1\right)}\,,
    \end{aligned}\\
    &P_{2}=\begin{aligned}[t]
        &\left(\frac{\partial E}{\partial V_{2}}\right)_{\ell,\mathcal{S},c_{2},\mathcal{Q},\mathcal{J}}\\
        =&\frac{\nu  \left(\alpha ^2-\left(4 \alpha ^2+1\right) z^2 \left(\alpha ^2 z^2-1\right)\right)}{4 \pi  G_{3} \ell  \left(\alpha ^2+\left(\chi ^2+3\right) z^2 \left(\alpha ^2 z^2-1\right)-1\right)^2}+\mathcal{O}(\nu^{2})\,,
    \end{aligned}\\
    &\mu_{2}=\begin{aligned}[t]\label{fjiope3qj}
        &\left(\frac{\partial E}{\partial c_{2}}\right)_{\ell,\mathcal{S},V_{2},\mathcal{Q},\mathcal{J}}\\
        =&-\bigg[\sqrt{-2\left(\left(\nu ^2+1\right) \left(-\nu ^2+\sqrt{\nu ^2+1}-1\right)\right)}\\
        &\quad\quad\times\big(\alpha ^2 \nu  z^5 \left(\left(\nu ^2+1\right) \chi ^2+1\right)-\left(2 \alpha ^2+1\right) \nu ^2 z^4-2 \nu  z^3\\
        &\quad\quad+2 \alpha ^2 z^2-\left(\alpha ^2+2\right) \nu  z-1\big)\\
        &\quad\quad\times\left(\alpha ^2+z^2 \left(\alpha ^2 z \left(-2 \nu +\nu  \chi ^2 (\nu  z+1)+z\right)-\nu  z-1\right)\right)\bigg]\\
        &\times\bigg[3\ell (\nu  z +1) \big(\alpha ^2+z^2 \big(-\chi ^2+z \big(-2 \nu  \left(2 \alpha ^2+\chi ^2+1\right)\\
        &\quad\quad+\chi ^2 z \left(\left(\alpha ^2-1\right) \nu ^2+\alpha ^2\right)+3 \alpha ^2 z\big)-3\big)-1\big)^2\bigg]^{-1}\,,
    \end{aligned}
\end{align}
where we have written $P_{2}$ to linear order in the backreaction parameter $\nu$ owing to its size; the full expression is   written in   appendix \ref{section:A}. In the above expressions, the parameter $\ell$ is kept fixed. This does not mean that the brane tension is kept constant, since in the boundary perspective, we let the three-dimensional Newton constant $G_{3}$ be a variable. This seems to be the only choice we have in order to obtain cohomogeneity thermodynamics. This means that the central charge on the brane $c_{3}$ is variable.

The above expressions satisfy the first law of black hole thermodynamics on the $\text{DCFT}_{2}$ as
\begin{equation}
    \mathrm{d}E=T\mathrm{d}\mathcal{S}-P_{2}\mathrm{d}V_{2}+\mu_{2}\mathrm{d}c_{2}+\Phi \mathrm{d}\mathcal{Q}+\Omega_H \mathrm{d}\mathcal{J}\,.
\end{equation}
Note the presence of a work term $P_{2}dV_{2}$ as opposed to $V_{2}dP_{2}$. This indicates that $E$ plays the role of the thermodynamic internal energy, in contrast to $M$ in the other perspectives, which played the role of enthalpy in the extended thermodynamic phase space, where pressure is a variable. The integral internal energy formula for the $\mathrm{DCFT}_2$ reads
\begin{equation}
    E=T\mathcal{S}+\mu_{2}c_{2}+\frac{1}{2}\mathcal{Q}\Phi+\Omega_H\mathcal{J}\,,
\end{equation}
where $V_{2}$ is absent since $V_{2}/G_{3}$ is dimensionless.

\section{Phase transitions and criticality}\label{je93p8923}

With the expressions for all the thermodynamic variables in the brane, bulk, and boundary perspectives respectively, at hand, we may investigate the phase transitions and critical points of the quantum black hole. Since our expressions are general, we may obtain functions for the charged-static, neutral-rotating, and neutral-static quantum black holes by setting $\alpha=0$, $\chi=0$, and $\alpha=\chi=0$, respectively. Our investigation of the phase structure of these black hole systems is carried out through the use of phase diagrams. In particular, we locate the second order critical points in phase space and calculate their associated critical exponents. Owing to the complicated nature of the functions we derived in the previous section, it is impossible to obtain closed-form equations of state, which means that the investigation must be carried out numerically. 

\subsection{Brane perspective}

We focus on the so-called canonical ensemble, in which the temperature $T$, pressure $P_{3}$, central charge $c_{3}$, electric charge $Q$, and angular momentum $J$ are all fixed. Looking at the first law of black hole thermodynamics on the brane (\ref{eq:first law on the brane}), and noting again that in the extended thermodynamic phase space, the black hole mass corresponds to enthalpy, and not the internal energy, this means that the relevant thermodynamic potential for the ensemble is the Gibbs free energy $G=M-TS_{\text{gen}}$. The path the system traces out in the phase space can be parameterized by the variable $z$. Given values for $P_{3}$, $c_{3}$, $Q$, $J$, and $z$, we are left with a system of four equations for four variables, $\nu$, $\chi$, $\alpha$, $\ell_{3}$. Adjusting the value of $z$, but keeping the same values for $P_{3}$, $c_{3}$, $Q$, and $J$, we can see how the other thermodynamic variables of the system, including the temperature and Gibbs free energy, evolve as we move from one system to another within the ensemble. We start with the simplest possible system, the neutral-static quantum BTZ black hole (so $Q=0$, and $J=0$). This is the only black hole that can transition to thermal $\text{AdS}$. The ensemble is one in which electric charge is fixed. This prevents any black hole carrying electric charge from making the transition. For the sake of consistency between different cases, we will not compare the neutral-static quantum BTZ black hole to quantum thermal $\text{AdS}$, (i.e., $\text{AdS}_{3}$ spacetime including backreaction from the $\text{CFT}_{3}$ on the brane), as was done in \cite{Frassino:2023wpc}. This will not affect our analysis of phase behavior or our calculation of critical exponents.

Keeping the central charge $c_{3}$ fixed at an arbitrary value, we may observe the behavior of the curves the system is constrained to be on in the $G-T$ plane, as we move from an ensemble with one value for the thermodynamic pressure $P_{3}$ to another. The cusps on the curve correspond to where $\frac{\partial G}{\partial z}=0$, and $\frac{\partial T}{\partial z}=0$. With the exception of the curve corresponding to the system being kept at the critical pressure $P_{c}$, each curve will have two cusps, and one point at which it self-intersects. As the critical pressure is approached (i.e., we move to copies of the system within the ensemble with fixed pressures that are closer and closer to $P_{c}$), these cusps will approach each other on the plane, until at the critical pressure, they merge and disappear, leaving a smooth curve with no self-intersections. Crucially, as the thermodynamic pressure of the neutral-static black hole on the brane exceeds $P_{c}$, the cusps will immediately reappear and the curve will self-intersect once more. On either side of the critical pressure, the self-intersection of the curve presents as an inverted swallowtail. This corresponds to the re-entrant phase transitions observed in \cite{Frassino:2023wpc}. The coexistence curve between two phases of the black hole will extend past the critical point in the $P_{3}-T$ plane, and not simply terminate, as it does in the case for ordinary phase behavior. We stress that this phenomenon is unique to the neutral-static quantum BTZ black hole. The addition of any electric charge or spin into the system removes the critical point associated with this re-entrant phase transition. This is shown in section \ref{section:Elimination}. 

\begin{figure}[ht]
\centering
\includegraphics[width=0.3\linewidth]{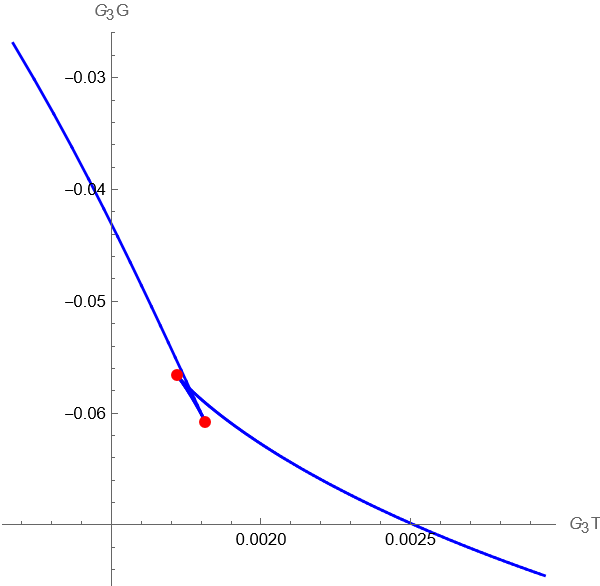} 
\includegraphics[width=0.3\linewidth]{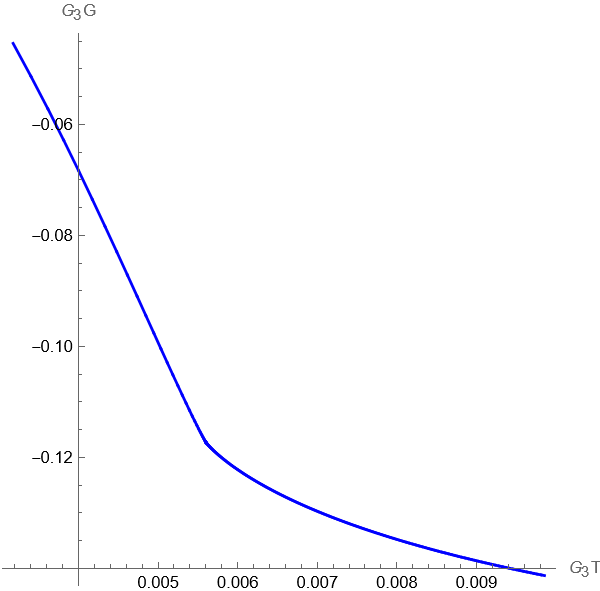}
\includegraphics[width=0.3\linewidth]{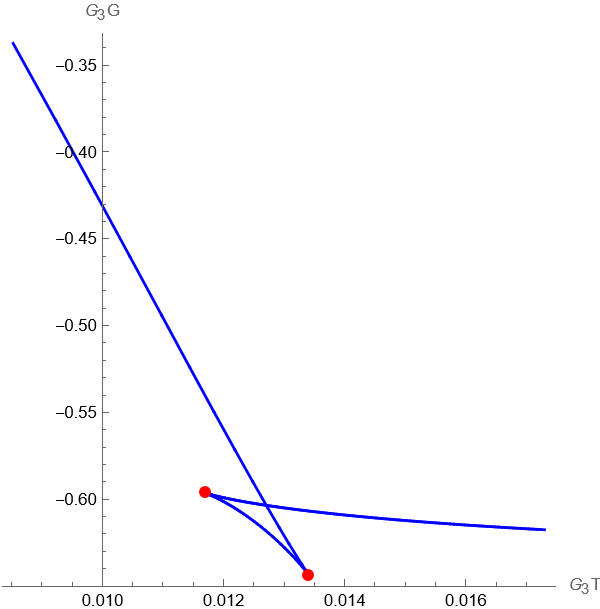}
\caption{Free energy-temperature curves for neutral-static systems in the canonical ensemble with different pressures. The central charge for each curve is kept at $c_{3}=10$. The cusps on each curve are highlighted in red. From left to right are curves at $P<P_{C}$, $P=P_{c}$, and $P>P_{c}$.
}
\label{fig:GT-Curves-1-V2}
\end{figure}

The second order critical points of the system in the canonical ensemble are those where
\begin{equation}\label{eq:critical points}
    \left(\frac{\partial T}{\partial S}\right)_{P_{3},c_{3},Q,J}=\left(\frac{\partial^{2}T}{\partial S^{2}}\right)_{P_{3},c_{3},Q,J}=0\,.
\end{equation}
Keeping $c_{3}$ fixed at some value, and setting the first two partial derivatives to zero gives us three equations to solve for the three parameters $(z_{c}, \nu_{c}, \ell_{3,c})$ (in the neutral-static limit, we know that $\chi=\alpha=0$). After numerically finding these parameters, we can calculate the critical values for the various thermodynamic variables. With this at hand, we will be able to define an order parameter and reduced temperature $\tilde{t}=\frac{T-T_{c}}{T_{c}}=\tau-1$. We select the reduced chemical potential $\frac{\mu_{3}-\mu_{3,c}}{\mu_{3,c}}$ as our order parameter, where the subscript $c$ indicates the value of the thermodynamic variable at criticality. In order to calculate the values of the critical exponents $\tilde{\alpha}$, $\beta$, and $\gamma$ (where $\tilde{\alpha}$ the critical exponent is not to be confused with the rotation parameter $\alpha$), the critical point must be approached in phase space along the coexistence curve. The critical exponent $\delta$ is found by approaching the critical point along the critical isotherm (so constrained to $t=0$ in phase space). Having chosen an order parameter, we can define the functions used to extract the critical exponents. Since we are working in the fixed $(P_{3}, c_{3})$ ensemble, the heat capacity (based on which $\tilde{\alpha}$ is calculated) is
\begin{equation}
    C_{P_{3},\mu_{3}}=T\left(\frac{\partial S}{\partial T}\right)_{P_{3},\mu_{3}} \propto |t|^{-\tilde{\alpha}}\,,
\end{equation}
where we have chosen to keep the order parameter constant when evaluating the heat capacity. The critical exponent $\beta$ is calculated based on the power-law behavior of the order parameter with respect to $t$ near the critical point,
\begin{equation}
    |\mu_{3}-\mu_{3,c}|\propto |t|^{\beta}\,.
\end{equation}
The analogue for isothermal compressibility that we use (the susceptibility) is
\begin{equation}
    \chi=-\frac{1}{\mu_{3}}\left(\frac{\partial\mu_{3}}{\partial c_{3}}\right)_{P_{3},T} \propto |t|^{-\gamma}\,.
\end{equation}
Finally, $\delta$ governs how the source function $\frac{c_{3}-c_{3,c}}{c_{3,c}}$ is related to the order parameter near the critical point, along the critical isotherm $T=T_{c}$ (we also constrain the approach to be along the critical isobar $P_{3}=P_{3,c}$),
\begin{equation}
    |c_{3}-c_{3,c}|\propto |\mu_{3}-\mu_{3,c}|^{\delta}\,.
\end{equation}

The coexistence curve corresponds to where the Gibbs free energy surface intersects itself in natural variables space, with coordinates $(S, P_{3}, c_{3})$. As stated previously, owing to the fact that our thermodynamic variables are expressed as complicated, non-invertible functions of independent parameters, we do not have the luxury of turning to an explicit equation of state to find the coexistence curve and critical exponents. With this in mind, we must resort to using numerical methods, just as we did when locating the critical point to begin with. We have used two such methods with consistent results. 

The first method uses the fact that along a coexistence curve, the two phases of the system will have equal temperature, pressure, and free-energy, but different thermodynamic volumes. The self-intersection of free energy-temperature curves will disappear at the critical point. Since we selected the  chemical potential $\mu_{3}$ to form the order parameter, we will keep the central charge constant for each of these curves (fixing it to be its critical value) so that the source function $\frac{c_{3}-c_{3,c}}{c_{3,c}}$ is kept at a value of zero. The pressure is varied. Since we want to focus on the coexistence curve very close to the critical point, in order to calculate the order of various divergences, the pressures we use will be very close to the critical pressure. The point of self-intersection is tracked, and the two separate sets of values of the parameters $(z,\nu,\ell_{3})$ that map on to the point of self-intersection are recorded. Then one list of values for the variables (corresponding to a single coexistence curve in $(z,\nu,\ell_{3})$ space) is used to calculate the heat capacity, order parameter, and susceptibility. Note that due to the phenomenon of re-entrant phase transitions for the quantum BTZ black hole, the coexistence curves do not actually terminate at the critical point, but extend beyond them. In practice when calculating the critical exponents $\alpha$, $\beta$, and $\gamma$, we will work on the side of the coexistence curve where the reduced temperature $t=\frac{T-T_{c}}{T_{c}}$ is negative.

The second method is simply to implement Maxwell's equal area construction numerically. We take the system and solve for the $P-V$ curves numerically, for temperatures close to the critical temperature (such that $t\rightarrow0^{-}$), again holding the central charge at its critical value. Along the $P-V$ isothermal curves, the places where the curve does not monotonically decrease (i.e., where $\frac{\partial P}{\partial V}>0$) must be corrected, and the metastable states removed, with an isobar that is positioned according to Maxwell's equal area construction. The ends of this isobar will be positioned at the two thermodynamic volumes of the two respective coexistent phases. As the temperature of the isotherms approaches the critical temperature, the isobar's position varies. Keeping track of the position of the isobar, and the values of the two coexistent volumes allows one to form a coexistence curve.

\begin{figure}
    \centering
    \includegraphics[width=0.5\linewidth]{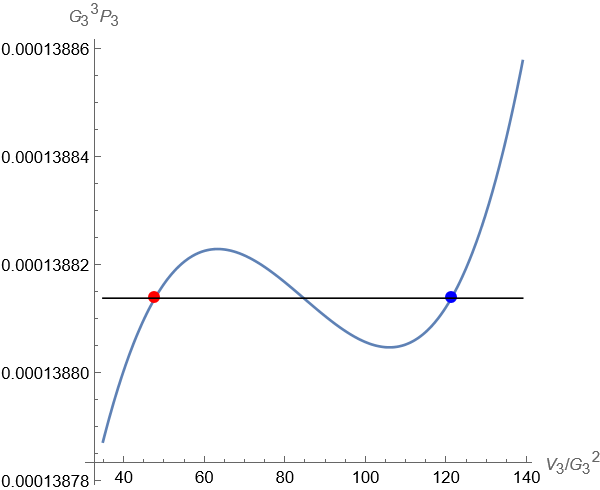}
    \caption{An isobar that corrects the presence of metastable states in the isotherm. The area above and below the isobar are equal. The smaller volume is marked in red while the larger volume is marked in blue. These are points on the coexistence curve. For the sake of visibility, this is far away from the critical point, and the curve corresponds to $\tilde{t}=-1/10$. When calculating the critical exponents, we take these points to be very close.}
    \label{fig:Maxwell}
\end{figure}

As stated previously, both numerical methods agree. We found values of the critical exponents for the second order critical point of the neutral-static quantum BTZ black hole in the canonical ensemble (up to some negligible numerical error) as
\begin{equation}\label{eq:neutral static critical exponents}
        \tilde{\alpha}=0,\quad
        \beta=1,\quad
        \gamma=2,\quad
        \delta=3
\end{equation}
that curiously are the
same values that occur for  a particular type of black hole in 3rd order Lovelock gravity \cite{Dolan:2014vba}.
These   unorthodox values   differ from those predicted by mean-field theory, but satisfy both the Rushbrooke inequality 
\begin{equation}
    \tilde{\alpha}+2\beta+\gamma\geq2\,,
\end{equation}
and the Widom relation
\begin{equation}
    \gamma=\beta(\delta-1)\,.
\end{equation}

Moving on to the charged-static quantum BTZ black hole, we find that the presence of any electric charge introduces an upper bound on the temperature. This is in contrast to the neutral-static black hole, where the temperature of the black hole grows unbounded as a function of $z$ past a second turning point \cite{Frassino:2023wpc}. The introduction of charge into the system gives rise to another turning point in the curve, past which the temperature decreases. The requirement that temperature be greater than or equal to zero means electric charge also introduces an upper bound on $z$. There is a critical point associated with a large value of electric charge for the quantum black hole. This point is found by solving (\ref{eq:critical points}) and $P_{3}=P_{3,0}$, $c_{3}=c_{3,0}$ for $z$, $\nu$, $\chi$, and $\ell_{3}$. As opposed to the neutral-static case, this critical point is not associated with a re-entrant phase transition, i.e., when the critical electric charge is exceeded, the cusps and self-intersection in the free energy-temperature curve do not reappear. The critical exponents for this second-order critical point were calculated numerically using the same methods as in the neutral-static case. They are
\begin{equation}\label{eq:van der Waals critical exponents}
        \tilde{\alpha}=0\,,\quad
        \beta=\frac{1}{2}\,,\quad
        \gamma=1\,,\quad
        \delta=3\,.
\end{equation}
Notably, these agree with the critical exponents found  in \cite{Kubiznak:2012wp} for the classical charged $\text{AdS}_{4}$ black hole, which has the same critical exponents as the van der Waals gas.

\begin{figure}[ht]
\centering
\includegraphics[width=0.49\linewidth]{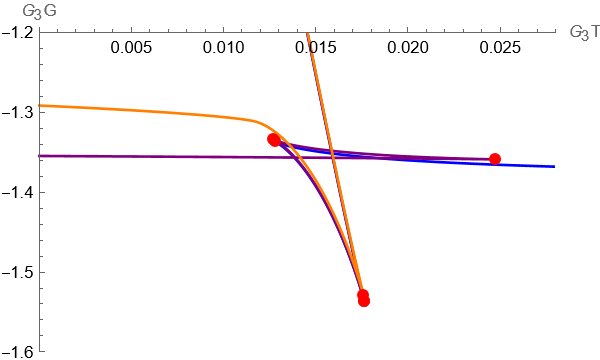} 
\includegraphics[width=0.49\linewidth]{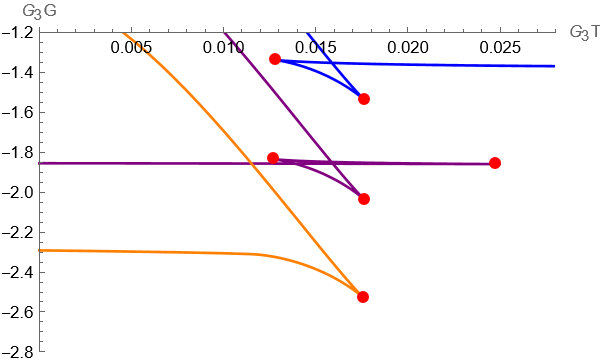}
\caption{(Left) Free energy-temperature curves for the charged-static quantum black hole, with central charge $c_{3}=10$ and pressure $G_{3}^{3}P_{3}=1.81\times 10^{-4}$, for three different charges: (blue) $\sqrt{G_{3}}Q=0$, (purple) $\sqrt{G_{3}}Q=\frac{1}{10}$, (orange) $\sqrt{G_{3}}Q=\frac{1}{3}$. The last of these has an electric charge that exceeds the critical electric charge value $Q_{c}$, and the cusp does not reappear, signifying a non-re-entrant phase transition. (Right) The same curves separated by vertical offsets for the sake of visibility. Note how the neutral curve extends to $T\rightarrow\infty$, but the charged curves have a maximum possible temperature. Note also how increasing the electric charge of the system pulls the new turning point/cusp towards another, until (at criticality) they merge and disappear.
}
\label{fig:Charged-GT-Curves}
\end{figure}

The case of the neutral-rotating black hole has a subtlety, in that its critical point (associated with a large spin) has an imaginary value for the rotation parameter $\alpha$. This remains a formal solution, but corresponds to a closed timelike curve formed from the rotation of a naked singularity, not a black hole \cite{Emparan:2020znc}. Attempting to restrict the search for critical points to real values of $\alpha$ leads to the location of an artificial critical point with a value of $\alpha_{c}$ that exceeds the upper bound (\ref{eq:upper bound}), which in turn means that the associated critical value of the angular momentum is imaginary. Treating the original critical point (with imaginary $\alpha$) as a formal solution, we find that it is associated with a phase transition that is not re-entrant, just as the critical point for the charged-static quantum black hole. The critical exponents for this are the orthodox ones (\ref{eq:van der Waals critical exponents}), and the corresponding free-energy curves have an identical form to those of the charged-static black hole, shown in Fig. \ref{fig:Charged-GT-Curves}. 

We restrict our treatment of the charged rotating quantum black hole to the boundary perspective for the sake of brevity, though an analysis from the brane and bulk perspectives was carried out and confirmed to give the same results.

\subsection{Bulk perspective}

The calculation of the critical exponents of the thermodynamic system from the bulk perspective proceeds in the same way as in the brane perspective, with minimal changes. We work in the canonical ensemble, with fixed temperature $T$, fixed charge $Q$, fixed angular momentum $J$, fixed pressure $P_{4}$, and fixed brane tension $\tau$. Calculations proceed using the bulk perspective formulas. Using $V_{4}$ as the order parameter, we can define the relevant heat capacity and susceptibility in the same manner as previously,
\begin{equation}
    C_{V_{4},P_{4}}=T\left(\frac{\partial S}{\partial T}\right)_{V_{4},P_{4}} \propto |t|^{-\tilde{\alpha}}\,,
\end{equation}
\begin{equation}
    \chi=-\frac{1}{V_{4}}\left(\frac{\partial V_{4}}{\partial P_{4}}\right)_{\tau,T} \propto |t|^{-\gamma}\,.
\end{equation}
The neutral-static system's critical exponents, calculated from the bulk perspective, match those of the brane perspective (\ref{eq:neutral static critical exponents}). This is insensitive to the choice of order parameter, as long as the susceptibility and heat capacity are defined consistently. The charged-static and neutral-rotating critical exponents are also consistent with the values found in the previous section (\ref{eq:van der Waals critical exponents}). We again stress   that the introduction of charge and spin into the system does not perturb the position of the re-entrant critical point in phase space or variable space; rather, it eliminates the point altogether. Instead, new critical points are found in the regime of large charge or large angular momentum. 

\subsection{Boundary perspective}

From the boundary perspective we work in the microcanonical ensemble, where $\mathcal{S}$, $V_{2}$, $c_{2}$, $\mathcal{Q}$ and $\mathcal{J}$ are fixed. The critical point is found using Maxwell's equal area construction in the $T-\mathcal{S}$ plane, where the presence of metastable states with negative heat capacity are replaced by an isotherm. The two coexisting phases of the $\text{DCFT}_{2}$ are distinguished by different values of the entropy $\mathcal{S}$. The coexistence curve is found numerically by adjusting the thermodynamic volume of the system and keeping track of the movement of the isotherm (based on the equal area requirement) and where it intersects the curve. The critical exponent values for the neutral-static, charged-static, and neutral-rotating quantum black holes are the same as in the brane and bulk perspectives.

The charged rotating black hole poses the same subtlety as the neutral-rotating black hole, in that the critical points in parameter space have imaginary values of $\alpha$, which means that the critical state is not a black hole. Treating it as a formal solution, the corresponding critical exponents are found to be the same as in the charged-static, and neutral-rotating cases (\ref{eq:van der Waals critical exponents}). 

\subsection{Elimination of re-entrant phase transitions}\label{section:Elimination}
The elimination of re-entrant phase transitions with the introduction of electric charge or spin in the thermodynamic system may be shown analytically as follows. We begin with electric charge, and for simplicity, focus on the charged-static quantum BTZ black hole. The following analysis is performed in the brane perspective, but is expected to hold true in the boundary and bulk perspectives as well. The partial derivatives $\big(\frac{\partial T}{\partial S}\big)|_{P_{3}, c_{3}, Q}$ and $\big(\frac{\partial^{2} T}{\partial S^{2}}\big)|_{P_{3}, c_{3}, Q}$ are expressed as functions of $z, \nu$, $\ell_{3}$, and $\chi$. The condition for criticality is that these functions vanish, and the critical points in parameter space are precisely those points where this occurs. If electric charge did not destroy the critical point associated with re-entrant phase transitions, then we would expect a very small charge to merely perturb its position. Instead, any non-zero electric charge (corresponding to any non-zero value of $\chi$) removes the point entirely.

The partial derivatives may be written as rational expressions of polynomials. We can extract the numerators of these rational expressions, which are multivariate polynomials. Any root of the original system of equations will obviously be a root of the system of polynomial equations. Due to the structure of the equations, $\ell_{3}$ is not one of the variables in these polynomials. The original critical point in the neutral-static case had the coordinates $z_{c}$, and $\nu_{c}$. Treating $\chi$ as a new parameter in the equations, it may be demonstrated that these roots take on complex values when $\chi\neq0$.

We may calculate the resultant of our two polynomials with respect to $\nu$. This leaves us with a very high-degree polynomial in $z$ and $\chi$. The resultant will only vanish when the two polynomials share a common root. We can factor the resultant into numerous polynomials. One of those factors has a root with $z=z_{c}$, when we set $\chi=0$. This is the factor that we shall focus on. We will make the assumption that $z$-coordinate of our original root varies smoothly with $\chi$. To that end, we can Taylor expand the relevant factor of the resultant to third-order in $z$, about $z=z_{c}$. This leaves us with a cubic equation in $z$, whose roots can be obtained analytically. This gives us three roots, but only two of them smoothly converge to $z=z_{c}$ as we take the limit $\chi\rightarrow0$. Any non-zero value of $\chi$ introduces an imaginary part to the roots. This can be visualized in the complex plane. At $\chi=0$, the root sits on the real axis. As $\chi$ takes on a very small value (it may be seen from the definition of $\chi$ that it must be real), the root splits into two, and each point moves off the real-axis and takes on an imaginary part. Since $z$ is physically restricted to take on real values, as may be seen from its definition (\ref{eq:dimensionless variables}), this means the critical point has been destroyed, as far as physics is concerned. This analytic approximation can be tested against a numerical solution to the original two equations, and the agreement is found to be excellent for small values of $\chi$. We conclude that electric charge gets rid of re-entrant phase transitions.

The introduction of angular momentum to the system also gets rid of the re-entrant critical point, but in a different manner than charge. It should be noted that the angular momentum function vanishes when we substitute in the values of the neutral-static critical coordinates, no matter what the value of the rotation parameter $\alpha$ is. This means that for an ensemble with non-zero angular momentum, the original critical point cannot be a valid solution. For the sake of simplicity, we will focus on the neutral-rotating case. We proceed as before, calculating the first two partial derivatives of $T$ with respect to $S$ as functions of $z$, $\nu$, $\alpha$, and $\ell_{3}$, and extract the numerator polynomials. Unlike in the charged case, where non-zero values of $\chi$ pulled the original root off the real-axis, a non-zero value of $\alpha$ does not add an imaginary part to the original root. This may be confirmed numerically. So at  first glance, it would appear that the introduction of angular momentum does not destroy our interesting critical point. The resultant method does give us one insight however: it has a factor of $(z-1)$. This means that $z_{c}=1$ will remain a root of the two polynomials for any value of $\alpha$. Substituting this into our two polynomials and solving for $\nu$, confirms that the $\nu$ coordinate of the critical point remains invariant here too, at $\nu_{c}=1$, for any value of $\alpha$. The original critical point remains a root insofar as the partial derivatives vanish there. It is only disqualified when we demand that the angular momentum be non-zero.

We proceed as follows. The relevant functions and variables, along with the conditions that we place on them, are
\begin{align}
    \left.\frac{\partial T}{\partial S}\right|_{P_{3}, c_{3}, J}&=0\,,\\
    \left.\frac{\partial^2 T}{\partial S^2}\right|_{P_{3}, c_{3}, J}&=0\,,\\
    J\in\mathbb{R}\,,\,
        J\neq 0\,,\,
        T&>0\,,\,
        z>0\,,\,
        \nu>0\,,\,
        \ell_{3}>0\,.
\end{align}
These conditions are obviously all physically motivated. We take these functions and expand everything to first order in $\alpha$ about $\alpha=0$. Due to the structure of the functions, this means that only the series for $J$ will be a function of $\alpha$. By analyzing the resulting system of equations, it may be seen that it is impossible for all of these conditions to be satisfied. This shows that for any small but non-zero angular momentum, there are no critical points. This should not be taken to imply that rotating black holes have no critical points. Once we increase the allowed angular momentum to be outside of the perturbative range of small $\alpha$, then the linear expansion in $\alpha$ is no longer valid, and the above conclusion no longer holds true. The fact that the above system of equations and conditions are inconsistent in the limit of small $\alpha$ however, demonstrates that angular momentum removes the original re-entrant critical point of the neutral-static system.

\section{Validity of semiclassical approximation}\label{jmvciope}

Braneworld methods allow for a convenient solution of the semiclassical problem incorporating backreaction of quantum matter to all orders. However, just like in any semiclassical approach, the results are expected to remain valid only when it is legitimate to treat the metric classically, that is, when quantum fluctuations of the geometry remain small. 

It has been recently understood that extremal black holes are highly quantum objects~\cite{Iliesiu:2020qvm}. In the case of extremal black holes in Einstein gravity, the low temperature dynamics is described by a Jackiw-Teitelboim theory that governs the fluctuations of the throat. In the absence of supersymmetry, these quantum fluctuations drive the extremal entropy from its semiclassical value down to zero.

The rotating quantum BTZ black hole and its charged generalization presented here admit extremal limits, akin to their higher-dimensional counterparts~\cite{Maulik:2025hax}. Here we will not study the quantum fluctuations of the geometry, but we will characterize when one can expect the semiclassical description to break down. On general grounds, the black hole entropy near extremality admits an expansion
\begin{equation}
    S = S_0 + \frac{4 \pi^2}{M_{\rm gap}} T + \cdots  \, .
\end{equation}
Here $S_0$ is extremal entropy in the semiclassical approximation, which is just given by the area of the four-dimensional bulk horizon. The energy scale $M_{\rm gap}$ determines when quantum fluctuations of the geometry become important. We can expect the semiclassical approximation to breakdown when  $T < |M_{\rm gap}|$. 

Here we will study the behaviour of $M_{\rm gap}$ for the charged and rotating quantum BTZ black hole, taking account of different cases in turn. We will define the $M_{\rm gap}$ as a simple derivative of the entropy,
\begin{equation} \label{eq:MgapDef}
M_{\rm gap} \equiv 4 \pi^2 \left[ \left(\frac{\partial S}{\partial T} \right)_{W_i; T=0} \right]^{-1} \, ,
\end{equation}
where here the subscript $W_i$ indicates the various thermodynamic potentials that are held fixed in the differentiation --- here we will be mostly interested in the case where the conserved charges $(Q, J)$ are held fixed, along with   theory specific parameters such as $\ell_3$ and $G_3$.

\subsection{Static charged quantum black holes}

Let us first consider the charged static quantum black hole with $\alpha = 0 \Rightarrow J = 0$. The behaviour of $M_{\rm gap}$ for this case was studied in~\cite{Climent:2024nuj}, and we use this as an opportunity to benchmark our results. 

In terms of the thermodynamic parameters, the temperature of the charged  quantum black hole vanishes when
\begin{equation}
    \chi^2_{\rm ext} = \frac{\nu  z^3+3 \nu  z+2}{\nu  z^3 (\nu  z+1)^2} \, .
\end{equation}
At fixed charge $Q$, we then obtain for $M_{\rm gap}$ the following
\begin{equation} 
M_{\rm gap} = \frac{4 G_3}{\ell_3^2} \frac{\left(\nu ^2 z^4+6 \nu ^2
   z^2+8 \nu  z+3\right)}{ \sqrt{\nu ^2+1} (2 \nu  z+1) } \, .
\end{equation}
Here it is to be understood that the quantities are evaluated for an extremal black hole with a specified charge.  This agrees with previous results ~\cite{Climent:2024nuj}  (see eq. (3.55))..

\subsection{Neutral rotating quantum black holes}

Let us now consider the neutral rotating quantum black hole, i.e. with $\chi = 0 \Rightarrow Q = 0$. Notably, there are now \textit{two} possible branches of zero temperature black holes. The condition $T = 0$ is satisfied when
\begin{equation}\label{eq:neutral extremal conditions}
\alpha^2_{\rm I} = \frac{z^2 (\nu  z+1)}{z^4-2 \nu  z^3+1} \, , \quad \text{or} \quad \alpha^2_{\rm II} = \frac{-\nu  z^3-3 \nu  z-2}{z \left(3 \nu +\nu  z^4-4 z\right)} \, .
\end{equation}
The first of these two possibilities preserves the classical rotating BTZ extremality condition, or three-dimensional ``Kerr bound''. Specifically, we have
\begin{equation} 
\frac{|J|}{M \ell_3} = 1 \quad \text{for} \quad \alpha^2 = \alpha_{\rm I}^2 \, .
\end{equation}
 As was explained already in~\cite{Emparan:2020znc}, for this branch of extremal black holes the quantum matter stress tensor vanishes and the CFT does not backreact on the geometry. These zero temperature black holes have $\bar{a}^2 = 1$, and it is because of this relationship that the temperature vanishes. The function $H(r)$ of the four-dimensional metric does not have a double root in this case. However, restricting to the brane perspective, the near horizon geometry of this extremal solution does indeed exhibit the usual features. 

On the other hand, the second branch of extremal solutions satisfies
\begin{equation}
    \frac{|J|}{M \ell_3} = 1 + \frac{\nu  \left(1-z^2\right)^2}{2 z} + \mathcal{O}(\nu^2) \quad \text{for} \quad \alpha^2 = \alpha_{\rm II}^2 \, .
\end{equation}
This branch saturates the three-dimensional Kerr bound in the limit of vanishing backreaction. For small but nonzero backreaction, these extremal solutions \textit{exceed} the three-dimensional Kerr bound. 

\begin{figure}[t]
    \centering
    \includegraphics[width=0.45\linewidth]{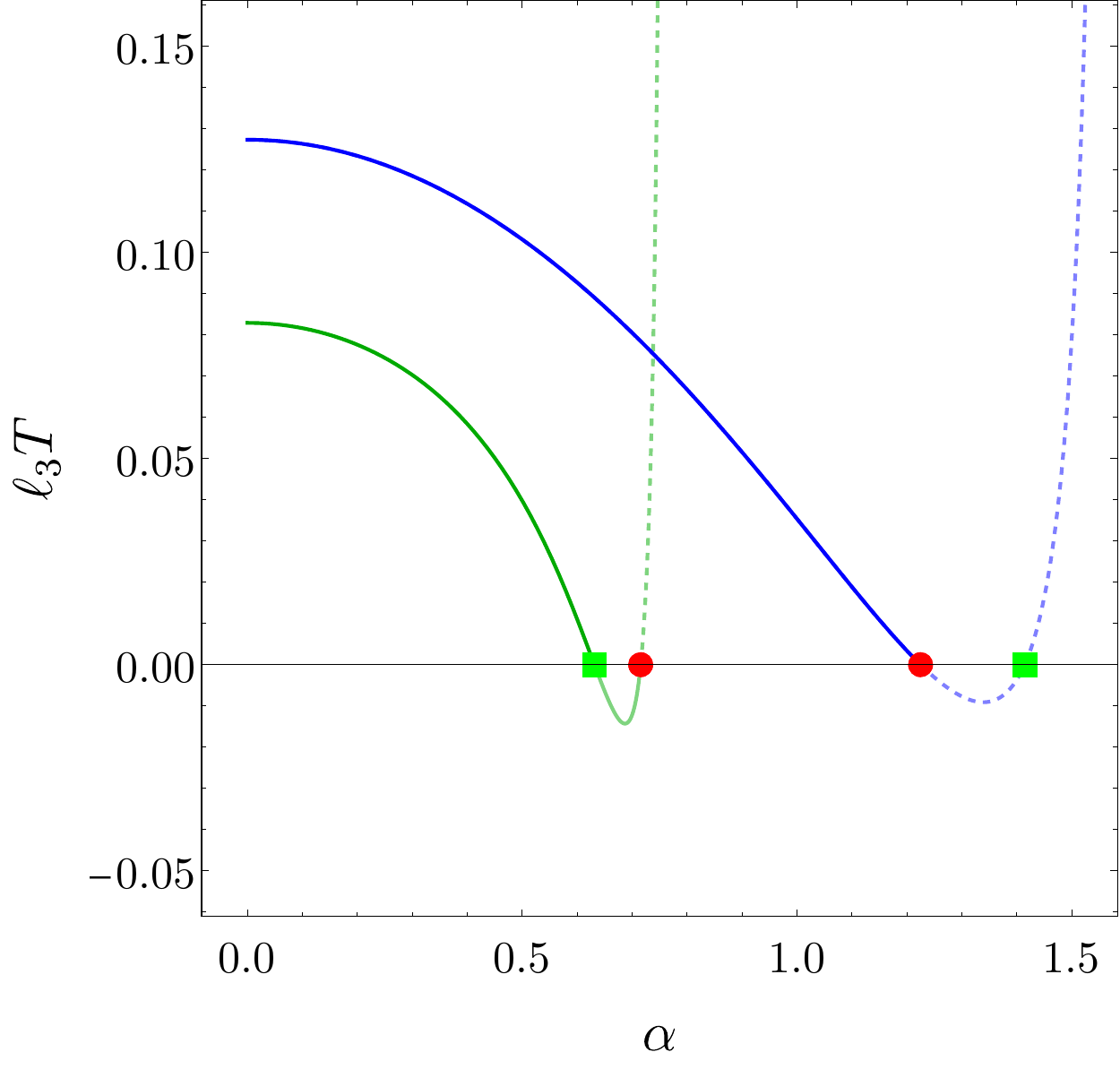}
    \quad 
    \includegraphics[width=0.45\linewidth]{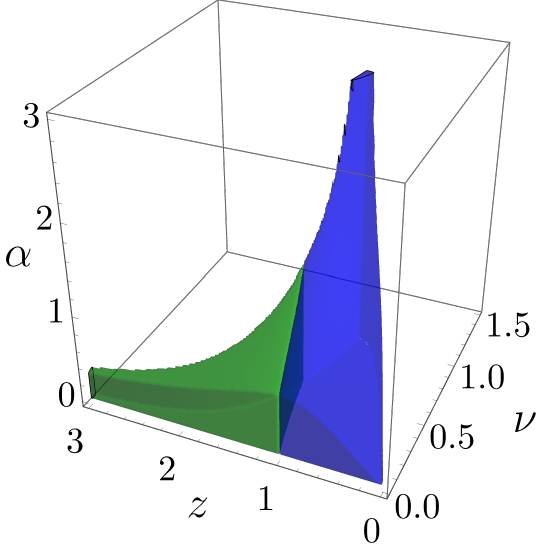}
    \caption{Left: A plot of the temperature as a function of $\alpha$ for the neutral rotating quantum black hole. Here the red dots indicate the extremal solutions with $\alpha = \alpha_{\rm I}$ while the green squares are the extremal solutions with $\alpha = \alpha_{\rm II}$. The blue curve corresponds to $(z, \nu) = (1, 1/2)$ while the green curve corresponds to $(z, \nu) = (1.5, 0.125)$. The dashed portions of the curves have $\bar{a}^2 > 1$ and are therefore unphysical. Right: The $(z, \nu, \alpha)$ parameter space color-coded according to the type of extremal solutons. For those parameters in the green region (leftmost side of the plot) the $\alpha_{\rm II}$ extremal solution is reached first. For those parameters in the blue region (rightmost side of the plot) the $\alpha_{\rm I}$ extremal solution is reached first, while the $\alpha_{\rm II}$ solution is unphysical with $\bar{a}^2 > 1$.}
    \label{fig:temp_plot}
\end{figure}

To allow for a better understanding of how these different zero temperature limits arise we   plot the temperature as a function of the parameter $\alpha$ for fixed values of $z$ and $\nu$ in Fig.~\ref{fig:temp_plot}. Given these parameters, we determined the range of $\alpha$ consistent with the physicality constraints of the metric. The behaviour of temperature and the set of physical solutions depends on the choice of $(\nu, z)$ parameters. We illustrate in  Fig.~\ref{fig:temp_plot} the two characteristic behaviours with  blue and green curves.

Let us consider first the blue curve. Increasing $\alpha$ from zero, the temperature begins to decrease and crosses zero when $\alpha = \alpha_{\rm I}$. This corresponds to $\bar{a}^2 = 1$ and all larger values of $\alpha$ are unphysical, corresponding to $\bar{a}^2 > 1$. Turning next to the green curve, the temperature once again decreases as $\alpha$ is increased. This time, the temperature vanishes first when $\alpha = \alpha_{\rm II}$. Larger values of $\alpha$ are consistent with the most natural physical criteria, i.e. demanding $\mu x_1 > 0$, $r_+ > 0$, $\bar{a}^2 < 1$, but possess negative temperature and are therefore discarded. Now, when $\alpha$ is further increased we reach $\alpha = \alpha_{\rm I}$, indicated by the red circle on the green curve. This is a completely isolated but otherwise physically acceptable solution. We cannot increase $\alpha$ any further beyond this point as doing so would result in $\bar{a}^2 > 1$. On the other hand, the solution cannot have its value of $\alpha$ sensibly decreased as doing so would result in a negative temperature configuration. 

Summarizing, for the neutral rotating black hole, there are two different choices of $\alpha$ that yield zero temperature quantum black holes, $\alpha_{\rm I}$ and $\alpha_{\rm II}$. Which of these extremal solutions exist depends on the combination of parameters. Generally, when $\alpha_{\rm I} < \alpha_{\rm II}$ then only $\alpha_{\rm I}$ is physical. When $\alpha_{\rm II} < \alpha_{\rm I}$ then both extremal solutions are technically physical, but the one corresponding to $\alpha_{\rm I}$ is highly exotic. It is completely isolated in parameter space, as the value of $\alpha$ cannot be made larger or smaller without violating physicality conditions. This exotic behaviour is consistent with the fact that, from the four-dimensional bulk perspective, the $\alpha_{\rm I}$ extremal solutions do not correspond to a double root of the $H(r)$ metric function. Moreover, in all cases, the $\alpha_{\rm I}$ solutions have completely vanishing backreaction, with vanishing quantum stress tensor. The most conservative perspective is that, while the $\alpha_{\rm II}$ extremal solutions are perfectly fine when physical, the $\alpha_{\rm I}$ must be regarded with care.

Let us now analyse when we expect the semiclassical approximation to break down in the vicinity of these black holes.  Holding fixed the angular momentum, we can compute $M_{\rm gap}$ for each of the two types of extremal solution. We find 
\begin{align}
    M_{\rm gap}^{\rm I} &= \frac{4 G_3}{\ell_3^2} \frac{\left(\nu  z^3+2 z^2-3 \nu  z-2\right)^2}{\sqrt{\nu ^2+1} 
     (\nu  z+1) \left(z^4-4 \nu  z^3-2 z^2+1\right)} \, , 
     \\
    M_{\rm gap}^{\rm II} &= \frac{G_3}{\ell_3^2} \frac{\left(\nu  z^3+2 z^2-3 \nu  z-2\right) \left(-3 \nu +\nu  z^4+6 \nu 
   z^2+8 z\right)}{\sqrt{\nu ^2+1} (z-1) z (z+1) (\nu  z+1)} \, .
\end{align}
In the limit of small backreaction, both $M_{\rm gap}^{\rm I}$ and $M_{\rm gap}^{\rm II}$ limit to the expected value for the classical rotating BTZ black hole,
\begin{align}
    M_{\rm gap}^{\rm I} &= \frac{16 G_3}{\ell_3^2} \left[1  +\frac{2 \nu  \left(z^3+z\right)}{\left(z^2-1\right)^2} +\mathcal{O}(\nu^2) \right] \, , 
     \\
    M_{\rm gap}^{\rm II} &= \frac{16 G_3}{\ell_3^2} \left[1  -\frac{\nu  \left(z^6+z^4-13 z^2+3\right)}{8 z-8 z^3} +\mathcal{O}(\nu^2) \right] \, .
\end{align}
We see that the corrections to the classical result differ. 

\begin{figure}
    \centering
    \includegraphics[width=0.5\linewidth]{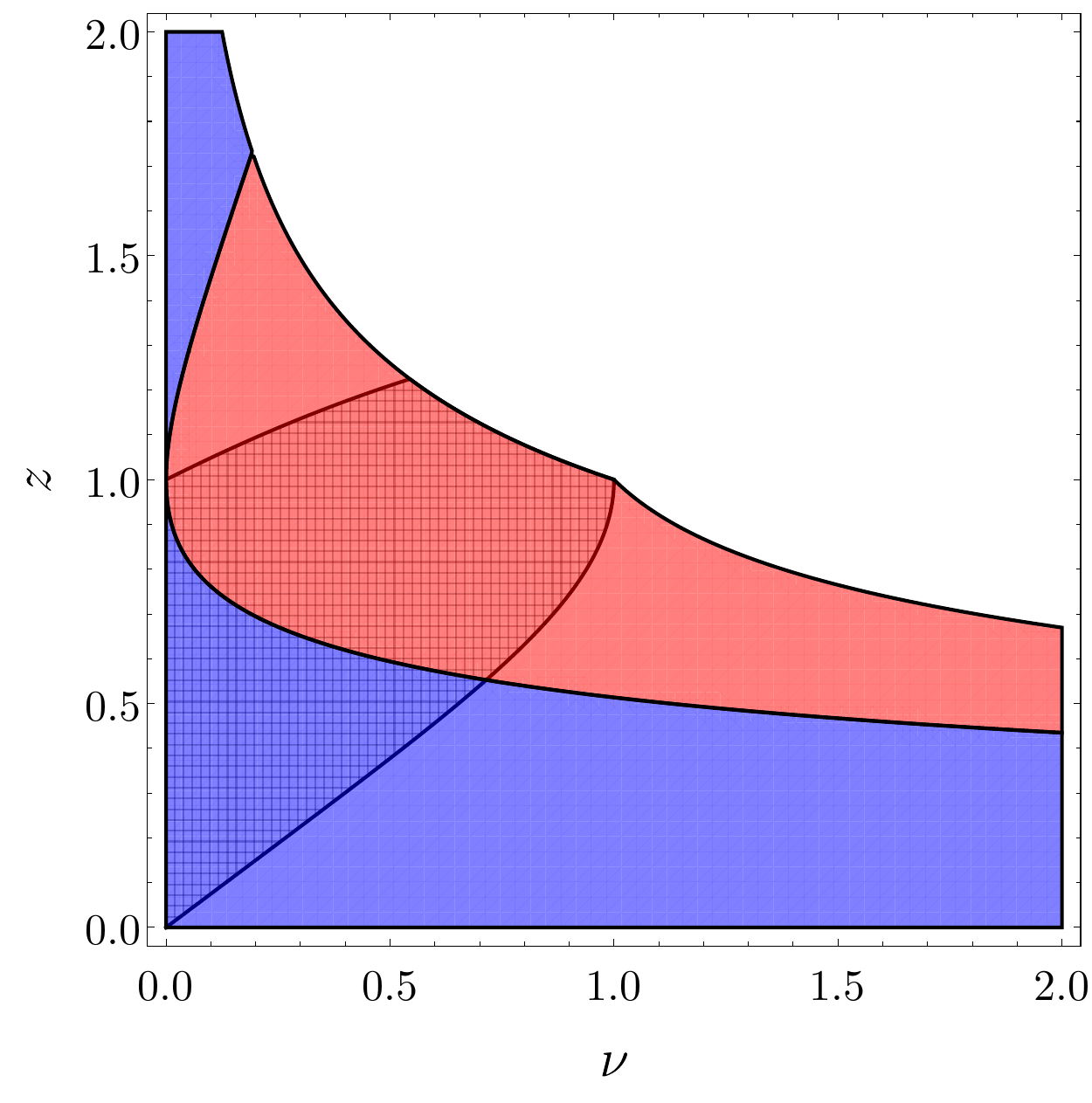}
    \caption{We show the $(\nu, z)$ parameter space for the neutral rotating quantum black hole characterizing the regions where $M_{\rm gap}^{\rm I}$ is negative (red) and positive (blue).  The black cross-hatched portion of the parameter space shows where $\alpha_{\rm I} < \alpha_{\rm II}$.}
    \label{fig:neg_mgap}
\end{figure}

Although it is not obvious from inspection, when implementing all physicality conditions $M_{\rm gap}^{\rm II}$ can be shown to be strictly positive. This is in line with all the well-known higher-dimensional black holes. On the other hand, $M_{\rm gap}^{\rm I}$ is \textit{not} always positive --- see Fig.~\ref{fig:neg_mgap}. This bizarre feature further solidifies the exotic nature of these solutions. We shall comment in more detail on this feature and its relation to weak cosmic censorship below. 

\begin{figure}
    \centering
    \includegraphics[width=\linewidth]
{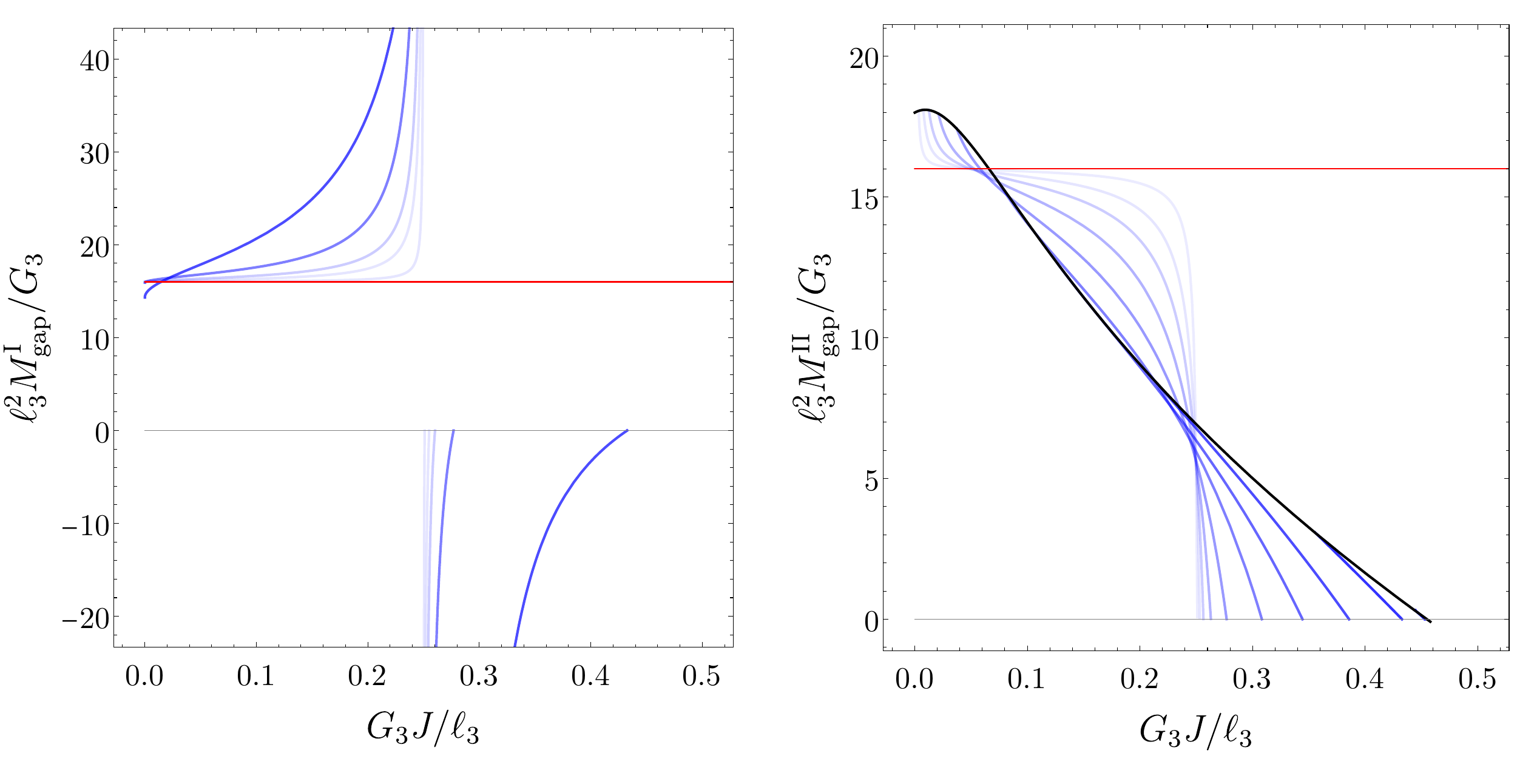}
    \caption{Left: $M_{\rm gap}^{\rm I}$ as a function of the angular momentum for difference choices of backreaction parameter $\nu$. Right: $M_{\rm gap}^{\rm II}$ as a function of the angular momentum for difference choices of backreaction parameter $\nu$. In each case, the darker the shade of blue, the larger the value of $\nu$ for that curve. The thin red lines in each plot show the value of $M_{\rm gap}^{\rm BTZ} = 16 G_3/\ell_3^2$ for the classical rotating BTZ black hole. On the right plot, the black line is $z = \nu^{-1/3}$ which marks the boundary of the physical parameter space. In both cases, $M_{\rm gap} \to 0$ when the angular momentum takes its largest value for a given $\nu$.}
    \label{fig:Mgap_vs_J}
\end{figure}

We consider the dependence of $M_{\rm gap}$ on the angular momentum in Fig.~\ref{fig:Mgap_vs_J}. The left plot shows the result for $M_{\rm gap}^{\rm I}$. As mentioned above, this extremal solution does not have a strictly positive $M_{\rm gap}$. For a given value of the backreaction parameter $\nu$, $M_{\rm gap}^{\rm I}$ initially increases as the angular momentum increases. This continues until, at a finite value of $J$, $M_{\rm gap}^{\rm I}$ diverges to infinity. For $J$ just larger than this critical value, $M_{\rm gap}^{\rm I} \to -\infty$. As the angular momentum is further increased, $M_{\rm gap}^{\rm I}$ increases but remains negative, terminating at $M_{\rm gap}^{\rm I} = 0$ for the maximum value of $J$. 

The behaviour of $M_{\rm gap}^{\rm II}$ is shown in the right panel of the same figure. The behaviour in this case is more reasonable, with $M_{\rm gap}^{\rm II}$ exhibiting a monotonic behaviour, decreasing as the angular momentum increases toward its maximum value. Note that in this case, the physical parameter space terminates at $z = 1/\nu^3$, which is indicated by the solid black line in the figure. For fixed $\nu$, the blue curves originate along this line and then tend toward zero as $J$ is increased. 

\subsection{Charged and rotating quantum black holes}

\begin{figure}
    \centering
    \includegraphics[width=\linewidth]{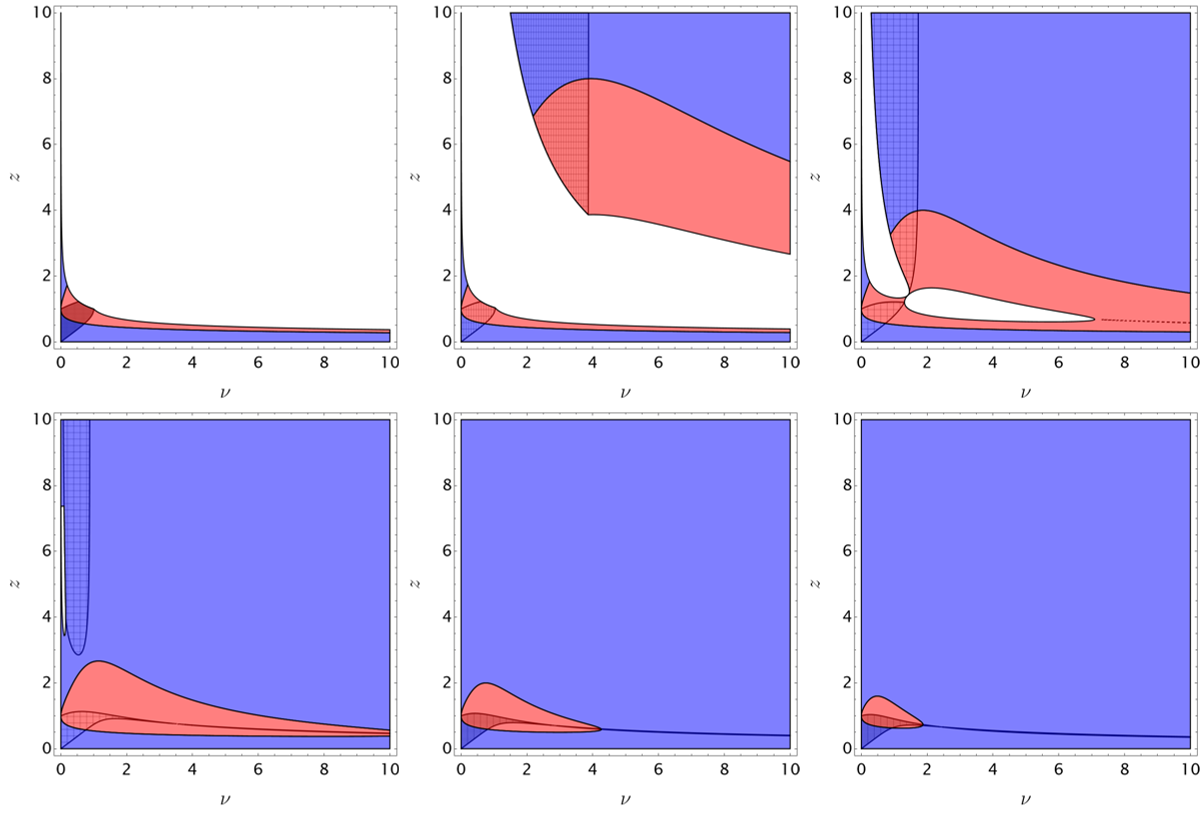}
    \caption{We extend Fig. \ref{fig:neg_mgap} to show how the regions in $(\nu, z)$ parameter space where $M_{\rm gap}^{\rm I}$ is negative evolves as $\chi$ increases. From the top left across and down, we have shown the regions for $\chi=0, 1/4, 1/2, 3/4, 1, 5/4.$ Just as before, the red-shaded portion of the parameter space corresponds to $M_{\rm gap}^{\rm I} < 0$ and the black cross-hatched portion of the parameter space shows where $\alpha_{\rm I} < \alpha_{\rm II}$. Note that increasing $\chi$ increases the portion of parameter space where there exist solutions satisfying physicality conditions, and shrinks the portion where $M_{\rm gap}^{\rm I}$ is negative. The apparent dashed curve is a thin section of meshed area.}
    \label{fig:neg_mgap_charged}
\end{figure}

We now extend the preceding results to the case with non-zero charge. The conditions for $T=0$ generalize from \eqref{eq:neutral extremal conditions} to
\begin{equation}
\begin{aligned}
\alpha^2_{\rm I} &= \frac{z^2 (\nu  z+1)}{z^3 \left(-2 \nu +\nu  \chi ^2 (\nu  z+1)+z\right)+1}\,,\\
\alpha^2_{\rm II} &= \frac{\nu  z \left(z^2 \left(1-(\chi +\nu  \chi  z)^2\right)+3\right)+2}{z \left(-3 \nu +\nu  z^4 \left(\left(\nu ^2+1\right) \chi ^2-1\right)+4 z\right)}\,,
\end{aligned}
\end{equation}
and the respective mass gaps for these extremal cases become
\begin{align}
    M_{\rm gap}^{\rm I} &= \frac{4G_{3}}{\ell_{3}^{2}}\frac{\Xi^2}{\sqrt{\nu ^2+1} \left(z^2+1\right)^2(\nu  z+1) \left(z^2 \left(z \left(-4 \nu +\nu  \chi ^2 (\nu  z+1)+z\right)-2\right)+1\right)} \, , 
     \\
    M_{\rm gap}^{\rm II} &= \frac{4G_{3}}{\ell_{3}^{2}}\frac{\Xi\left(3 \nu +\nu  \left(\nu ^2+1\right) \chi ^2 z^4-z \left(\nu  z \left(z^2+6\right)+8\right)\right)}{\sqrt{\nu ^2+1} z (\nu  z+1) \left(z^2 \left(\nu  \chi ^2 z+2\right)-2\right) \left(z^2 \left(\nu  \chi ^2 z (\nu  z+1)-2\right)-2\right)} \, ,
\end{align}
where $\Xi=z^3 \left(-2 \nu +\nu  (\chi +\nu  \chi  z)^2+z (\nu  z+2)\right)-3 \nu  z-2$. Just as for the neutral rotating quantum black holes, $M_{\rm gap}^{\rm II}$ is strictly positive under physical conditions, while $M_{\rm gap}^{\rm I}$ is not always positive. Fig. \ref{fig:neg_mgap} can be shown to evolve under non-zero $\chi$, with the effect that in the limit of large positive $\chi$, the region in $(\nu, z)$ parameter space where $M_{\rm gap}^{\rm I}$ is negative shrinks --- see Fig. \ref{fig:neg_mgap_charged}.

Note that the extremal solution parameters $\alpha_{\rm I}^{2}$ and $\alpha_{\rm II}^{2}$ are equal when
\begin{equation}\label{eq:neutral-rotating extremal conditions}
\chi^2_{1} = \frac{\nu  z^3-1}{\nu  z^3 (\nu  z+1)} \, , \quad \text{or} \quad \chi^2_{2} = \frac{-\nu  z^5-2 z^4+2 \nu  z^3+3 \nu  z+2}{\nu  z^3 (\nu  z+1)^2} \, .
\end{equation}
The corresponding values that the extremal rotation parameter takes are
\begin{equation}
\alpha^2_{1} = \frac{\nu  z+1}{z^2-\nu  z} \, , \quad \text{or} \quad \alpha^2_{2} = -\frac{z^2 (\nu  z+1)^2}{\left(2 \nu ^2+1\right) z^4-4 \nu  z-3} \, .
\end{equation}
The pair $(\chi_{1}, \alpha_{1})$ can be dismissed, since their substitution yields divergent $x_{1}^{2}$ and $r_{+}^{2}$. However, the second pair $(\chi_{2}, \alpha_{2})$ is physically acceptable, since it admits solutions that satisfy all physical conditions. The (coincident) mass gap that results from this value of $\chi$ is
\begin{equation}
    M_{\rm gap}^{\rm I}(\chi_{2}, \alpha_{2})=M_{\rm gap}^{\rm II}(\chi_{2}, \alpha_{2})=0\,.
\end{equation}
This is an extremal value of the mass gap itself, as under physical conditionsis impossible for $M_{\rm gap}^{\rm II}$ to be negative. The mass gap vanishing at extremality only occurs in this coincident limit for the charged and rotating black hole.  This implies that quantum fluctuations of the geometry become arbitrarily important even at infinitesimally small temperatures, which may signal a deeper, nontrivial quantum gravitational regime, distinct from conventional semiclassical black hole thermodynamics. Specifically, the coincident limit marks a boundary in parameter space at which the usual semiclassical picture fails entirely, suggesting that these particular extremal charged rotating quantum black holes could serve as a window into a novel quantum gravitational phase.

\subsection{Connection with weak cosmic censorship}

Weak cosmic censorship (WCC) --- the idea that there are no ``naked'' singularities --- is a cornerstone of our modern perspective on gravitational collapse and black holes. However, there is to date no \textit{proof} that weak cosmic censorship is true. Instead, over the last half century, WCC has been subjected to a battery of consistency tests. Largely, these tests have shown that WCC is robust, and in those cases where counterexamples have been found in the classical theory, one is forced into a regime where quantum gravitational corrections ultimately cannot be ignored~\cite{Emparan:2020vyf}. 

A popular class of tests of WCC are the \textit{gedankenexperiments} where one attempts to overcharge or overspin an (near) extremal black hole by throwing in a test particle carrying charge or angular momentum~\cite{Wald:1974hkz}. It was once believed that these thought experiments \textit{could} succeed in creating a naked singularity~\cite{Hubeny:1998ga}. However, in recent years it has been understood that this is an artifact of failing to account for higher-order perturbations induced by the test particle~\cite{Sorce:2017dst}. 

Taking this idea further, a series of recent studies have identified a connection between the robustness of WCC in gedankenexperiments and the positivity of a single parameter ---  called ``$W$'' in their notation~\cite{Wu:2024ucf, Lu:2025ntu}. As defined in, e.g., eq.~(9) of~\cite{Wu:2024ucf},
\begin{equation}
    W \equiv \lim_{T \to 0} \left(\frac{\partial S}{\partial T} \right)_{Q_i} \, ,
\end{equation}
where $Q_i$ denote conserved charges which are held fixed during the physical process.  These investigations find that gedankenexperiments cannot succeed if $W > 0$, while if $W < 0$ then it may be the case that a black hole can be overcharged/overspun. Obviously, the parameter $W$ is, up to unimportant factors, the same as $M_{\rm gap}$ as we defined it in~\eqref{eq:MgapDef}. We will make several comments about this, though a full analysis of these ideas would take us well outside the scope of the present article. 

First let us note the following. In~\cite{Frassino:2024fin}, the authors studied gedankenexperiments for the neutral rotating quantum black hole. Their work focused entirely on the physically sensible $\alpha_{\rm II}$ extremal solutions. The results presented there are consistent with the neutral, rotating quantum BTZ respecting WCC. It is not immediately obvious that the results of~\cite{Wu:2024ucf, Lu:2025ntu} apply directly to quantum black holes, and it would be interesting to analyse this carefully. That is, to assess whether $W > 0$ remains a sufficient condition for robustness of WCC against perturbations due to classical test particles.  Our finding that $M_{\rm gap}^{\rm II}$ is strictly positive for the neutral rotating quantum black hole is fully consistent with the explicit results obtained in~\cite{Frassino:2024fin}. We also observed that $M_{\rm gap}$ is strictly positive for the charged static quantum black hole, suggesting that WCC will be robust to overcharging experiments. 

On the other hand, we observed that  for the rotating solutions it is possible for $M_{\rm gap}^{\rm I}$ to be \textit{negative}. As we have explained at several instances, these black holes are highly exotic. The extremality condition is met due to $\bar{a}^2 = 1$ and the four-dimensional bulk solution does not have a double root for the metric function $H(r)$ in this case. The stress tensor of the quantum matter vanishes for these extremal solutions, and the brane geometry is identical to the extremal classical rotating BTZ black hole.  Nonetheless, supposing that a connection such as that outlined in~\cite{Wu:2024ucf, Lu:2025ntu} remains valid, then this suggests that these black holes are candidates for overspinning/overcharging. However, what is less clear is whether such a gedankenexperiment would potentially result in the creation of a naked singularity or if this would signal the onset of a dynamical instability instead. 

A closely related point is the following. Since it is now known that quantum fluctuations of the geometry are non-negligible for $T < M_{\rm gap}$, it is not sensible to treat black holes very close to extremality as having a well-defined semiclassical geometry. This calls for a reconsideration of the classical gedankenexperiments to see whether they continue to hold up in this regime. Supposing that the exotic quantum black holes continue to admit a Schwarzian description at low temperatures, then the fact that $M_{\rm gap}^{\rm I} < 0$ is suggestive of a dynamical instability. This is because in the one-dimensional effective description provided by the Schwarzian, $M_{\rm gap}$ plays the role of a coupling ---  see, for example, eq.~(5.2) of~\cite{Iliesiu:2020qvm} or eq.~(3.13) of~\cite{Emparan:2023ypa}. Thus changing the sign of $M_{\rm gap}$ suggests the Schwarzian mode becomes a ghost.\footnote{RAH is grateful to Roberto Emparan for discussions on this point, and suggesting the appearance of ghost instabilities.} 

We can speculate a bit further on how such a potential instability could manifest. Let us recall the structure of the semiclassical theory living on the brane. This consists of an infinite tower of higher-curvature terms coupled to a CFT. As explained in~\cite{Aguilar-Gutierrez:2023kfn, Bueno:2023dpl}, in the absence of the CFT, the spectrum of the gravitational theory on the brane consists of an infinite tower of massive ghost modes. Combining this action with the CFT cures the otherwise pathological behaviour of the purely gravitational theory. However, as mentioned above, the exotic extremal solutions have $\bar{a}^2 = 1$ and there is therefore no coupling with the CFT. These black holes are therefore solutions of the corresponding purely gravitational theory on the brane. It thus seems at least plausible that perturbations of these exotic extremal solutons that do not excite the CFT could succumb to ghost instabilities. It would be interesting, though rather complicated, to understand the stability of these solutions from both a bulk and brane perspective. Since the four-dimensional configuration is a solution of the Einstein (Maxwell) equations, there could be an interesting and intricate connection between the two perspectives.

\section{Quantum inequalities}\label{jfoip3928}

\subsection{Quantum Penrose inequality}

The classical black holes of general relativity obey a number of geometric inequalities that place bounds on their physical properties. Perhaps the most famous of these is the Penrose inequality, which is a conjectured bound on the mass of an asymptotically flat spacetime in terms of the area of black hole horizons it contains~\cite{Penrose:1973um},
\begin{equation}\label{eq:flat_pen} 
G_4 M_{\rm ADM} \ge \sqrt{\frac{A}{16 \pi}} \, .
\end{equation}

Besides providing a lower bound on the mass, the Penrose inequality is important for the role it plays in the weak cosmic censorship hypothesis. Specifically, assuming weak cosmic censorship and assuming gravitational collapse produces a Kerr black hole, then~\eqref{eq:flat_pen} must follow. Therefore, establishing a counterexample to the Penrose inequality would suggest that at least one of these two assumptions must be wrong. In turn, a proof of the Penrose inequality is generally expected to greatly inform our understanding of weak cosmic censorship. 

A more subtle question concerns what happens when quantum effects become important. Since many expect that quantum gravity will cure the singularity problem entirely, it is perhaps reasonable to expect that weak quantum effects should not make the classical singularity problem \textit{worse}. A sharp question is whether there exists a version of weak cosmic censorship that holds in semiclassical gravity. While this is at present not clear, there have been attempts to construct generalizations of the Penrose inequality valid in semiclassical gravity, called \textit{quantum} Penrose inequalities. The hope is that these inequalities, if true, will contribute to our understanding of quantum cosmic censorship~\cite{Emparan:2002px, Engelhardt:2024hpe, Frassino:2025buh}.

The first such construction was carried out in \cite{Bousso:2019var, Bousso:2019bkg}, where it was also shown that the classical Penrose inequality is \textit{badly} violated by quantum effects. Instead, it was argued, a valid inequality can be obtained in $D \ge 4$ by replacing the area term in the Penrose inequality by the corresponding generalized entropy,
\begin{equation} 
A \to 4\mathcal{G}_{\rm D} S_{\rm gen} \, .
\end{equation}
There is evidence that the resulting inequalities are indeed satisfied, but this is limited to the perturbative regime due to the general complexities of semiclassical gravity.

Braneworld holography offers a promising avenue for testing such inequalities beyond the perturbative regime. Indeed, already in the context where the bulk is classical, strong holographic arguments have been obtained in favor of the validity of the Penrose inequality, e.g.,~\cite{Engelhardt:2019btp, Engelhardt:2020mme, Folkestad:2022dse}. Considering semiclassical gravity, a quantum Penrose inequality was recently proposed in three-dimensions~\cite{Frassino:2024bjg},
\begin{equation}\label{eq:quant_pen}
8 \mathcal{G}_3 \left(M - M_{\rm cas} \right) \ge \frac{1}{\ell_3^2} \left(\frac{2 \mathcal{G}_3 S_{\rm gen}}{\pi}\right)^2 \, ,
\end{equation}
where $M_{\rm cas}$ is the Casimir contribution to the mass. In~\cite{Frassino:2024bjg}, it was demonstrated that~\eqref{eq:quant_pen} holds for all known quantum black holes constructed via braneworld holography, even nonperturbatively in the backreaction. This does not prove that a quantum Penrose inequality exists in three-dimensions, but it does provide strong evidence in favour of it. 

To further stress-test this proposal, here we examine its validity for the charged and rotating quantum black hole. This turns out to be a more complicated endeavor than any of the quantum black holes considered in~\cite{Frassino:2024bjg}. In that case, the solutions and parameter constraints were sufficiently simple that Mathematica could be used to prove the validity of~\eqref{eq:quant_pen} exactly. Here, we find this to be impossible, and we resort to a combination of numerical techniques. 

The first technique makes use of Mathematica's ``\textsc{NMinimize}'' routine. We specify particular values of the charge and rotation parameters $(\chi, \alpha)$ and then require Mathematica to find a minimum of the functional
\begin{equation}
    8 \mathcal{G}_3 \left(M - M_{\rm cas} \right) - \frac{1}{\ell_3^2} \left(\frac{2 \mathcal{G}_3 S_{\rm gen}}{\pi}\right)^2 \, ,
\end{equation}
subject to constraints that ensure the physical viability of the solution. Specifically, we require the positivity of $x_1^2, r_+^2, \mu x_1$, as given in eq.~\eqref{eq:dimensionless expressions}, along with the positivity of the temperature, $z$, and $\nu$. We repeat the minimization procedure for a large set of $(\chi, \alpha)$ parameters. We find in all cases that the minimum of the functional is positive, indicating no violation of the conjectured quantum Penrose inequality.

The second method makes use of Mathematica's ``\textsc{FindInstance}'' routine. We specify two parameters of the solution, e.g., $(\alpha, \chi)$ and then use \textsc{FindInstance} to generate a large set of the remaining parameters such that all the constraints just described are satisfied. We then test the validity of~\eqref{eq:quant_pen} on the numerically constructed data. We repeat the same procedure for different choices of the manually specified parameter values. In no cases do we find violations of~\eqref{eq:quant_pen}. 

In summary: our numerical investigations are consistent with the validity of~\eqref{eq:quant_pen} for charged and rotating quantum black holes.

\subsubsection{Penrose inequality with angular momentum}

The Penrose inequality is one member of a hierarchy of conjectured inequalities that bound the mass in terms of the horizon area and additional conserved charges~\cite{Mars:2009cj}. For example, in four-dimensional Einstein-Maxwell theory, the following generalized Penrose inequality is expected to hold~\cite{Dain:2013qia},
\begin{equation}\label{eq:q_J_pen}
    m^2 \ge \frac{A}{16 \pi} + \frac{q^2}{2} + \frac{\pi \left(q^4 + 4J^2\right)}{A} \quad \text{when} \quad A \ge 4 \pi \sqrt{q^4 + 4 J^2} \, .
\end{equation}
(We refer the reader to~\cite{Dain:2013qia} for a discussion of the conventions in this formula). The inequality is saturated by the Kerr-Newman solution. The constraint on the area arises because it is only when $A \ge 4 \pi \sqrt{q^4 + 4 J^2}$ that the functional on the right-hand-side of the inequality is monotonically increasing as a function of $A$. When the charge and rotation vanish, the inequality reduces to the familiar Penrose inequality but otherwise is a stricter constraint.

It is natural to wonder whether there exist quantum inequalities that incorporate additional conserved charges. The charged and rotating quantum black hole provides us with a means to check the viability of such statements. To obtain one such inequality, we will follow the same chain of reasoning that led to the inequality~\eqref{eq:q_J_pen}. 

We shall pursue an inequality that incorporates the angular momentum. The first step is to obtain an identity that holds for the \textit{classical} rotating BTZ geometry. It is straightforward to verify that
\begin{equation}
    8 \mathcal{G}_3 M = \frac{A^2}{4 \pi^2 \ell_3^2} + \frac{64 \pi^2 \mathcal{G}_3^2 J^2}{A^2} \quad \text{for classical rotating BTZ\,.}
\end{equation}
The function on the right-hand-side is an increasing function of $A$ provided that
\begin{equation}
    A \ge 4 \pi \sqrt{\mathcal{G}_3 \ell_3 J} \, .
\end{equation}
This suggests the following ``naive'' quantum Penrose inequality with angular momentum
\begin{equation}
    8 \mathcal{G}_3 M \ge \frac{1}{\ell_3^2} \left(\frac{2 \mathcal{G}_3 S_{\rm gen}}{\pi} \right)^2 + \left(\frac{ 2\pi J}{S_{\rm gen}} \right)^2 \quad \text{when} \quad S_{\rm gen} \ge \pi \sqrt{\frac{\ell_3 J}{\mathcal{G}_3}}  \quad \text{(naive)} \, .
\end{equation}

The above naive inequality does not hold for quantum black holes. The reason for this is simple: saturation of the inequality by the classical solution is too strong a requirement in three dimensions. In four and higher dimensions, the case $A = 0$ corresponds to the vacuum. But this is not so in three dimensions, as there is a mass gap: there exist naked conical singularities whenever~\cite{Casals:2019jfo}
\begin{equation}
    - \sqrt{1 + \left(8 \mathcal{G}_3 J\right)^2} \le 8 \mathcal{G}_3 M < 0 \, .
\end{equation}
These naked singularities become cloaked by event horizons when quantum matter backreacts on the geometry. This generates horizons of non-vanishing area in regions of parameter space where they would be classically forbidden. The naive inequality fails to account for this fact, and this is the reason why it fails to hold. To remedy this, the mass appearing in the inequality must be shifted so that the zero point now coincides with the boundary of the parameter space containing naked conical singularities.  The same reasoning applies to the nonrotating case, as explained in~\cite{Frassino:2024bjg}. The result is then,
\begin{equation}
    8 \mathcal{G}_3 \left(M + \frac{\sqrt{1 + \left(8 \mathcal{G}_3 J\right)^2}}{8 \mathcal{G}_3}\right) \ge \frac{1}{\ell_3^2} \left(\frac{2 \mathcal{G}_3 S_{\rm gen}}{\pi} \right)^2 + \left(\frac{ 2\pi J}{S_{\rm gen}} \right)^2 \quad \text{when} \quad S_{\rm gen} \ge \pi \sqrt{\frac{\ell_3 J}{\mathcal{G}_3}}  \, .
\end{equation}
When $J \to 0$, this inequality recovers the quantum Penrose inequality~\eqref{eq:quant_pen} first presented in~\cite{Frassino:2024bjg}.  With the mass shifted as above, the inequality is no longer saturated in the classical limit, but it is still satisfied. 

If the above proposal is to be considered viable, then it must hold for all quantum black holes. We have numerically investigated the validity of the inequality for the neutral and charged rotating quantum black holes using the same techniques described in the previous section. We have found no counterexamples within the scanned parameter space. 

In fact we have found that a slightly stronger inequality holds for the charged and rotating quantum black hole
\begin{equation}
    8 \mathcal{G}_3 \left(M - M_{\rm cas}\right) \ge \frac{1}{\ell_3^2} \left(\frac{2 \mathcal{G}_3 S_{\rm gen}}{\pi} \right)^2 + \left(\frac{ 2\pi J}{S_{\rm gen}} \right)^2 \quad \text{when} \quad S_{\rm gen} \ge \pi \sqrt{\frac{\ell_3 J}{\mathcal{G}_3}}  \, ,
\end{equation}
where 
\begin{equation}
    M_{\rm cas} = - \frac{1}{8 \mathcal{G}_3} \, .
\end{equation}
To assess whether this stronger inequality holds in general will require a study of the $\kappa = +1$ quantum black holes. This would be a good target for future work.

\subsection{Quantum reverse isoperimetric inequality}

Within the context of extended phase space thermodynamics, it was conjectured that asymptotically AdS black holes obey the \textit{reverse} of the familiar isoperimetric inequality for surfaces in Euclidean space~\cite{Cvetic:2010jb}, 
\begin{equation}
    \mathcal{R} \equiv \left( \frac{(D-1) V_{\rm th}}{\Omega_{D-2}}\right)^{\frac{1}{D-3}} \left(\frac{\Omega_{D-2}}{A} \right)^{\frac{1}{D-2}} \ge 1 \, .
\end{equation}
Physically, this conjecture captures the idea that, at fixed thermodynamic volume, the entropy is maximized for the Schwarzschild-AdS black hole. 

Over the last decade, this inequality has been subjected to a variety of tests and has largely held up. In four and higher dimensions, there exists no robust counterexample of the conjecture for asymptotically AdS black holes in Einstein gravity.\footnote{The `super entropic' black holes first studied in~\cite{Hennigar:2014cfa} were initially thought to be a counterexample to the inequality. However, it was later realized that these black holes have degenerate thermodynamics, calling into question their status as a counterexample to the inequality~\cite{Hennigar:2018cnh, Appels:2019vow}. Moreover, beside this point, those solutions are actually asymptotically \textit{locally} AdS.} In three-dimensions, the classical charged BTZ metric is a potential counterexample~\cite{Johnson:2019wcq}. However, as already explained in~\cite{Frassino:2015oca}, this case is ambiguous and depends on how the length scale associated with the charge is defined. 

In~\cite{Frassino:2024bjg}, a quantum generalization of the reverse isoperimetric inequality was proposed and tested against all known three-dimensional quantum black holes. The quantum reverse isoperimetric conjecture states,
\begin{equation}\label{eq:quantum_RII}
    \mathcal{R}_{\mathcal{Q}} \equiv \left( \frac{(D-1) V_{\rm th}}{\Omega_{D-2}}\right)^{\frac{1}{D-3}} \left(\frac{\Omega_{D-2}}{4 \mathcal{G}_D S_{\rm gen}} \right)^{\frac{1}{D-2}} \ge 1 \,  \quad \text{for thermally stable BHs.}
\end{equation}
In understanding the ingredients in this formula, it is important to recall how the thermodynamic volume is defined. The thermodynamic volume has a zero point ambiguity that is directly tied to the zero point ambiguity of the mass. In higher dimensions, the ``interesting'' thermodynamic volume is the one that is obtained when the mass is defined according to the Ashtekar-Magnon-Das (AMD) prescription, which assigns zero mass to the global AdS vacuum~\cite{Cvetic:2010jb}. This is equivalent to assigning zero thermodynamic volume to the vacuum. The conventions used in three-dimensions do not adhere to this prescription, and we must subtract a Casimir contribution from the volume,
\begin{equation} 
V_{\rm th} = V - V_{\rm cas} \, , \quad V_{\rm cas} = \left(\frac{\partial M_{\rm cas}}{\partial P_D} \right)_{S_{\rm gen}, \dots} \, .
\end{equation}
For the quantum black holes considered here, we have
\begin{equation} 
M_{\rm cas} = -\frac{1}{8 \mathcal{G}_3} \quad  \Rightarrow  \quad V_{\rm cas} = - \frac{\pi \nu^2 \ell_3^2}{2} \, .
\end{equation}

We have tested the quantum reverise isoperimetric inequality for the charged and rotating quantum black holes. We have done this numerically, using the same constraints as for the Penrose inequality. That is, we demand positive temperature, and select our data such that $z > 0$, $\nu > 0$ along with positivity of the quantities in eq.~\eqref{eq:dimensionless expressions}. We also demand that $\bar{a}^2 < 1$, since this is implicitly assumed in the derivation of the coordinate chart and subsequent thermodynamic expressions. 

In~\cite{Frassino:2024bjg} it was found that there exist some black holes that violate~\eqref{eq:quantum_RII} but are thermally unstable (adhering to Johnson's revised version of the classical reverse isoperimetric conjecture~\cite{Johnson:2019mdp}). Here we find that such examples actually have $\bar{a}^2 > 1$, and are therefore pathological.  In summary, we find strong numerical evidence that the quantum reverse isoperimetric inequality holds robustly for these solutions. When $\bar{a}^2 < 1$ and the other physicality constraints hold, we find not a single violation of the inequality. 

\section{Gyromagnetic ratio}\label{ejkrio34}

It is possible to make a preliminary investigation into the gyromagnetic ratio of the quantum black hole. In the appropriate coordinates $(\bar{t},\bar{r},\bar{\phi})$ where the metric is manifestly asymptotically $\mathrm{AdS}_3$, the expression for the magnetic dipole moment may be extracted from the asymptotic expansion of the angular component of the gauge field on the brane as
\begin{align}
    \bar{\mathcal{A}}_{\bar{\phi}}\sim \frac{\mu_{m}}{\bar{r}}\,,\quad \bar{r}\rightarrow\infty\,.
\end{align}
Using the gauge field around the black hole (\ref{eq:gauge field on brane}), we find that 
\begin{equation}\label{eq:magnetic dipole}
    \mu_{m}=\frac{2\bar{a}\sqrt{1-\bar{a}^{2}}q\ell\ell_{3}\Delta^{2}}{\ell_{*}}\,.
\end{equation}

We define the gyromagnetic ratio $g_m$ of the quantum black hole via 
\begin{equation}
    \mu_{m}=g_{m}\frac{g_{4}QJ}{8\pi(M-M_{\text{cas}})}\,,
\end{equation}
where $M, J, Q$ are given in \eqref{eq:mass}, \eqref{eq:angular momentum}, and \eqref{eq:charge}, respectively. This formula takes into account the fact that the definition for the electric charge of the black hole used in this paper differs from that of the standard literature by a factor of $4\pi/g_{4}$. The standard formula $\mu_{m}=g_{m}QJ/2M$ is appropriate when the mass of the black hole is defined such that the vacuum has zero energy. This is not the case here, where the vacuum has an energy contribution of $M_{\text{cas}}=-1/8\mathcal{G}_{3}$ which must be substracted to ensure the ratio is positive. 

We have verified that the above definition of the gyromagnetic ratio passes a non-trivial consistency check. Specifically, when one carefully takes the $\ell \to \infty$ limit of the charged and rotating C-metric, the result is the Kerr-Newman(-AdS) black hole --- see Appendix~\ref{app:knads}. In this limit, our expression for the gyromagnetic ratio reproduces well-established results in the literature, e.g. limiting to $g_m = 2$ for the Kerr-Newman solution~\cite{Wald:1974np}.

Evaluating the above ratio for the thermodynamic quantities we have computed earlier, we obtain
\begin{equation}
    g_{m}=\frac{2 g_4 \sqrt{1-\bar{a}^2} \left(4 \Delta ^2 \bar{a}^2+x_1^2 \left(1-\Delta ^2 \kappa  \left(\bar{a}^2+1\right)\right)\right)}{\Delta  x_1^4 \left(\mu +q^2 x_1\right)}.
\end{equation}
This in turn can be written in terms of the dimensionless variables $z$, $\nu$, $\chi$, and $\alpha$, as well as the gauge constant on the brane $g_{3}$. In the limit of large backreaction, this expression diverges as $\sqrt{\nu}$. Holding fixed the `bare' parameters $\mu, \bar{a}, q$ and examining each term in the expression shows that while $Q$ and $\mu_{m}$ are well behaved and finite in the limit of large backreaction, the mass and spin diverge like $M\sim\nu$, and $J\sim\sqrt{\nu}$. 

\begin{figure}
    \centering
    \includegraphics[width=0.6\linewidth]{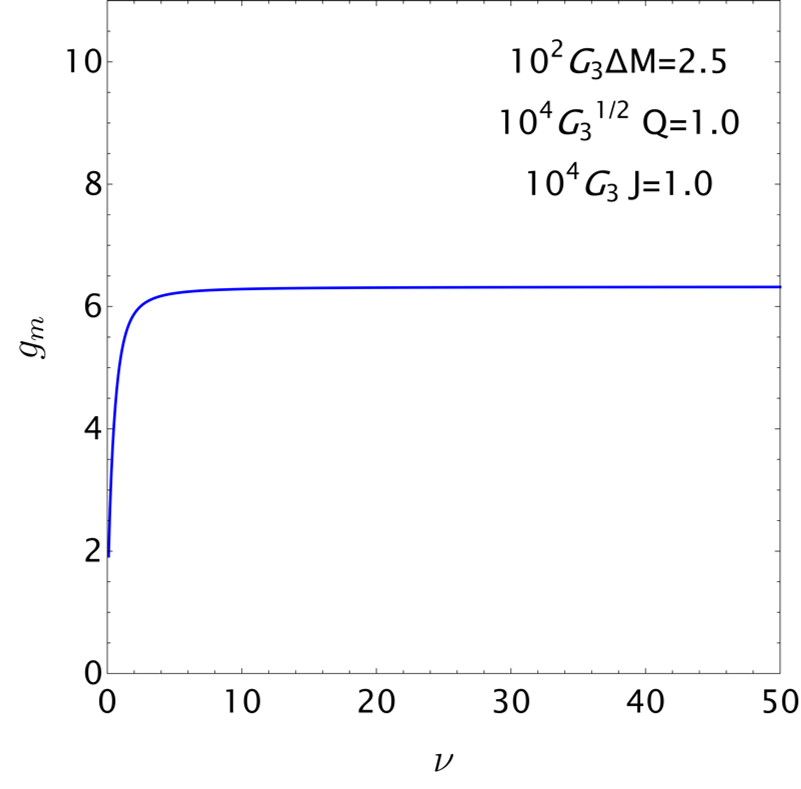}
    \caption{The renormalized gyromagnetic ratio of the quantum black hole on the brane as a function of the backreaction parameter $\nu$ for the displayed values of the Casimir subtracted mass $\Delta M$, charge $Q$ and spin $J$. The structure of this curve is insensitive to the fixed physical values that an observer would measure, of the black hole on the brane. The gauge coupling $g_{3}$ has been set to unity.}
    \label{fig:gyromagnetic ratio}
\end{figure}

Of course, the `bare' quantities are not physical, and so we should instead examine the behaviour of the gyromagnetic ratio as a function of the physical mass, charge and angular momentum as the backreaction varies. Unfortunately, the necessary analysis cannot be carried out analytically, and therefore we study this problem numerically. By fixing physically measurable quantities such as the mass difference $M-M_{\text{cas}}$, electric charge $Q$, and angular momentum $J$, we let the underlying geometric parameters $z$, $\chi$, and $\alpha$ vary with the backreaction parameter $\nu$. 
% In this sense, $\nu$ plays the role of a renormalization scale, analogous to energy scales in quantum field theory, albeit of an unorthodox type, since it is also itself a scale of physical significance. 
Due to the algebraic complexity of the parameter dependencies, the necessary inverse relationships, e.g. $z(\nu, Q, J, M)$, etc., must be obtained numerically.  The resulting numerically computed gyromagnetic ratio as a function of the backreaction parameter is presented in Fig. \ref{fig:gyromagnetic ratio}. A significant result of this analysis is that the structure of the $g_m$ curve proves robust against variations of fixed physical parameters. Importantly, for large values of backreaction parameter $\nu$, the renormalized gyromagnetic ratio exhibits a clear universal behavior, asymptotic to $2\pi$. A good approximation to the curve is $2 \pi  \sqrt{\nu }/{(\nu ^2+1)^{1/4}}$.

\section{Discussion}\label{jfi389}

We studied aspects of the thermodynamic properties of the charged rotating quantum black holes in the braneworld. The charged rotating quantum black hole can be obtained by incorporating an ETW brane in the charged rotating AdS C-metric. The quantum black hole then serves as an exact solution to the semi-classical gravitational field equation on the ETW brane, which includes the quantum backreaction from the quantum matter on the brane at all orders.

First, we calculated the thermodynamic quantities, such as the mass, and generalized entropy for the black holes. The mass of the black hole is determined by the  conical deficit of the three-dimensional geometry. The generalized entropy of the black hole incorporates all-order of backreaction from the holographic CFT on the brane. Working with the methods of black hole chemistry, where the cosmological constant is treated as a thermodynamic variable, we found both the thermodynamic first law and its integral form (Smarr formula) for the quantum black hole, from three perspectives: the brane, the bulk, and the boundary. From the brane perspective, where there is a quantum black hole coupled to quantum matter, the central charge of the holographic quantum field was considered as a thermodynamic variable; from the bulk perspective, where there is a tensional brane, the tension of the brane was taken to be variable; from the boundary perspective, we studied the thermodynamics of the two-dimensional DCFT.

A numerical analysis of thermodynamic phase diagrams allowed us to identify second-order critical points and calculate their associated critical exponents. The neutral-static quantum BTZ black hole's re-entrant phase transitions were found to be characterized by unorthodox critical exponents $\tilde{\alpha}=0,\, \beta=1,\, \gamma=2$ and $\delta=3$, which, however, still satisfy classical thermodynamic scaling relations--the Rushbrooke inequality and the Widom relation. Notably, to the best of our knowledge, this is the first and only example of a black hole in General Relativity with critical exponents outside of the mean field theory class. While similar non-standard critical exponents have indeed been observed elsewhere, these examples always require modifications of Einstein's theory~\cite{Dolan:2014vba, Hennigar:2016ekz, Dykaar:2017mba, Hu:2024ldp}. However, it must be emphasized that here the corresponding black holes do not dominate the thermodynamic ensemble, as they have positive free energy. We found that the introduction of any electric charge or spin into the system changes the thermodynamics of the black hole system such that to eliminate the re-entrant phase transitions, and introduce non-re-entrant phase transitions which have conventional critical behavior, with the van der Waals type orthodox critical exponents of mean-field theory $\tilde{\alpha}=0,\, \beta=1/2,\, \gamma=1$ and $\delta=3$. This was confirmed across all three perspectives with consistent  results. 

The braneworld approach offers a mechanism by which one can solve the equations of semiclassical gravity via holography. However, the semiclassical approximation will remain valid only insofar as the quantum fluctuations of the metric remain under control. This is particularly relevant in our case because we have seen it is possible to construct \textit{extremal} quantum black holes. Generally speaking, a key development of the last decade has been the understanding that the semiclassical approximation cannot be fully trusted for low temperature horizons. To study the limitations of the semiclassical approximation, we used the mass gap $M_{\rm gap}$ near extremality which defines a temperature scale above which the semiclassical approximation can be relied on. 

Concerning extremal solutions, we showed that in the presence of rotation there are two branches of extremal horizons while in the presence of charge alone there is a single extremal branch. Of the two branches of extremal solutions for (charged)-rotating black holes, one behaves analogous to familiar higher-dimensional extremal rotating solutions, while the other is highly exotic. In particular, while we demonstrated that (upon imposing physicality conditions) $M_{\rm gap}$ is always positive for the `sensible' extremal solutions, it can become negative for the exotic ones. Since $M_{\rm gap}$ plays the role of a coupling in the Schwarzian action that governs the fluctuations of quantum zero modes of extremal black hole throats, this suggests likely semiclassical instabilities of the exotic extremal solutions. We also speculated about the connection between this observation and recent results in weak cosmic censorship `Gedankenexperiments'. In that context, it has been shown via completely unrelated methods, that the positivity of $M_{\rm gap}$ is a sufficient condition to ensure that \textit{classical} black holes cannot be overspun or overcharged. Hence, there may be a connection between the outcome of classical weak cosmic censorship gedankenexperiments and the quantum stability of extremal horizons.

In the same vein, we also used the charged-rotating quantum black hole as a testbed for recent proposals of quantum inequalities that generalize certain geometric inequalities in classical general relativity. A three-dimensional quantum Penrose inequality was proposed in~\cite{Frassino:2024bjg} and tested against charged and rotating quantum black holes, finding agreement. Here, for the charged-rotating quantum black hole,  the complexity of the expressions involved precluded the ability to directly confirm that the inequality is always satisfied, but a wide-spread sampling of the parameter space found no violations. More significantly, here we proposed and tested a \textit{generalized} quantum Penrose inequality that bounds the mass in terms of both the generalized entropy \textit{and} the angular momentum. While such generalized Penrose inequalities have long been studied classically, we believe our proposal is the first of its kind for semiclassical gravity (certainly the first to be tested \textit{nonperturbatively}). In classical general relativity, the Penrose inequality plays an important role in the weak cosmic censorship conjecture. The hope is that their quantum generalizations can serve as a guiding light for quantum weak cosmic censorship~\cite{Frassino:2025buh}. Beyond the Penrose inequality and its generalized version, we also examined the quantum reverse isoperimetric inequality, proposed in \cite{Frassino:2024bjg}, with the inclusion of charge and spin. Numerical tests confirm the validity of the inequality for all physically admissible quantum black holes. Violations of this inequality were found in \cite{Frassino:2024bjg}, but noted to be thermally unstable. We find that these violations actually occur within the unphysical region of the parameter space and, hence, the quantum reverse isoperimetric inequality is true in all known examples.

Finally, we investigated the gyromagnetic ratio $g_m$ of the charged and rotating quantum black hole, defined by obtaining the magnetic dipole moment from the asymptotic behavior of the gauge field on the brane, and studied its behaviour as a function of the backreaction parameter. To ensure physicality of the result, we fixed the Casimir subtracted mass, electric charge, and spin leaving only the backreaction parameter to vary. Under these constraints, $g_m$ is finite as $\nu\rightarrow\infty$, with the limiting value of $2\pi$. This is a robust feature, in that adjusting the Casimir subtracted mass, charge, and spin of the black hole does not change the limiting behaviour of the curve, which is approximated well by $2 \pi  \sqrt{\nu }/{(\nu ^2+1)^{-1/4}}$.

\paragraph{Note.} When we were in the final stages of preparing this manuscript, Ref.~\cite{Cui:2025qdy} appeared on the arXiv which has some overlap with our section 3.

\acknowledgments

RAH is grateful to Roberto Emparan, Antonia Frassino, Hong L{\"u}, and Andrew Svesko for helpful discussions and correspondence. This work was supported in part by the Natural Sciences and Engineering Research Council of Canada. The work of MZ received support from National Natural Science Foundation of China (Grant No. 12365010) and Chinese Scholarship Council Scholarship. Research at Perimeter Institute is supported in part by the Government of Canada through the Department of Innovation, Science and Economic Development and by the Province of Ontario through the Ministry of Colleges and Universities.

\appendix 

\section{Explicit expressions}\label{section:A}

\subsection{Brane volume and chemical potential}
The thermodynamic volume on the brane, $V_{3}$ is given by
\begin{align}
    V_{3}=&\frac{2 \pi  \ell_{3}^{2}}{O_{V}^2}\sum_{n=0}^{8}N_{V,n}z^{n}\,,
\end{align}
where
\begin{align}
N_{V, 0}= & \alpha^2 \nu^2 \\
N_{V, 1}= & -2 \alpha^2\left(\alpha^2+1\right) \nu \\
N_{V, 2}= & 2 \alpha^2\left(\alpha^2+\nu^2+2\right)-\nu^2+2 \\
N_{V, 3}= & \nu\left(2 \alpha^4\left(\nu^2+2\right)+\alpha^2\left(2\left(\nu^2+1\right) \chi^2-\nu^2+6\right)-\nu^2+4\right) \\
N_{V, 4}= & \alpha^2\left(2 \alpha^2\left(\nu^2\left(1-\chi^2\right)-2\right)+2 \nu^4 \chi^2+7 \nu^2-4\right) \\
N_{V, 5}= & \nu\left(2 \alpha^4\left(-\nu^2 \chi^2+\chi^2-3\right)+\alpha^2\left(-6 \nu^2+4 \chi^2-8\right)-\nu^2\left(\chi^2+3\right)\right) \\
N_{V, 6}= & -\nu^4\left(2 \chi^2+1\right)+2 \alpha^4\left(\nu^4\left(\chi^2-2\right)+\nu^2\left(3 \chi^2-2\right)+1\right) \\
& +\alpha^2 \nu^2\left(\nu^2\left(\chi^4+\chi^2-4\right)+2\left(\chi^2+1\right)^2\right)\nonumber \\
N_{V, 7}= & \nu\left(2 \alpha^4\left(3 \nu^2+\left(\nu^4+\nu^2-2\right) \chi^2+2\right)\right. \\
& \left.+\alpha^2 \nu^2\left(2\left(\nu^2+2\right) \chi^4-\left(\nu^2-1\right) \chi^2+3\right)-\nu^4 \chi^2\right)\nonumber \\
N_{V, 8}= & -\alpha^2 \nu^2\left(\left(\nu^2+2\right) \chi^2+1\right)\left(2 \alpha^2-\nu^2 \chi^2\right)
\end{align}
and
\begin{equation}
\begin{aligned}
O_V= & \alpha^2+z^2\left(\chi^2+z^2\left(\nu^2 \chi^2-\alpha^2\left(\left(\nu^2+1\right) \chi^2+3\right)\right)\right. \\
& \left.+2 \nu z\left(2 \alpha^2+\chi^2+1\right)+3\right)+1
\end{aligned}
\end{equation}
Let $f(x)=\sqrt{1+x^{2}}$. The chemical potential on the brane $\mu_{3}$ is
\begin{align}
    \mu_{3}=&\frac{1}{\ell_{3}O_{\mu}}\sum_{n=0}^{12}N_{\mu,n}z^{n}\,,
\end{align}
where
\begin{align}
    N_{\mu,0}=&2 \alpha ^2 \nu  f(\alpha )^2 f(\nu )^2 (f(\nu )+1)\,,\\
    N_{\mu,1}=& \begin{aligned}[t]&\alpha ^2 f(\alpha )^2 f(\nu )^2 \left(\alpha ^2 \left(-2 f(\nu )+\nu ^2-2\right)+\nu ^2 (2 f(\nu )+3)-2 (f(\nu )+1)\right)\,,
    \end{aligned}\\
    N_{\mu,2}=&\begin{aligned}[t]-2 \nu  f(\nu )^2 \left(\alpha ^6 (2 f(\nu )+3)+3 \alpha ^4 (f(\nu )+2)+\alpha ^2 (f(\nu )+4)+f(\nu )+2\right)\,,
    \end{aligned}\\
    N_{\mu,3}=&\begin{aligned}[t]&f(\nu )^2 \big(\alpha ^6 \left(\left(4 \nu ^2+6\right) f(\nu )-\nu ^2+6\right)+\alpha ^4 \big(\nu ^2 \left(\chi ^2 (6 f(\nu )+5)+4 f(\nu )-9\right)\\
    &+2 \left(\chi ^2+5\right) (f(\nu )+1)\big)+\alpha ^2 \big(\nu ^2 \left(\left(4 \chi ^2-2\right) f(\nu )+\chi ^2-15\right)\\
    &-2 \left(\chi ^2-3\right) (f(\nu )+1)\big)-\nu ^2 (4 f(\nu )+9)+2 (f(\nu )+1)\big)\,,
    \end{aligned}\\
    N_{\mu,4}=&\begin{aligned}[t]&\nu  f(\nu )^2 \big(2 \alpha ^6 \left(2 f(\nu ) \left(\nu ^2-\chi ^2+2\right)+\left(\nu ^2-2\right) \chi ^2+7\right)\\
    &+\alpha ^4 \left(2 f(\nu ) \left(5 \left(\nu ^2-1\right) \chi ^2+\nu ^2+8\right)+\nu ^2 \left(7 \chi ^2-9\right)-10 \chi ^2+28\right)\\
    &+\alpha ^2 \left(\nu ^2 \left(\left(8 \chi ^2-4\right) f(\nu )+3 \chi ^2-14\right)-6 \chi ^2 (f(\nu )+1)+16 f(\nu )+22\right)\\
    &-\nu ^2 (2 f(\nu )+5)+2 (f(\nu )+1)\big)\,,
    \end{aligned}\\
    N_{\mu,5}=&\begin{aligned}[t]&f(\nu )^2 \big(-\big(\alpha ^6 \big(\nu ^2 \left(\chi ^2 (16 f(\nu )+21)-12\right)\\
    &+2 \left(\chi ^2+4\right) (f(\nu )+1)+2 \nu ^4 \chi ^2\big)\big)\\
    &+\alpha ^4 \big(4 \nu ^4 \chi ^2 f(\nu )+\nu ^2 \left(-2 \chi ^2 (10 f(\nu )+13)+4 f(\nu )+29\right)\\
    &+2 \left(2 \chi ^2-7\right) (f(\nu )+1)\big)+\alpha ^2 \big(2 \nu ^4 \chi ^2 (2 f(\nu )+1)\\
    &+\nu ^2 \left(13-\chi ^2 (8 f(\nu )+15)\right)\\
    &-2 \left(\chi ^2+5\right) (f(\nu )+1)\big)-\nu ^2 \left(2 \left(\chi ^2+3\right) f(\nu )+\chi ^2+7\right)\\
    &+2 \left(\chi ^2-1\right) (f(\nu )+1)\big)\,,
    \end{aligned}\\
    N_{\mu,6}=&\begin{aligned}[t]&\nu  f(\nu )^2 \big(\alpha ^6 \big(\nu ^2 \left(4 \left(\chi ^2-3\right) f(\nu )-3 \chi ^2-4\right)\\
    &+2 \left(9 \chi ^2 (f(\nu )+1)-8 f(\nu )-11\right)\big)\\
    &+\alpha ^4 \big(\nu ^2 \left(\left(6 \chi ^4-4 \chi ^2-26\right) f(\nu )+4 \chi ^4-22 \chi ^2-13\right)\\
    &+4 \chi ^2 \left(\chi ^2+3\right) (f(\nu )+1)-32 f(\nu )-38\big)+\alpha ^2 \big(\nu ^2 \big(2 \left(\chi ^4-4 \chi ^2-13\right) f(\nu )\\
    &-\left(\left(\chi ^2+17\right) \chi ^2\right)-22\big)-2 \left(\chi ^4-5 \chi ^2+6\right) (f(\nu )+1)\big)\\
    &-\nu ^2 \left(\chi ^2 (6 f(\nu )+3)+8 f(\nu )+9\right)+2 \left(3 \chi ^2-1\right) (f(\nu )+1)\big)\,,
    \end{aligned}\\
     N_{\mu,7}=&\begin{aligned}[t]&f(\nu )^2 \big(\alpha ^6 \big(\nu ^4 \left(2 \chi ^2 (6 f(\nu )+1)-8 (f(\nu )+1)+\chi ^4\right)\\
     &+\nu ^2 \left(-2 \chi ^4 (f(\nu )+1)+\chi ^2 (16 f(\nu )+25)-12\right)\\
     &-2 \left(\chi ^2-4\right) (f(\nu )+1)\big)\\
     &+\alpha ^4 \big(\nu ^4 \left(2 \left(7 \chi ^4+\chi ^2-8\right) f(\nu )+6 \chi ^4-7 \chi ^2-16\right)\\
     &+\nu ^2 \left(\chi ^2 (34 f(\nu )+48)+12 f(\nu )+1\right)+2 \left(2 \chi ^2+7\right) (f(\nu )+1)\big)\\
     &+\alpha ^2 \big(nu ^4 \left(\chi ^4 (6 f(\nu )-3)-4 \chi ^2 (2 f(\nu )+1)-10 (f(\nu )+1)\right)\\
     &+\nu ^2 \left(-6 \chi ^4 (f(\nu )+1)+\chi ^2 (28 f(\nu )+25)+16 f(\nu )+19\right)\\
     &-6 \left(\chi ^2-1\right) (f(\nu )+1)\big)-\nu ^4 \left(\left(6 \chi ^2+2\right) f(\nu )+3 \chi ^2+2\right)\\
     &+6 \nu ^2 \chi ^2 (f(\nu )+1)\big)\,,
    \end{aligned}\\
    N_{\mu,8}=&\begin{aligned}[t]&\nu  f(\nu )^2 \big(\alpha ^6 \big(\nu ^4 \chi ^2 \left(4 f(\nu )-3 \chi ^2+8\right)\\
    &+\nu ^2 \left(\chi ^2 \left(-2 \chi ^2 (7 f(\nu )+9)+12 f(\nu )+35\right)+4 (5 f(\nu )+4)\right)\\
    &-2 \chi ^2 \left(2 \chi ^2+3\right) (f(\nu )+1)+16 f(\nu )+18\big)\\
    &+\alpha ^4 \big(\nu ^4 \chi ^2 \left(2 \chi ^2 (5 f(\nu )-1)+2 f(\nu )+11\right)\\
    &+\nu ^2 \left(2 \chi ^2 \left(-7 \chi ^2 (f(\nu )+1)+11 f(\nu )+18\right)+38 f(\nu )+37\right)\\
    &+2 \left(2 \left(\chi ^2-7\right) \chi ^2+7\right) (f(\nu )+1)\big)\\
    &+\alpha ^2 \big(\nu ^4 \chi ^2 \left(3 \chi ^2-2\right) (2 f(\nu )-1)\\
    &+\nu ^2 \left(2 \left(-4 \chi ^4+8 \chi ^2+7\right) f(\nu )-9 \chi ^4+9 \chi ^2+16\right)\\
    &-2 \left(\chi ^4+7 \chi ^2-2\right) (f(\nu )+1)\big)-\nu ^2 \chi ^2 \left(2 \left(\nu ^2-1\right) f(\nu )+\nu ^2-2\right)\big)\,,
    \end{aligned}\\
    N_{\mu,9}=&\begin{aligned}[t]&-\alpha ^2 f(\nu )^2 \big(\alpha ^4 \big(\nu ^4 \left(\left(6 \chi ^4+20 \chi ^2-8\right) f(\nu )+9 \left(\chi ^2+2\right) \chi ^2-8\right)\\
    &+\nu ^2 \left(-10 \chi ^4 (f(\nu )+1)+\chi ^2 (40 f(\nu )+47)+8 f(\nu )+5\right)\\
    &+2 \left(\chi ^2+3\right) (f(\nu )+1)+3 \nu ^6 \chi ^4\big)+\alpha ^2 \big(-2 \nu ^6 \chi ^4 (f(\nu )-2)\\
    &-\nu ^4 \left(2 \left(\chi ^6-6 \chi ^4-11 \chi ^2+4\right) f(\nu )+\chi ^6-11 \chi ^4-27 \chi ^2+8\right)\\
    &+\nu ^2 \left(40 \chi ^2 f(\nu )-2 \left(\chi ^2+3\right) \chi ^4 (f(\nu )+1)+14 f(\nu )+37 \chi ^2+17\right)\\
    &-6 \left(\chi ^2-1\right) (f(\nu )+1)\big)+\nu ^6 \chi ^4 (1-2 f(\nu ))+\nu ^4 \big(\chi ^4 (8 f(\nu )+11)\\
    &-2 (f(\nu )+1)+3 \chi ^2\big)+6 \nu ^2 \chi ^2 (f(\nu )+1) f(\chi )^2\big)\,,
    \end{aligned}\\
    N_{\mu,10}=&\begin{aligned}[t]&\alpha ^2 \nu  f(\nu )^2 \big(\alpha ^4 \big(\nu ^4 \chi ^2 \left(\chi ^2 (8 f(\nu )+7)-12 (f(\nu )+1)\right)\\
    &+\nu ^2 \left(\chi ^2 \left(\chi ^2 (14 f(\nu )+17)+4 f(\nu )-1\right)-12 (f(\nu )+1)\right)\\
    &-2 \left(\chi ^4-9 \chi ^2+2\right) (f(\nu )+1)\big)\\
    &+\alpha ^2 \big(\nu ^4 \chi ^2 \left(6 \left(\chi ^4-\chi ^2-1\right) f(\nu )+3 \chi ^4-5 \chi ^2-5\right)\\
    &+\nu ^2 \left(6 \chi ^6 (f(\nu )+1)+\chi ^2 (8 f(\nu )+13)-6 f(\nu )+2 \chi ^4-7\right)\\
    &+2 \left(2 \chi ^4+5 \chi ^2-1\right) (f(\nu )+1)\big)\\
    &+\nu ^2 \chi ^2 \left(-6 \chi ^2 f(\nu )^3+2 f(\nu )-3 \left(3 \nu ^2+2\right) \chi ^2+\nu ^2+2\right)\big)\,,
    \end{aligned}\\
    N_{\mu,11}=&\begin{aligned}[t]&\alpha ^2 f(\nu )^2 \big(\alpha ^4 \big(\nu ^6 \chi ^4 \left(4 f(\nu )-\chi ^2+5\right)\\
    &+\nu ^4 \chi ^2 \left(2 \left(-\chi ^4+\chi ^2+8\right) f(\nu )-3 \chi ^4+10 \chi ^2+14\right)\\
    &+\nu ^2 \left(-2 \chi ^6 (f(\nu )+1)+\chi ^2 (8 f(\nu )+7)+4 f(\nu )+5\right)\\
    &-2 \left(\chi ^2-1\right) (f(\nu )+1)\big)+\alpha ^2 \nu ^2 \chi ^2 \big(\nu ^4 \chi ^2 \big(\chi ^2 (6 f(\nu )+3)\\
    &-4 (f(\nu )+1)\big)+\nu ^2 \left(6 \chi ^4 (f(\nu )+1)+4 (f(\nu )+1)+3 \chi ^2\right)\\
    &+6 \chi ^2 (f(\nu )+1)\big)-\nu ^4 \chi ^4 \left(2 f(\nu )^3+3 \nu ^2+2\right)\big)\,,
    \end{aligned}\\
    N_{\mu,12}=&\begin{aligned}[t]&-\alpha ^4 \nu  \chi ^2 f(\nu )^2 \big(\nu ^2 \big(\left(\nu ^2+2\right) \chi ^4 \left(\left(\alpha ^2-1\right) \nu ^2+\alpha ^2\right)\\
    &+\chi ^2 \left(\alpha ^2 \left(2 \nu ^2+5\right)-3 \nu ^2-2\right)+5 \alpha ^2\big)\\
    &+2 f(\nu ) \left(\nu ^2 \chi ^2+1\right) \left(\alpha ^2 \left(2 \nu ^2+1\right)+\chi ^2 (\alpha -\nu ) (\alpha +\nu ) f(\nu )^2\right)+2 \alpha ^2 f(\chi )^2\big)\,,
    \end{aligned}
\end{align}
and
\begin{align}
    O_{\mu}=&\begin{aligned}[t]&(f(\nu )+1)^2 (\nu  z+1) \left(\alpha ^2 \left(z^2 \left(\nu  \chi ^2 z-1\right)+1\right)+1\right)\big(\alpha ^2+\alpha ^2 \chi ^2 z^4 f(\nu )^2\\
    &-z^2 \left(-3 \alpha ^2 z^2+4 \alpha ^2 \nu  z+(\chi +\nu  \chi  z)^2+2 \nu  z+3\right)-1\big)^2\,.
    \end{aligned}
\end{align}
\subsection{Boundary pressure}
The boundary pressure $P_{2}$ is given by
\begin{align}
    P_{2}=&\frac{1}{2\sqrt{2}G_{3}\ell\nu O_{P}}\sum_{n=0}^{12}N_{P,n}z^{n}\,,
\end{align}
where
\begin{align}
    N_{P,0}=&\begin{aligned}[t]&\alpha ^2 \nu ^2 f(\alpha )^2 f(\nu )^3 \left(7 f(\nu )-4 \nu ^2-7\right)\,,
    \end{aligned}\\
    N_{P,1}=&\begin{aligned}[t]&-\alpha ^2 \nu  f(\alpha )^2 f(\nu )^2 \big(\alpha ^2 \left(\nu ^2 (9-5 f(\nu ))-8 f(\nu )+\nu ^4+8\right)\\
    &+2 \nu ^2 \left(\nu ^2 (2 f(\nu )-3)+f(\nu )+1\right)-8 f(\nu )+8\big)\,,
    \end{aligned}\\
    N_{P,2}=&\begin{aligned}[t]&f(\nu )^2 \big(\nu ^2 \big(\alpha ^2 \left(\alpha ^4 \left(2-6 \nu ^2\right)+3 \alpha ^2 \left(\nu ^2+9\right)+17 \nu ^2+41\right)\\
    &+\left(\alpha ^2 \left(4 \left(\alpha ^4-2\right) \nu ^2+2 \alpha ^4-15 \alpha ^2-29\right)-5\right) f(\nu )+\nu ^2+9\big)\\
    &-8 f(\alpha )^6 (f(\nu )-1)\big)\,,
    \end{aligned}\\
    N_{P,3}=&\begin{aligned}[t]&\nu  f(\nu )^4 \big(\alpha ^6 \left(25 \nu ^2+32\right)+\alpha ^4 \left(\nu ^2 \left(22 \chi ^2+51\right)+8 \left(\chi ^2+9\right)\right)\\
    &+\alpha ^2 \left(\nu ^2 \left(11 \chi ^2+36\right)+8 \left(\chi ^2+8\right)\right)+3 \left(\nu ^2+8\right)\big)\\
    &-\nu  f(\nu )^3 \big(\alpha ^6 \left(12 \nu ^4+41 \nu ^2+32\right)+\alpha ^4 \big(\nu ^2 \left(12 \nu ^2 \left(\chi ^2+2\right)+26 \chi ^2+87\right)\\
    &+8 \left(\chi ^2+9\right)\big)+\alpha ^2 \left(\nu ^2 \left(4 \nu ^2 \left(\chi ^2+4\right)+15 \chi ^2+68\right)+8 \left(\chi ^2+8\right)\right)\\
    &+3 \left(5 \nu ^2+8\right)\big)\,,
    \end{aligned}\\
    N_{P,4}=&\begin{aligned}[t]&f(\nu )^4 \big(-2 \alpha ^6 \left(\nu ^4 \left(\chi ^2-12\right)+\nu ^2 \left(8 \chi ^2-7\right)+12\right)\\
    &+\alpha ^4 \left(\nu ^4 \left(26 \chi ^2+36\right)+\nu ^2 \left(37-8 \chi ^2\right)-48\right)+\alpha ^2 \big(7 \nu ^4 \left(3 \chi ^2+2\right)\\
    &+\nu ^2 \left(8 \chi ^2+60\right)-24\big)+2 \nu ^2 \left(\nu ^2+8\right)\big)-f(\nu )^3 \big(2 \alpha ^6 \big(6 \nu ^6+\nu ^4 \left(16-5 \chi ^2\right)\\
    &+\nu ^2 \left(1-8 \chi ^2\right)-12\big)+\alpha ^4 \big(4 \nu ^6 \left(3 \chi ^2+4\right)+\nu ^4 \left(22 \chi ^2+56\right)\\
    &+\nu ^2 \left(13-8 \chi ^2\right)-48\big)+\alpha ^2 \big(\nu ^6 \left(8 \chi ^2+4\right)+\nu ^4 \left(25 \chi ^2+46\right)\\
    &+8 \nu ^2 \left(\chi ^2+6\right)-24\big)+2 \nu ^2 \left(5 \nu ^2+8\right)\big)\,,
    \end{aligned}\\
    N_{P,5}=&\begin{aligned}[t]&\nu  f(\nu )^3 \big(\alpha ^6 \left(16 \nu ^4+68 \nu ^2+\left(4 \nu ^6+16 \nu ^4+11 \nu ^2-16\right) \chi ^2+64\right)\\
    &+\alpha ^4 \left(28 \nu ^4-2 \left(7 \left(\nu ^2+2\right) \nu ^2+16\right) \chi ^2+129 \nu ^2+128\right)\\
    &+\alpha ^2 \left(16 \nu ^4+73 \nu ^2-\left(4 \nu ^6+18 \nu ^4+29 \nu ^2+24\right) \chi ^2+72\right)\\
    &-f(\nu ) \big(\alpha ^6 \left(36 \nu ^2+\left(4 \nu ^4+19 \nu ^2-16\right) \chi ^2+64\right)\\
    &+\alpha ^4 \left(65 \nu ^2-2 \left(3 \nu ^4+6 \nu ^2+16\right) \chi ^2+128\right)\\
    &+\alpha ^2 \left(37 \nu ^2-\left(10 \nu ^4+17 \nu ^2+24\right) \chi ^2+72\right)\\
    &+4 \nu ^2 \chi ^2 f(\nu )+2 \nu ^2 \left(7-2 \chi ^2\right)\big)+8 \nu ^4+14 \nu ^2\big)\,,
    \end{aligned}
\end{align}
\newpage
\begin{align}
    N_{P,6}=&\begin{aligned}[t]&f(\nu )^4 \big(\alpha ^6 \left(\nu ^4 \left(51 \chi ^2-56\right)+\nu ^2 \left(64 \chi ^2-46\right)+24\right)\\
    &+\alpha ^4 \left(\nu ^4 \left(23 \chi ^4+76 \chi ^2-110\right)+\nu ^2 \left(16 \chi ^2 \left(\chi ^2+6\right)-107\right)+24\right)\\
    &+\alpha ^2 \nu ^2 \left(\nu ^2 \left(4 \chi ^4+65 \chi ^2-89\right)+8 \left(\chi ^4+7 \chi ^2-5\right)\right)+3 \nu ^4 \left(4 \chi ^2-7\right)\big)\\
    &-f(\nu )^3 \big(\alpha ^6 \left(28 \nu ^6 \left(\chi ^2-1\right)+\nu ^4 \left(83 \chi ^2-80\right)+\nu ^2 \left(64 \chi ^2-34\right)+24\right)\\
    &+\alpha ^4 \big(4 \nu ^6 \left(3 \chi ^4+10 \chi ^2-14\right)+\nu ^4 \left(31 \chi ^2 \left(\chi ^2+4\right)-166\right)\\
    &+\nu ^2 \left(16 \chi ^2 \left(\chi ^2+6\right)-95\right)+24\big)\\
    &+\alpha ^2 \nu ^2 \left(8 \nu ^4 \left(5 \chi ^2-6\right)+\nu ^2 \left(8 \chi ^4+93 \chi ^2-109\right)+8 \left(\chi ^4+7 \chi ^2-5\right)\right)\\
    &+3 \nu ^4 \left(4 \nu ^2 \left(\chi ^2-1\right)+4 \chi ^2-7\right)\big)\,,
    \end{aligned}\\
    N_{P,7}=&\begin{aligned}[t]&-\nu  f(\nu )^3 \big(\alpha ^6 \big(4 \nu ^6 \left(9 \chi ^2-4\right)+\nu ^4 \left(-5 \chi ^4+86 \chi ^2-12\right)\\
    &+\nu ^2 \left(-8 \chi ^4+13 \chi ^2+68\right)-40 \chi ^2+64\big)+\alpha ^4 \big(4 \nu ^6 \left(5 \chi ^2 \left(\chi ^2+3\right)-8\right)\\
    &+\nu ^4 \left(61 \chi ^4+132 \chi ^2-24\right)+\nu ^2 \left(32 \chi ^4+36 \chi ^2+101\right)-48 \chi ^2+72\big)\\
    &+\alpha ^2 \nu ^2 \left(24 \chi ^4 f(\nu )^2+\left(52 \nu ^4+74 \nu ^2+19\right) \chi ^2-5 \left(4 \nu ^4+3 \nu ^2-7\right)\right)\\
    &+f(\nu ) \big(\alpha ^6 \left(\nu ^4 \left(\chi ^4-70 \chi ^2+28\right)+\nu ^2 \left(8 \chi ^4-33 \chi ^2-36\right)+8 \left(5 \chi ^2-8\right)\right)\\
    &-\alpha ^4 \left(\nu ^4 \left(45 \chi ^4+100 \chi ^2-56\right)+\nu ^2 \left(32 \chi ^4+60 \chi ^2+65\right)-48 \chi ^2+72\right)\\
    &-\alpha ^2 \nu ^2 \left(\nu ^2 \left(12 \chi ^4+70 \chi ^2-35\right)+24 \chi ^4+19 \chi ^2+35\right)\\
    &+\nu ^4 \left(7-12 \chi ^2\right)\big)+\nu ^4 \left(12 \chi ^2 f(\nu )^2-4 \nu ^2-7\right)\big)\,,
    \end{aligned}\\
    N_{P,8}=&\begin{aligned}[t]&\alpha ^2 f(\nu )^4 \big(\alpha ^4 \big(\nu ^2 \big(\left(8-3 \nu ^2 \left(\nu ^2+4\right)\right) \chi ^4+68 \nu ^2\\
    &+\left(28 \nu ^4-63 \nu ^2-104\right) \chi ^2+46\big)-8\big)\\
    &+\alpha ^2 \nu ^2 \left(23 \nu ^4 \chi ^2 \left(\chi ^2+2\right)+\nu ^2 \left(41 \chi ^4-132 \chi ^2+116\right)+8 \left(\chi ^4-14 \chi ^2+4\right)\right)\\
    &+\nu ^4 \left(12 \left(\nu ^2+2\right) \chi ^4+\left(22 \nu ^2-39\right) \chi ^2+35\right)\big)\\
    &-f(\nu )^3 \big(\alpha ^6 \big(\nu ^2 \big(36 \nu ^4+92 \nu ^2+\left(20 \nu ^6-115 \nu ^2-104\right) \chi ^2\\
    &-\left(4 \nu ^6+11 \nu ^4+8 \nu ^2-8\right) \chi ^4+42\big)-8\big)+\alpha ^4 \nu ^2 \big(4 \left(16 \nu ^4+33 \nu ^2+8\right)\\
    &+\left(4 \nu ^6+47 \nu ^4+45 \nu ^2+8\right) \chi ^4+2 \left(18 \nu ^6-17 \nu ^4-94 \nu ^2-56\right) \chi ^2\big)\\
    &+\alpha ^2 \nu ^4 \left(24 \chi ^4 f(\nu )^2+10 \nu ^2 \left(\left(2 \nu ^2-1\right) \chi ^2+2\right)-39 \chi ^2+35\right)\\
    &+4 \nu ^6 \chi ^2 (f(\nu )-1) f(\nu )\big)\,,
    \end{aligned}\\
    N_{P,9}=&\begin{aligned}[t]&-\alpha ^2 \nu  f(\nu )^3 \big(\alpha ^4 \big(\nu ^2 \big(16 \nu ^4+12 \nu ^2-3 \left(24 \nu ^4+52 \nu ^2+19\right) \chi ^2\\
    &+\left(-4 \nu ^6+13 \nu ^4+43 \nu ^2+32\right) \chi ^4-43\big)+24 \left(\chi ^2-1\right)\big)\\
    &+\alpha ^2 \nu ^2 \big(4 \left(\nu ^4+3 \nu ^2+2\right) \chi ^6-2 \left(50 \nu ^4+86 \nu ^2+21\right) \chi ^2+4 \left(4 \nu ^4+3 \nu ^2-7\right)\\
    &+\left(-4 \nu ^6+33 \nu ^4+51 \nu ^2+8\right) \chi ^4\big)\\
    &+f(\nu ) \big(\alpha ^4 \big(-28 \nu ^4+31 \nu ^2+\left(3 \nu ^4-27 \nu ^2-32\right) \nu ^2 \chi ^4\\
    &+3 \left(40 \nu ^4+23 \nu ^2-8\right) \chi ^2+24\big)-\alpha ^2 \nu ^2 \big(8 \chi ^6 f(\nu )^2-6 \left(26 \nu ^2+7\right) \chi ^2\\
    &+28 \left(\nu ^2-1\right)+\left(\nu ^4+47 \nu ^2+8\right) \chi ^4\big)-\nu ^4 \left(4 \left(\nu ^2+2\right) \chi ^4-33 \chi ^2+7\right)\big)\\
    &+\nu ^4 \left(8 \chi ^4 f(\nu )^2-3 \left(8 \nu ^2+11\right) \chi ^2+4 \nu ^2+7\right)\big)\,,
    \end{aligned}\\
    N_{P,10}=&\begin{aligned}[t]&-\alpha ^2 \nu ^2 f(\nu )^2 \big(\alpha ^4 \big(\nu ^6 \chi ^2 \left(4 \left(8 \chi ^2-7\right) f(\nu )-53 \chi ^2+48\right)\\
    &+\nu ^4 \left(\chi ^2 \left(\chi ^2 (57 f(\nu )-64)-4 f(\nu )-25\right)-20 f(\nu )+36\right)\\
    &+\nu ^2 \left(\chi ^2 \left(\chi ^2 (3 f(\nu )+5)+101 f(\nu )-129\right)-40 f(\nu )+44\right)\\
    &-8 \left(2 \chi ^4-7 \chi ^2+1\right) (f(\nu )-1)\big)\\
    &+\alpha ^2 \nu ^2 \big(2 \nu ^4 \chi ^2 \left(6 \chi ^4 (f(\nu )-2)+2 \chi ^2 (4 f(\nu )-5)-10 f(\nu )+17\right)\\
    &+\nu ^2 \left(2 \left(18 \chi ^6+8 \chi ^4-5 \chi ^2-4\right) f(\nu )-48 \chi ^6-11 \chi ^4+14\right)\\
    &+\left(24 \chi ^6-9 \chi ^4+34 \chi ^2-14\right) (f(\nu )-1)\big)\\
    &+\nu ^4 \chi ^2 \left(\nu ^2 (7-4 f(\nu ))-7 f(\nu )+7\right)\big)\,,
    \end{aligned}\\
    N_{P,11}=&\begin{aligned}[t]&-\alpha ^4 \nu ^3 f(\nu )^3 \big(\alpha ^2 \big(4 \nu ^2+\left(3 \nu ^2+2\right) \left(4 \nu ^4-3 \nu ^2-16\right) \chi ^4\\
    &+\left(32 \nu ^4+66 \nu ^2+19\right) \chi ^2+7\big)\\
    &+f(\nu ) \big(-\alpha ^2 \left(\left(19 \nu ^2 \left(\nu ^2-2\right)-32\right) \chi ^4+\left(58 \nu ^2+19\right) \chi ^2+7\right)\\
    &+\nu ^2 \chi ^6 f(\nu ) \left(\alpha ^2-\left(\alpha ^2+24\right) f(\nu )\right)+\nu ^2 \chi ^2 \left(\left(2 \nu ^2+9\right) \chi ^2-14\right)\big)\\
    &+\nu ^2 \chi ^2 \left(-3 \left(2 \nu ^2+3\right) \chi ^2+8 \nu ^2+12 \left(\nu ^4+3 \nu ^2+2\right) \chi ^4+14\right)\big)\,,
    \end{aligned}\\
    N_{P,12}=&\begin{aligned}[t]&\alpha ^6 \left(-\nu ^4\right) \chi ^2 f(\nu )^3 \big(f(\nu ) \left(-\nu ^2 \chi ^4 (f(\nu )-1) f(\nu )+\left(22 \nu ^2+15\right) \chi ^2+7\right)\\
    &-4 \nu ^2-3 \left(4 \nu ^4+10 \nu ^2+5\right) \chi ^2-7\big)-4 \alpha ^4 \nu ^6 \chi ^6 f(\nu )^5 \left(-2 f(\nu )+\nu ^2+2\right)\,,
    \end{aligned}
\end{align}
and
\begin{align}
    O_{P}=&\begin{aligned}[t]&(f(\nu )-1) \sqrt{f(\nu ) (f(\nu )+1)} \left(f(\nu )-2 f(\nu )^2\right) (\nu  z+1)\\
    &\times\left(\alpha ^2 \left(z^2 \left(\nu  \chi ^2 z-1\right)+1\right)+1\right) \big(\alpha ^2+\alpha ^2 \chi ^2 z^4 f(\nu )^2\\
    &-z^2 \left(-3 \alpha ^2 z^2+4 \alpha ^2 \nu  z+(\chi +\nu  \chi  z)^2+2 \nu  z+3\right)-1\big)^2\,.
    \end{aligned}
\end{align}

\section{Tensionless limit: Kerr-Newman-AdS}
\label{app:knads}

In the tensionless limit $\ell \to \infty$ the acceleration of the C-metric vanishes and the metric limits to the corresponding member of the Kerr-Newman-AdS family of solutions. In the static case, this limit is straightforward and was discussed in~\cite{Emparan:2020znc}. The limit is more subtle in the rotating case and it is the purpose of this appendix to discuss the relevant details. We have used this limit to make contact between our calculations of the gyromagnetic ratio and the well-established result $g = 2$ for four-dimensional Kerr-Newman-AdS black holes. 

We begin with the C-metric and reproduce the metric and its functions here for convenience:
\begin{equation}
\begin{aligned}
{\rm d} s^2=
\frac{1}{\Omega^{2}}&\left[  -\frac{H(r)}{\Sigma(x, r)}\left(\mathrm{d}t+a x^2 \mathrm{d} \phi\right)^2+\frac{\Sigma(x, r)}{H(r)} \mathrm{d} r^2\right. \\
& \left.\,\,+r^2\left(\frac{\Sigma(x, r)}{G(x)} \mathrm{d} x^2+\frac{G(x)}{\Sigma(x, r)}\left(\mathrm{d} \phi-\frac{a}{r^2} \mathrm{d} t\right)^2\right)\right]\,,
\end{aligned}
\end{equation}
where
\begin{align}
    H(r)& =\frac{a^2}{r^2}+\kappa +\frac{\ell^2 q^2}{r^2}+\frac{r^2}{\ell_3^2}-\frac{\ell \mu }{r} \, , \quad 
    G(x)=\frac{a^2 x^4}{\ell_3^2}-q^2 x^4-\mu  x^3-\kappa  x^2+1\,,
    \nonumber\\
    \Sigma(x, r)&=1+\frac{a^2 x^2}{r^2}\,, \quad \Omega=1+\frac{xr}{\ell}\,,
\end{align}
and the gauge field takes the form
\begin{equation}
    A_\mu dx^\mu=\frac{2}{\ell_\star}\left(-\frac{\ell q r}{a^2 x^2+r^2}\,,\,0\,,\,0\,,\,-\frac{a \ell q r x^2}{a^2 x^2+r^2}\right)\,.
\end{equation}

It is only in the case $\kappa = +1$ the Kerr-Newman-AdS metric will be recovered, and henceforth we make this restriction. We rescale the parameters of the solution in the following way
\begin{align}
    \mu &= \frac{2m}{\ell x_1^3} \, ,\quad q = \frac{Q}{\ell x_1^2} \, , \quad a = \frac{\hat{a}}{x_1^2} 
\end{align}
and transform some of the coordinates,
\begin{align}
    t &= x_1 \hat{t} - \hat{a} \hat{\phi} \, , \quad r = \frac{\hat{r}}{x_1} \,, \quad \phi =  \frac{x_1 \hat{\phi}}{\Xi} \, .
\end{align}
We can then take the limit $\ell \to \infty$ and obtain a finite metric and gauge field. When taking the limit, we keep the four-dimensional parameters $\ell_4, G_4$ and $g_4$ fixed. To express the final result in Boyer-Lindquist form, we define $x = x_1 \cos \theta$ in terms of which the metric and gauge field read
\begin{align} \label{knads}
    {\rm d}s^2 &= - \frac{\Delta_r}{\Sigma} \left[{\rm d}{\hat t} - \frac{\hat{a} \sin^2 \theta}{\Xi} {\rm d}\hat{\phi} \right]^2 + \frac{\Sigma}{\Delta_r} {\rm d}\hat{r}^2 + \frac{\Sigma}{\Delta_\theta} {\rm d} \theta^2 + \frac{\Delta_\theta\sin^2\theta}{\Sigma} \left[\hat{a} {\rm d} \hat{t} - \frac{\hat{r}^2 +\hat{a}^2}{\Xi} {\rm d} \hat{\phi} \right]^2 \, ,
    \\
    A &= - \frac{2 Q \hat{r}}{\ell_\star \Sigma} \left[{\rm d}\hat{t} - \frac{\hat{a} \sin^2 \theta}{\Xi} {\rm d} \hat{\phi} \right] \, ,
\end{align}
where
\begin{align}
    \Delta_r = (\hat{r}^2 +\hat{a}^2) \left(1+\frac{\hat{r}^2}{\ell_4^2} \right) - 2 m \hat{r} + Q^2 \, , \quad \Delta_\theta = 1 - \frac{\hat{a}^2}{\ell_4^2} \cos^2 \theta \, , \quad \Sigma = \hat{r}^2 + \hat{a}^2 \cos^2\theta \, ,
\end{align}
and
\begin{align}
    \Xi = 1 - \frac{\hat{a}^2}{\ell_4^2} \, , \quad x_1 = \sqrt{1 + \frac{\hat{a}^2}{\ell_4^2}} \, .
\end{align}

The above parameter rescalings also allow one to recover the Kerr-Newman-AdS thermodynamics from the quantum black hole thermodynamics discussed in this manuscript. The only subtlety is that one must subtract the Casimir energy contribution from the mass,
\begin{equation} 
M_{\rm KNAdS} = \lim_{\ell \to \infty} M - M_{\rm cas} = \frac{m}{\Xi^2} \, .
\end{equation}
In terms of this Casimir subtracted mass, we find (upon using the above parameter rescalings) 
\begin{equation}
\lim_{\ell \to \infty} \frac{J}{M - M_{\rm cas}} = \hat{a} \, ,
\end{equation}
which, as required, matches the ratio of angular momentum to mass of the Kerr-Newman-AdS black hole.

Let us further note the following. If instead of starting from the metric~\eqref{eq:metric2} we start from the metric written in the `barred' coordinates of~\eqref{eq:coordinate transformation} then the following differences must be accounted for. The transformation of the $(t,\phi)$ coordinates of~\eqref{eq:coordinate transformation} result in the Kerr-Newman-AdS metric in a frame which does not rotate at infinity. That is, taking the limit in the barred $(t,\phi)$-coordinates yields a metric that is equivalent to \eqref{knads} up to the following coordinate transformation,
\begin{equation}
\hat{t} = \hat{T} \, , \quad \hat{\phi} = \hat{\Phi} - \frac{\hat{a}}{\ell_4^2} \hat{T}\,.
\end{equation} 
To understand the implications of the new radial coordinate introduced in~\eqref{eq:coordinate transformation}, we must recall that the Kerr-Newman-AdS metric written in the Boyer-Lindquist coordinates~\eqref{knads} is not manifestly asymptotically AdS as the metric components do not have the required fall off properties~\cite{Henneaux:1985tv}. Instead, to bring the metric into a manifestly asymptotically AdS form requires performing the following transformations to the Boyer-Lindquist coordinates as appearing in~\eqref{knads}: 
\begin{align}
    \hat{t} &= \hat{T} \, , \quad \hat{\phi} = \hat{\Phi} - \frac{\hat{a}}{\ell_4^2} \hat{T} \,, \quad 
    \hat{R} \cos \hat{\Theta} = \hat{r} \cos \theta  
    \,,
    \\
    \Xi \hat{R}^2 &= \hat{r}^2 + \hat{a}^2 \sin^2 \theta - \frac{\hat{a}^2}{\ell_4^2} \hat{r}^2 \cos^2 \theta \, ,
\end{align}
where $(\hat{T}, \hat{R}, \hat{\Theta}, \hat{\Phi})$ are the coordinates in the manifestly asymptotically AdS chart. Given this, one recognizes that $\bar{r}$ from~\eqref{eq:coordinate transformation} limits to $\hat{R}$ when restricting to the location of the brane, $\theta = \pi/2$.

\bibliographystyle{JHEP}

\bibliography{biblio.bib}

\end{document}